\documentclass{aa}  

\usepackage{graphicx}
\usepackage{xcolor}
\usepackage{colortbl}
\usepackage{tabu}
\usepackage{float}
\usepackage{lscape}

\usepackage{txfonts}
\usepackage[breaklinks=true]{hyperref} 

\usepackage{siunitx}
\DeclareSIUnit\year{yr}
\DeclareSIUnit\erg{erg}
\DeclareSIUnit\angstrom{\AA}
\DeclareSIUnit\parsec{pc}
\DeclareSIUnit\zsun{{\it Z}_\odot}
\DeclareSIUnit\lsun{{\it L}_\odot}
\DeclareSIUnit\msun{{\it M}_\odot}
\DeclareSIUnit\rsun{{\it R}_\odot}

\usepackage{tabularx}
\usepackage{tablefootnote}
\usepackage{tabto}
\usepackage{booktabs}

\newcommand{\ts}[1]{\textsuperscript{#1}}

\newcommand{\halpha}{H$\alpha$}

\newcommand{\hdelta}{H$\delta$}

\newcommand{\civ}{C\,{\sc iv}}
\newcommand{\ciii}{C\,{\sc iii}}
\newcommand{\siiv}{Si\,{\sc iv}}
\newcommand{\niv}{N\,{\sc iv}}
\newcommand{\niii}{N\,{\sc iii}}
\newcommand{\nv}{N\,{\sc v}}
\newcommand{\oiv}{O\,{\sc iv}}
\newcommand{\heii}{He\,{\sc ii}}
\newcommand{\hei}{He\,{\sc i}}

\newcommand{\tick}{\multicolumn{1}{c}{$\blacksquare$}}
\newcommand{\eks}{\multicolumn{1}{c}{$\square$}}

\newcommand{\fw}{\textsc{Fastwind}}
\newcommand{\powr}{\textsc{PoWR}}
\newcommand{\kiwi}{\textsc{Kiwi-GA}}
\newcommand{\boost}{\textsc{BoOST}}

\newcommand{\teff}{$T_{\rm eff}$}
\newcommand{\logg}{$\log g$}
\newcommand{\vrot}{$v\sin i$}
\newcommand{\vrad}{$v_{\rm rad}$}

\newcommand{\mdot}{$\dot{M}$}
\newcommand{\dmom}{$D_{\rm mom}$}
\newcommand{\vinf}{$v_{\infty}$}
\newcommand{\msun}{$M_\odot$}
\newcommand{\fcl}{$f_{\rm cl}$}
\newcommand{\microturb}{$\xi$}
\newcommand{\vclstart}{$v_{\rm cl,start}$}
\newcommand{\vclmax}{$v_{\rm cl,max}$}

\defcitealias{castor1975radiation}{CAK}

\usepackage{natbib,twoopt}
\bibpunct{(}{)}{;}{a}{}{,}
\makeatletter
\newcommandtwoopt{\citeads}[3][][]{\href{http://adsabs.harvard.edu/abs/#3}
{\def\hyper@linkstart##1##2{}
\let\hyper@linkend\@empty\citealp[#1][#2]{#3}}}
\newcommandtwoopt{\citepads}[3][][]{\href{http://adsabs.harvard.edu/abs/#3}
{\def\hyper@linkstart##1##2{}
\let\hyper@linkend\@empty\citep[#1][#2]{#3}}}
\newcommandtwoopt{\citetads}[3][][]{\href{http://adsabs.harvard.edu/abs/#3}
{\def\hyper@linkstart##1##2{}
\let\hyper@linkend\@empty\citet[#1][#2]{#3}}}
\newcommandtwoopt{\citeyearads}[3][][]
{\href{http://adsabs.harvard.edu/abs/#3}
{\def\hyper@linkstart##1##2{}
\let\hyper@linkend\@empty\citeyear[#1][#2]{#3}}}
\makeatother

\begin{document} 

   \title{The wind properties of O-type stars at sub-SMC metallicity}

   \author{Ciar\'{a}n Furey
          \inst{1,2}
          \and
          O. Grace Telford\inst{3,4}
          \and
          Alex de Koter\inst{2,5}
          \and
          Frank Backs\inst{5}
          \and
          Sarah A. Brands\inst{2}
          \and
          Jorick S. Vink\inst{1}
          \and
          Lex Kaper\inst{2}
          \and
          Jes\'{u}s Gomez-Mantecon\inst{6}
          \and
          Frank Tramper\inst{6}
          \and
          Miriam Garcia\inst{6}
          }

   \institute{Armagh Observatory, College Hill, Armagh BT61 9DG, UK\\
              \email{ciaran.furey@armagh.ac.uk}
         \and
             Anton Pannekoek Institute for Astronomy, University of Amsterdam, 1090 GE Amsterdam, The Netherlands
        \and
             Department of Astrophysical Sciences, Princeton University, 4 Ivy Lane, Princeton, NJ 08544, USA
        \and
            The Observatories of the Carnegie Institution for Science, 813 Santa Barbara Street, Pasadena, CA 91101, USA
        \and
            Institute of Astronomy, KU Leuven, Celestijnenlaan 200D, Leuven, Belgium
        \and
            Centro de Astrobiolog\'{i}a (CAB), CSIC-INTA, Carretera de Ajalvir km 4, 28850 Torrej\'{o}n de Ardoz, Madrid, Spain
             }

   \date{Received 22 November 2024; accepted 24 March 2025}
 
  \abstract
   {Powerful, radiation-driven winds heavily influence the evolution and end-of-life products of massive stars. Feedback processes from these winds strongly impact the thermal and dynamical properties of the interstellar medium of their host galaxies. The dependence of mass loss on stellar properties is poorly understood, particularly at low metallicity (Z).}
   {We aim to characterise global, photospheric and wind properties of hot massive stars in Local Group dwarf galaxies with metal contents below that of the Small Magellanic Cloud and confront our findings to theories of radiation-driven winds.}
   {We perform quantitative optical and ultraviolet spectroscopy on a sample of 11 O-type stars in nearby dwarf galaxies with $Z<0.2\,Z_\odot$. The stellar atmosphere code \fw{} in combination with the genetic algorithm \kiwi{} are used to determine the stellar and wind parameters. Clumpy structures present in the wind outflow are assumed to be optically thin.} 
   {The winds of the sample stars are very weak, with mass loss rates $\sim 10^{-9}-10^{-7}\,M_\odot\,{\rm yr}^{-1}$. Such feeble winds can only be constrained if ultraviolet spectra are available. The modified wind momentum as a function of luminosity ($L$) for stars in this $Z$ regime is in agreement with extrapolations to lower $Z$ of a recently established empirical relation for this quantity as a function of both $L$ and $Z$. However, theoretical prescriptions do not match either our results or those of other recent analyses at low luminosity ($L \lesssim 10^{5.2}\,L_{\odot}$) and low $Z$; in this regime, they predict winds that are stronger by an order of magnitude or more.}   
   {For our sample stars at $Z \sim 0.14\,Z_\odot$, with masses $\sim 30 - 50\,M_{\odot}$, stellar winds strip only little mass during the bulk of the main-sequence evolution. However, if the steep dependence of mass loss on luminosity found here also holds for (so far undiscovered) much more massive stars at these metallicities, these may suffer (almost) as severely from main-sequence mass stripping as well-known very massive stars in the Large Magellanic Cloud and Milky Way.}
   
   \keywords{Stars: early-type; Stars: massive; Stars: mass loss; Stars: winds, outflows; Galaxies: dwarf; Ultraviolet: stars}

   \maketitle

\section{Introduction}
\label{sec:intro}

\begin{table*}
\centering
{\caption{The sample of stars studied in this work and their photometric information.}\label{tab:phot}}
\begin{small}
\begin{tabular}
{l l l c c c c c c c}
\midrule\midrule
& & & & & \multicolumn{5}{c}{Photometry} \\ \cmidrule(lr){6-10}
Galaxy & Star & Spectral Type & RA(J2000) & DEC(J2000) & Band & $m_\lambda$ &  $E(B-V)$ & $A_\lambda$ & $M_\lambda$ \\
\midrule
IC1613 & 64066 & O3\,III((f)) & 01\ts{h}05\ts{m}02.07\ts{s} & $\ang[parse-numbers = false]{+02;09;28.10}$ & $V$ & 19.03\ts{a} & 0.07\ts{f} & 0.21 & $-5.47 \pm 0.03$ \\
   & A13 & O3-4\,V((f)) & 01\ts{h}05\ts{m}06.25\ts{s} & $\ang[parse-numbers = false]{+02;10;43.00}$ & $I$ & 19.26\ts{b} & 0.03\ts{g} & 0.05 & $-5.08 \pm 0.03$ \\
   & 62024 & O6.5\,IIIf & 01\ts{h}05\ts{m}00.65\ts{s} & $\ang[parse-numbers = false]{+02;08;49.26}$ & $V$ & 19.60\ts{c} & 0.15\ts{h} & 0.45 & $-5.14 \pm 0.03$ \\
   & B2 & O7.5\,III-V((f)) & 01\ts{h}05\ts{m}03.07\ts{s} & $\ang[parse-numbers = false]{+02;10;04.54}$ & $V$ & 19.62\ts{c} & 0.12\ts{h} & 0.36 & $-5.02 \pm 0.03$ \\
   & B11 & O9.5\,I & 01\ts{h}04\ts{m}43.80\ts{s} & $\ang[parse-numbers = false]{+02;06;44.75}$ & $I$ & 18.78\ts{b} & 0.08\ts{g} & 0.14 & $-5.64 \pm 0.03$ \\
WLM & A15 & O7\,V((f)) & 00\ts{h}02\ts{m}00.53\ts{s} & $\ang[parse-numbers = false]{-15;29;52.41}$ & F814W & 20.52\ts{d} & -- & 0.03\ts{i} & $-4.46 \pm 0.09$ \\    
   & A11 & O9.7\,Ia & 00\ts{h}01\ts{m}59.97\ts{s} & $\ang[parse-numbers = false]{-15;28;19.20}$ & $I$ & 18.57\ts{b} & 0.04\ts{g} & 0.07 & $-6.45 \pm 0.09$ \\
NGC\,3109 & N20$^\dagger$ & O8\,I & 10\ts{h}03\ts{m}03.22\ts{s} & $\ang[parse-numbers = false]{-26;09;21.41}$ & $I$ & 19.47\ts{e} & -- & 0.06 & $-6.17 \pm 0.03$ \\
  & N34$^\dagger$ & O8\,I(f) & 10\ts{h}03\ts{m}14.24\ts{s} & $\ang[parse-numbers = false]{-26;09;16.96}$ & $I$ & 19.86\ts{e} & -- & 0.22 & $-5.93 \pm 0.03$ \\
Sextans\,A & S3$^*$ & O9\,V & 10\ts{h}10\ts{m}58.19\ts{s} & $\ang[parse-numbers = false]{-04;43;18.45}$ & F153M & 21.40\ts{d} & -- & 0.05\ts{i} & $-4.32 \pm 0.08$ \\
Leo\,P & LP26 & O7-8\,V & 10\ts{h}21\ts{m}45.12\ts{s} & $\ang[parse-numbers = false]{+18;05;16.93}$ & F153M & 22.34\ts{d} & -- & 0.01\ts{i} & $-3.71 \pm 0.20$ \\
\midrule	 
\end{tabular}
\end{small}
\tablefoot{SIMBAD-resolvable identifiers for each star are given in Table \ref{tab:simbad_names}. $^\dagger$ The prefixes of the IDs of NGC\,3109 stars have been shortened from `NGC-3109-' as in, for example, \cite{tramper2011mass, tramper2014properties} to `N' in this work. $^*$ s029 in \cite{2022MNRAS.516.4164L}.}
\tablebib{
\textit{Magnitudes:} (a) \cite{garcia2013young}; (b) \cite{bouret2015spectro}; (c) \cite{garcia2014winds}; (d) \cite{telford2021far}; (e)  \cite{evans2007araucaria}. \textit{Colour excesses:} (f) \cite{garcia2013young}, \cite{martins2006ubvjhk}; (g) \cite{bouret2015spectro}; (h) \cite{garcia2013young}. $A_\lambda$: (i) $A_V$ for each star from \cite{telford2021far}.}
\end{table*}

Most stars that have accumulated at least 8\,$M_{\odot}$ at the end of the formation process will end as core-collapse supernovae \citep{poelarends2008supernova}. They are termed `massive stars' and are further characterised by strong feedback on their surroundings during the entirety of their evolution. For the majority of their lives, they release copious amounts of H-ionising photons, forming circumstellar H\,{\sc ii} regions of large dimensions that expand into their surroundings and sweep up interstellar gas \citep[e.g.][]{geen2015detailed}. Core collapse supernovae, especially those that leave neutron stars, instantaneously inject enormous amounts of energy and gas into the interstellar medium \citep[e.g.][]{2024ApJ...969...57S}, enriching it with products of nucleosynthesis, driving galactic chemical evolution \citep[e.g.][]{2020ApJ...900..179K}. Moreover, massive stars drive strong stellar winds that inject kinetic energy and nuclear-processed materials (predominantly He and CNO processed elements) into their environment, creating hot X-ray emitting bubbles \citep[e.g.][]{2008Sci...319..309G}. Mass lost through stellar winds also strongly affects the evolution of the massive star itself, potentially impacting the series of morphological phases the star experiences and, ultimately, the type and properties of the supernova and nature of the compact remnant \citep[e.g.][]{renzo2017systematic, vink2022theory}.

It is mainly radiation pressure on iron-group elements that drives the stellar wind mass loss, therefore these winds are anticipated to depend on metallicity ($Z$). The lowest metallicity regimes accessible for quantitative spectroscopy of isolated sources are those in (irregular) dwarf galaxies in the Local Group. Several of these have metallicities below that of the Small Magellanic Cloud \citep[SMC; which has $Z = 1/5\,Z_{\odot}$; see e.g.][]{2007A&A...473..603M}. Unfortunately, the star-formation activity in these galaxies is typically low, consequently, massive stars are rare (but see \citealp{2022MNRAS.516.4164L} for the case of Sextans\,A at $\sim 1/10\,Z_{\odot}$). Although analyses of sometimes dozens of B and A-type supergiants have been carried out \citep[e.g.][]{bresolin2006araucaria, bresolin2007vlt, hosek2014quantitative,berger2018quantitative,urbaneja2023metallicity}, detailed studies of the optical spectra of only approximately 10 O-type stars have been performed across all sub-SMC metallicity galaxies combined \citep{herrero2012peculiar,tramper2014properties,bouret2015spectro, telford2024observations}. 

Constraints on the mass loss rate ($\dot{M}$) of these stars are very difficult to obtain. First, because they are at distances of $\sim$\,1\,Mpc, the signal-to-noise ratios of the spectra are modest. Second, at sub-SMC metal content, O-type star winds are expected to be very weak. Empirically, $\dot{M}$ is found to scale with $Z^{0.5-0.8}$ \citep[e.g.][]{mokiem2007empirical,marcolino2022wind}, while theoretical predictions suggest a scaling with $Z^{0.5-0.7}$ \citep[e.g.][]{vink2001mass,2014A&A...567A..63K}. The winds are so weak, in fact, that the standard optical wind diagnostics (H$\alpha$ and He\,{\sc ii}\,$\lambda 4686$) become photospheric in nature and lose their diagnostic power \citep[at $\dot{M} \lesssim 10^{-7}\,M_{\odot}\,{\rm yr}^{-1}$;][]{bouret2015spectro}. To obtain accurate wind parameters for these stars, it is thus essential to simultaneously analyse both the optical and ultraviolet (UV) spectral regimes, the latter containing the extremely wind-sensitive resonance lines of C\,{\sc iv}, N\,{\sc iv}, and Si\,{\sc iv} (probing rates as low as $10^{-9}\,M_{\odot}\,{\rm yr}^{-1}$).  So far, this has only been attempted on small samples of $\sim 3$ stars \citep{bouret2015spectro, telford2024observations}. The main goal of this study is to constrain the wind properties of a sample of 11 of these sub-SMC $Z$ O-stars (with $M \gtrsim \SI{20}{\msun}$) using quantitative optical and UV spectroscopy.

An extra complication of the analysis of hot star winds is that the outflows may not be completely smooth and can instead contain overdense `clumps' of material. The contrast of the square of the density in the clumps and the mean wind density squared (i.e. the clumping factor, $f_{\rm cl}$) is found to range between $\sim 10-40$ for Galactic and LMC stars \citep[e.g.][]{hawcroft2021empirical,brands2022r136}. In fact, when determined from $\rho^2$ diagnostics alone (e.g. optical recombination lines), there is a degeneracy between \mdot{} and \fcl{} \citep{fullerton2006discordance}. This means that relying only on H$\alpha$ (and, if present, He\,{\sc ii}\,$\lambda 4686$) and {\em not} accounting for wind clumping would lead to an over-estimation of \mdot{} by about a factor $\sqrt{f_{\rm cl}} \sim 3-7$. Including the UV resonance lines -- as we will do in this study -- allows one to break this degeneracy, as the strength of these lines depend only linearly on density, thus yielding both clumping properties and clumping-corrected mass loss rates. 

Recent empirical findings on the dependence of mass loss on metallicity suggest that a $Z^{x}$ dependence may not adequately capture this relationship at relatively low luminosities ($L$) and metallicities, rather an $L$ dependence on $x$ should also be considered \citep{ramachandran2019testing,rickard2022stellar,backs2024smc}. This could possibly explain the findings of previous studies \citep[e.g.][]{martins2004puzzling, martins2005stars, marcolino2009analysis} that found, in this low $L$ regime, discrepancies between observation and theoretical prediction. These authors discuss this in the context of the `weak-wind' problem, which manifests as a steepening of the $\dot{M}(L)$ relation compared to theory at low $L$ ($L \la 10^{5.2}\,L_\odot$; \citealp{vink2022theory}) in Milky Way and SMC O stars.

The impact of this added complexity in mass loss behaviour may be large, strongly impacting stellar evolution and wind feedback processes of the most massive stars in metal-poor dwarf galaxies, potentially even of the very first stars that formed in the universe \citep{hirano2014one}. We therefore here also aim to test the proposed $\dot{M} \propto Z^{x(L)}$ behaviour in the metallicity range $\sim 0.13 - \SI{0.16}{\zsun}$.

This paper is outlined as follows: in Sect. \ref{sec:data} we present the sample of 11 O-type stars in five different dwarf galaxies and the optical and UV data used. In Sect. \ref{sec:methods} we describe our methodology. In Sect. \ref{sec:results} we present the results we find; a discussion of which is given in Sect. \ref{sec:discussion}, and we conclude this study in Sect. \ref{sec:summary}.

\section{Data acquisition and preparation}
\label{sec:data}

In this section, we introduce the sample and discuss the data quality.  We provide photometric information about the stars in Sect. \ref{subsec:phot}, as well as outline the quality and sources of the UV and optical data in Sects. \ref{subsec:uv_spectra} and \ref{subsec:optical_spectra}, respectively. We conclude in Sect. \ref{subsec:contnorm} by outlining the methods employed to normalise the spectra.

\subsection{The sample}
\label{subsec:sample}
In Table \ref{tab:phot} we present the sample of stars (for SIMBAD-resolvable identifiers, see Table \ref{tab:simbad_names}). They are located in dwarf galaxies in and even beyond the Local Group with estimated metal contents lower than that of the SMC. The host galaxy properties are given in Table \ref{tab:gals}. The majority of the sample is located in IC\,1613, the closest galaxy to the Milky Way of those in the sample. The stars are all of spectral class O, are mostly of late spectral type, and sample the entire range of luminosity classes, from supergiants to dwarfs.

In Table \ref{tab:gals}, we present the metallicity ($Z$) of each galaxy. For the stars in each galaxy, we adopt this value as their metallicity for the rest of this work. However, the methodologies employed to determine the value of the metallicity that we quote are different between galaxies. For IC\,1613 and NGC\,3109, the abundances quoted are those of oxygen derived from spectroscopic analyses of blue supergiants (BSGs), while for WLM, Sextans\,A, and Leo\,P, we cite values derived from nebular oxygen emission. As a full accounting of all metallic species is not available for these galaxies, we adopt the $Z$ values given in Table \ref{tab:gals} as representative of the stars in our sample. We note that this introduces uncertainty, as metallicity gradients may be present in some of these galaxies (as has been found by \citealp{berger2018quantitative} in IC\,1613) and the oxygen-to-iron abundance ratio may not be the solar value \citep{garcia2014winds}. Additionally, the metallicity of O-type stars in IC\,1613 and WLM is a topic of debate, with \citet{garcia2014winds} and \citet{bouret2015spectro} motivating that the Fe abundance of these galaxies is possibly more SMC-like.  Though we adopt the value listed in Table \ref{tab:gals}, we revisit this assumption in Sects\,\ref{subsec:lowz_talk} and\,\ref{subsec:assumptions}.

\subsection{Photometry}
\label{subsec:phot}

\begin{table}
 \centering
{\caption{Host galaxy properties and the number of sample stars in each.}\label{tab:gals}}
\begin{tabular}
{l c c c c c}
\midrule
\midrule
Galaxy &Distance (Mpc)& $Z/Z_{\odot}$     & \# Stars     \\
\midrule
IC\,1613 &$\num{0.72(1)}$\ts{a} & \num{0.16(3)}\ts{f}  & 5    \\
 WLM  &$\num{0.98(4)}$\ts{b} & \num{0.14(2)}\ts{g} & 2  \\
 NGC\,3109 &$\num{1.30(2)}$\ts{c}  &  \num{0.12(2)}\ts{h} &2 	\\
 Sextans\,A  &$\num{1.38(5)}$\ts{d} &  \num{0.06(1)}\ts{i} &  1  \\
 Leo\,P  &$\num{1.62(15)}$\ts{e}  & \num{0.03(1)}\ts{j}  & 1  \\
\midrule	 
\end{tabular}
\tablebib{\textit{Distances}: (a) \citet[cepheids]{pietrzynski2006araucaria}; (b) \citet[TRGB]{jacobs2009extragalactic}; (c) \citet[cepheids]{soszynski2006araucaria}; (d) \citet[TRGB]{dalcanton2009acs}; (e) \citet[HB+RR Lyrae]{mcquinn2015leo}. \textit{Metallicities}: (f) \citet[BSGs]{bresolin2007vlt}; (g) \citet[nebular O]{lee2005investigating}; (h) \citet[BSGs]{evans2007araucaria}; (i) \citet[nebular O]{skillman1989oxygen}; (j) \citet[nebular O]{skillman2013alfalfa}.}
\end{table}
To determine the luminosity of our stars, we need an anchor magnitude. In order to minimise the effects of extinction, we used, for each star, the reddest possible filter that was available in the literature. In a few cases, magnitudes in the reddest available filter (F153M or F814W) of Hubble Space Telescope (HST) photometry were used. Otherwise, $I$ (equivalent to F814W magnitude), and, if not available, $V$ band magnitudes were used. Table \ref{tab:phot} provides the magnitudes in these bands, $m_\lambda$.

For the absolute magnitude, $M_\lambda$, the extinction towards the star, $A_\lambda$, is required. In the cases where $A_\lambda$ was not directly available from the literature, values for the colour excess $E(B-V)$ were used with a value $R_V = 3.1$ to determine $A_V$. These were converted to $A_\lambda$ using the \cite{fitzpatrick1999correcting} extinction law. Values for $A_V$ for A15, S3, and LP26 determined by \cite{telford2021far} were used to derive $A_\lambda$ using the same extinction law. An exception is 64066, where $E(B-V)$ was calculated using the $B-V$ value given by \cite{garcia2013young} and the intrinsic value for the spectral type O3 III given by \cite{martins2006ubvjhk}.

For the stars in NGC\,3109, no $E(B-V)$ values were available in the literature. Therefore, extinction $A_V$ was determined by calculating the difference between the observed $V$ magnitude $m_V$ and the intrinsic $V$ magnitude, $m_{V_0}$. For both stars the latter was determined using intrinsic $M_V$ values from \cite{martins2006ubvjhk} for Milky Way stars of spectral type O8 I scaled to the distance to NGC\,3109 quoted in Table \ref{tab:gals}. $A_V$ was then converted to $A_\lambda$ as explained above.

To calculate $M_\lambda$, we used these derived values for $A_\lambda$ and assumed the distance to the host galaxy, given in Table \ref{tab:gals}, as the distance to the star. The uncertainties in $A_\lambda$ were neglected in the determination of those of $M_\lambda$, meaning only the distance uncertainties were considered. The final values for $M_\lambda$ of each star are given in the rightmost column of Table \ref{tab:phot}. Overall, the extinction towards these stars is quite low, as expected for metal-poor dwarf galaxies (but see \citealp{2022MNRAS.516.4164L} for Sextans\,A, and \citealp{garcia2009young} for IC\,1613).

\subsection{UV spectra}
\label{subsec:uv_spectra}

\begin{table*}
\centering
\caption{The quality and coverage of the UV spectra of the stars in the sample.}
\label{tab:uv_spec}
\begin{tabular}
{l c c c l c c c l}
\midrule\midrule
& \multicolumn{3}{c}{HST COS Gratings}& & \multicolumn{3}{c}{Continuum Signal-to-Noise} & \\\cmidrule(lr){2 - 4}\cmidrule(lr){6 - 8}
Star & G130M & G160M & G140L & $t_{\rm exp}$ (s)  & \ciii\,$\lambda 1176$ & \civ\,$\lambda 1550$ & \heii\,$\lambda 1640$ & Prog. ID \\ \midrule
64066 & \tick & \eks & \eks & 7976 & 7  & -- & -- & 15880 \\
A13 & \tick & \tick & \eks & 16326 & 6  & 6  & 5  & 12867 \\
62024 & \eks & \eks & \tick & 11485 & 13  & 8  & 6  & 12587 \\
B2 & \eks & \eks & \tick & 8567 & 12  & 7  & 6  & 12587 \\
B11 & \tick & \tick & \eks & 18720 & 6  & 5 & 4  & 12867 \\
A15 & \tick & \tick & \eks & 23580 & 4  & 4  & 3  & 15967 \\
A11 & \tick & \tick & \eks & 14618 &4  & 5  & 4  & 12867 \\
N20 & \eks & \eks & \tick & 9631 & 7  & 5  & 4  & 16511 \\
N34 & \eks & \eks & \tick & 7163 & 8  & 5  & 4  & 16511 \\
S3 & \tick & \tick & \eks & 37633 & 2  & 2  & 1  & 15967 \\
LP26 & \tick & \tick & \eks & 69901 & 5  & 5  & 3  & 15967 \\
\midrule
\end{tabular}
\end{table*}

The UV spectra were obtained from MAST\footnote{The Barbara A. Mikulski Archive for Space Telescopes, \url{https://mast.stsci.edu/search/ui/\#/ullyses/results?telescope=hst&targettype=lowz}, accessed September 2023.} using the predefined query of the `Low-Z' subsample\footnote{\url{https://ullyses.stsci.edu/ullyses-targets-lowz.html}, accessed September 2023.} of the Hubble UV Legacy Library of Young Stars as Essential Standards (ULLYSES) sample \citep{roman2020ultraviolet}. This is a compilation of observations of 31 OB stars in metal-poor dwarf galaxies in the Local Group. However, only eight of these stars are of spectral type O and have optical data publicly available (described in the following subsection). N20 and N34 come from the ULLYSES core sample, while 64066, A13, 62024, B2, B11, and A11 are of the archival stars in the sample. The UV spectra of A15, S3, and LP26 are those analysed in \cite{telford2021far}, thus defining the size of the sample in this work, 11.

The spectra downloaded from MAST were already processed into high level science products, meaning they were fit for analysis upon retrieval. The quality and coverage of these spectra is outlined in Table \ref{tab:uv_spec}. All spectra were taken with the Cosmic Origins Spectrograph (COS) on the HST using either the lower resolution ($R$) G140L ($R \sim 2600$) grating or a combination of the medium resolution G130M and G160M gratings ($R \sim 15000$). These cover the wavelength range \SIrange{1100}{1800}{\angstrom} that contains useful diagnostic spectral lines. The only exception to this is 64066 in IC\,1613, which has only been observed with the G130M grating, so only data at wavelengths $\lesssim$ \SI{1450}{\angstrom} are available. This means that the useful \civ\,$\lambda 1550$ line is not available for this star. The assumptions necessary to deal with this are explained in Appendix \ref{a_sec:comments}.

Listed in Table \ref{tab:uv_spec} is the continuum signal-to-noise ratio (SNR) around different UV spectral lines for each star. In general, they are all $\sim 5-10$ and are lower at longer wavelengths.

\subsection{Optical spectra}
\label{subsec:optical_spectra}
\begin{table*}
\centering
\caption{The quality of the optical spectra of the stars in the sample.}
\label{tab:optical_spec}
\begin{tabular}
{l l c l c c c l}
\midrule\midrule
& & & & \multicolumn{3}{c}{Signal-to-Noise} & \\\cmidrule(lr){5 - 7}
Star & Instrument & $N_{\rm exp}$ & $t_{\rm exp}$ (s) & \hdelta & \heii\,$\lambda 4686$ & \halpha & Prog. ID \\ \midrule
64066 & X-Shooter & 4 & 5982 & 51  & 52  & 22 & 094.D-0130(A) \\
A13 & X-Shooter & 1 & 5400 & 27  & 28  & 10  & 085.D-0741(B) \\
62024 & X-Shooter & 6 & 8981 & 38  & 39  & 15  & 090.D-0298(A), 094.D-0130(A) \\
B2 & X-Shooter & 6 & 8973  & 45  & 47  & 18  & 094.D-0130(A) \\
B11 & X-Shooter & 4 & 7200  & 23  & 24  & 10  & 085.D-0741(B) \\
A15 & KCWI & 6 & 7200 & 61  & 54  & -- &  2021B\_N194 \\ 
A11 & X-Shooter & 3 & 5400 &  21  & 23  & 11  & 085.D-0741(B) \\
N20 & X-Shooter & 1 & 3600 & 11  & 11  & 5 & 085.D-0741(A) \\
N34 & X-Shooter & 4 & 14637 & 26  & 26  & 8  & 090.D-0212(A) \\
S3 & KCWI & 8 & 9600 & 54  & 41  & -- &  2021B\_N194\\ 
LP26 & KCWI & 10 & 12000 & 54  & 57  & -- &  2020B\_N194 \\ 
\midrule
\end{tabular}
\end{table*}

The optical spectra for all but three of the stars in the sample were taken with the X-Shooter spectrograph on the Very Large Telescope \citep{vernet2011xshooter}. These were obtained from the ESO archive\footnote{\url{https://archive.eso.org/wdb/wdb/adp/phase3_spectral/form}, accessed October 2023.}. The observing programmes under which they were taken are listed in Table \ref{tab:optical_spec}. The reduction was done automatically with the ESO pipeline and all three X-Shooter arms (UVB, VIS, and NIR) were available. However, the data quality of the VIS and NIR arms was poor. For our analysis, we did not use any data from the NIR arm, and from the VIS arm, we only analysed \halpha. All other optical features are in the UVB arm, spanning the wavelength range 3000 to $5595\, {\rm \AA}$. The resolutions of the UVB and VIS arms are 5400 and 6500, respectively.

In order to achieve the best signal possible, the multi-epoch spectra were stacked following  the methodology of \citet{backs2024properties}. This was done by first matching the flux calibration between each epoch to ensure the continuum levels were the same and then combining the flux of each epoch, weighting by the inverse of the flux uncertainties. In Table \ref{tab:optical_spec}, we provide an overview of the quality of the optical spectra. The UVB has typical SNRs of $\sim 20 - 30$, reaching as low as 10 for one star and as high as 50 for others. The SNR in the VIS arm is low, with typical values at \halpha{} of $\sim 10$.

The spectra of A15, S3, and LP26 were observed using the Keck Cosmic Web Imager (KCWI) on the 10-m Keck II Telescope. For more information about the observations and data reduction of these spectra, see \cite{telford2023ionizing, telford2024observations}. The Keck spectra have a typical SNR of $50 -60$ and have $R \sim 4000$ as they were taken with the medium slicer and BM grating. Since the Keck spectra were taken with a central wavelength of $\SI{4500}{\angstrom}$, the wavelength coverage is $\sim$ \SIrange{4050}{4950}{\angstrom}, meaning the \halpha{} line is not available for these stars.

All 11 stars exhibit contamination in the form of narrow emission features near the centres of hydrogen Balmer lines. These features were clipped from the final spectrum. Finally, we cross-correlate the spectra with a template spectrum of hydrogen and helium lines to find a best-fit radial velocity ($v_{\rm rad}$), which we show in Table \ref{tab:powrmods}.

\subsection{Binarity check}
\label{subsec:binarity_check}
As the optical spectra were taken over multiple epochs, this allowed us to inspect the features present in the individual spectra for radial velocity shifts. Shifts like this would indicate that the observed system is composed of two or more stars rather than just one. It is important to check for this as most massive stars are formed in multiple systems \citep{sana2012binary}. Furthermore, many of the stars in the ULLYSES sample have turned out to be binaries, despite efforts to avoid such systems during target selection \citep[e.g.][]{pauli2022binary,sana2024xshootu,ramachandran2024bebinaries}. Therefore, it is likely that several of the stars in our sample are binaries.

For stars with more than one exposure available (i.e. all but A13 and N20), the spectra were visually examined for evidence of radial velocity shifts and, by extension, binarity. No such shifts were detected. This may be attributed to the low SNR of the individual epochs and the limited observing periods of the spectra, as, for each star, most were taken over the course of one or two nights. Even for stars observed over longer periods -- B2 (with spectra collected in November and December 2014) and 62024 (observed in November and December 2012, as well as December 2014) -- no radial velocity shifts were observed, despite expectations of greater phase coverage. 

We also checked for signatures of an SB2 system (i.e. double lines in the spectrum) and no obvious signatures were found. Therefore, we treat all targets as single stars in this work. This is also a reasonable assumption for binaries, provided the luminosity ratio of the two components is not near unity. More observations over longer periods would be required to further investigate the potential binary nature of these stars.

\subsection{Continuum normalisation}
\label{subsec:contnorm} 
\begin{figure}
   \centering
   \includegraphics[width=\hsize]{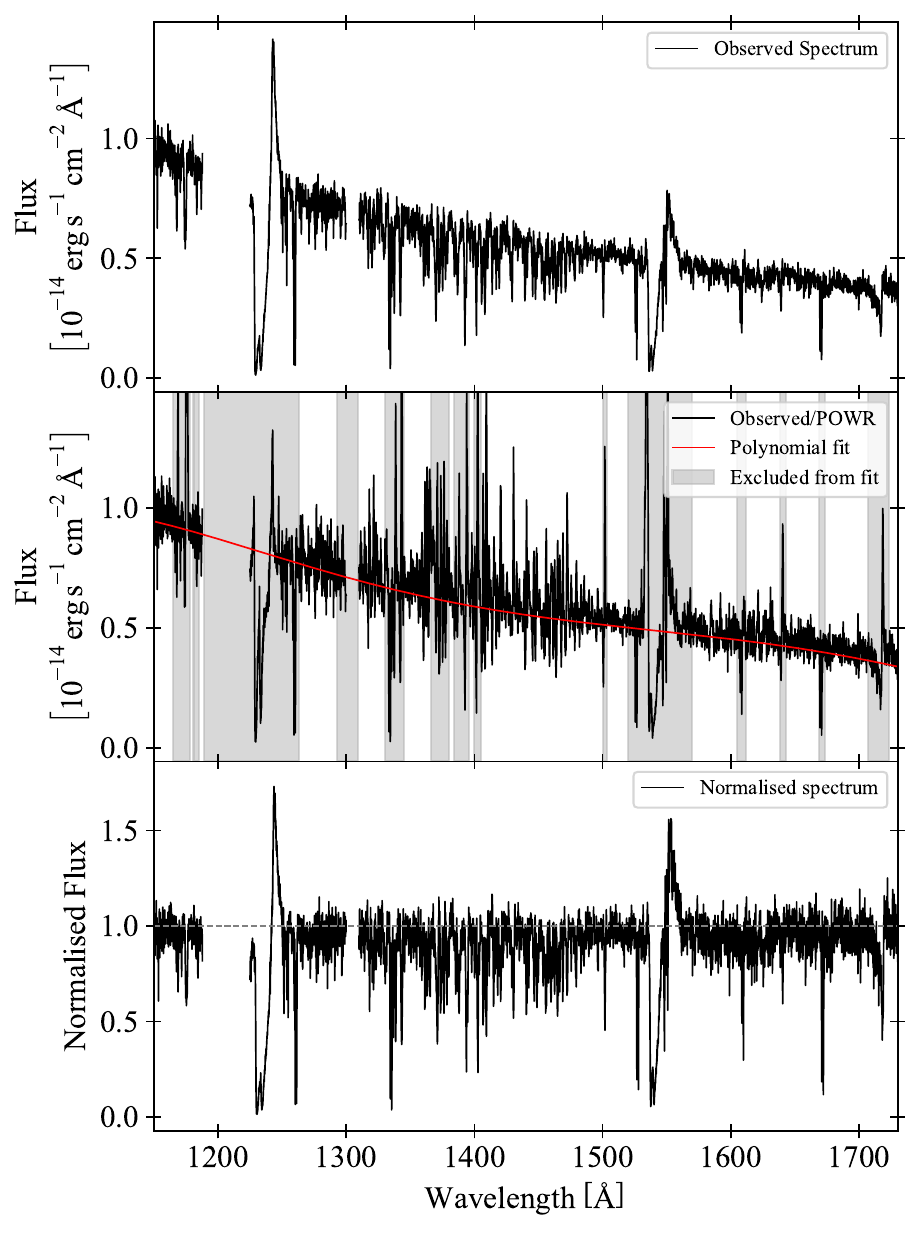}
      \caption{An example of the normalisation process. Top: the UV spectrum of A13 in IC\,1613. Middle: The observed spectrum divided by the best-fitting normalised \powr{} model, and a polynomial fit through this. The shaded regions indicate parts of the spectrum that are excluded from this polynomial fit process. Bottom: the normalised spectrum, which is the result of dividing the original spectrum by the polynomial fit.}
         \label{fig:normalising}
\end{figure}

As optical O-type star spectra show relatively few spectral lines, we normalise our diagnostic lines locally. For a given spectral line, this is done by masking any features in the local continuum around this line, fitting a straight line through this masked continuum and then dividing through by this fit.  

Even though our stars are metal-poor, the normalisation of their UV spectra is not straightforward. This is due to the thousands of Fe-group lines (henceforth, Fe-lines) that overlap, thus removing almost all information about the original continuum, with themselves forming a so-called `pseudo-continuum'. The properties of this pseudo-continuum depend on stellar parameters, such as \teff{}, \logg{}, \vrot{}, and the micro-turbulent velocity, $\xi$ \citep{heap2006fundamental}, as these affect the relative strengths and the shapes of the profiles of these Fe lines. 

To normalise the UV spectra, we follow the methodology of \cite{brands2022r136}. This involves the use of line-blanketed model atmospheres. To this end we downloaded from the web interface\footnote{\url{https://www.astro.physik.uni-potsdam.de/PoWR}} and used the `normalized line spectrum' products of the LMC and SMC-Vd3 OB model grids of The Potsdam Wolf-Rayet models \citep[\powr{}; ][]{hainich2019powr}. Specifically, we used the subset of models that cover the range of \teff{} of \SIrange{28}{45}{\kilo\kelvin} and \logg{} of 3 to 4.2 in cgs units. We also explored the dimensions of \vrad{} and \vrot{} in increments of \SI{10}{\kilo\meter\per\second} and \SI{15}{\kilo\meter\per\second}, respectively. These models have a fixed $\xi = \SI{10}{\kilo\meter\per\second}$, which we note may introduce some uncertainties in this process, as this parameter can affect the shape and strength of the Fe lines \citep[e.g. ][]{bouret2015spectro}.

\begin{table*}
\centering
\caption{The parameters of the normalised \powr{} models used to normalise the UV spectra and the radial velocity determinations.} 
\label{tab:powrmods}
\begin{tabular}
{l l c c c c | c c}
\midrule\midrule
Star & Grid & $T_{\rm eff}$ & $\log g$ & $v\sin i$ & $v_{\rm rad, UV}$ & $v_{\rm rad, optical}$ & $v_{\rm rad, adopted}$ \\
 & & [kK] & $[{\rm cm\,s}^{-2}]$ & $[{\rm km\,s}^{-1}]$ & $[{\rm km\,s}^{-1}]$ & $[{\rm km\,s}^{-1}]$ & $[{\rm km\,s}^{-1}]$ \\
\midrule
64066 &  LMC & 38 & 3.6 & 179 & $-$220 & $-$201 & $-$220 \\
A13 &  LMC & 42 & 3.8 & 93 & $-$240 & $-$249 & $-$240 \\
62024 &  SMC & 37 & 3.6 & 136 & $-$270 & $-$182 & $-$270 \\
B2 &  SMC & 38 & 3.6 & 150 & $-$240 & $-$193 & $-$240/$-$193 \\
B11 &  SMC & 31 & 3.6 & 93 & $-$240 & $-$224 & $-$240 \\
A15 &  SMC & 38 & 4.0 & 64 & $-$140 & $-$87 & $-$140 \\
A11 &  SMC & 29 & 3.6 & 107 & $-$110 & $-$148 & $-$110 \\
N20 &  SMC & 28 & 3.0 & 207 & 310 & 432 & 310/432 \\
N34 &  SMC & 31 & 3.2 & 164 & 330 & 407 & 330/407 \\
S3 &  SMC & 28 & 3.6 & 400 & 180 & 352 & 352/300 \\
LP26 &  SMC & 40 & 4.4 & 400 & 220 & 305 & 283* \\
\midrule
\end{tabular}
\tablefoot{For the radial velocities, $v_{\rm rad, UV}$ is that determined in the normalisation process, $v_{\rm rad, optical}$ is that found through cross-correlation of the optical spectrum, and $v_{\rm rad,adopted}$ is the final adopted $v_{\rm rad}$. In the case where two values separated by a slash are presented for $v_{\rm rad,adopted}$, values to the left and right show the adopted $v_{\rm rad}$ for the UV and optical spectrum, respectively.}
\tablebib{*Adopted $v_{\rm rad}$ from \cite{telford2021far}.}
\end{table*}
Figure \ref{fig:normalising} shows the spectrum of A13 in IC\,1613 as an example of how we normalise the UV spectra. First, the observed spectrum is divided by the normalised \powr{} model. In an ideal case where the Fe lines of each spectrum match, this will result in a featureless curve. Because parameters like CNO abundances or wind parameters in the model spectrum may not exactly match those of the observed spectrum, many spectral lines may not match up. This can be seen in the middle panel of Fig. \ref{fig:normalising}. For example, due to different terminal velocities between the model and the observed spectra, there is a spurious increase in flux at the blue edge of the \civ\,$\lambda 1550$ feature. In order to find the pseudo-continuum generated by the Fe lines, we mask these regions where the observation and models may differ and fit a polynomial of degree 5 through the resulting curve. Finally, the observed spectrum is divided through by this polynomial, thus resulting in a normalised spectrum. This polynomial fit accounts for the overall shape of the spectral energy distribution in this wavelength range and, consequently, the extinction towards the star is captured in the fit and is divided out.

This process is repeated for all models in the grid and the normalised \powr{} model whose \teff, \logg, \vrad, and \vrot{} best matches the observed spectrum (i.e. that with the lowest $\chi^2$ value) is adopted as the best-fit normalised spectrum. The parameters of the best-fit normalised \powr{} models are shown in Table \ref{tab:powrmods}. We chose not to use this as an Fe abundance determination due to the coarseness of the grid in $Z$-space and due to the degeneracy between \microturb{} and Fe abundances.

For most cases, the UV \vrad{} determination was sufficient for both the optical and UV spectra. Other cases required that the two spectra have their own \vrad, while LP26 and S3 required manual inspection (see Table~\ref{tab:powrmods}). The UV $v_{\rm rad}$ determinations are constrained to within $\sim \SI{10}{\kilo\meter\per\second}$, while those determined from the optical spectra have uncertainties of $\sim 30$ to $\SI{70}{\kilo\meter\per\second}$ for slower to faster rotators, respectively. The latter uncertainties were estimated by eye. We performed a test calculation for source S3, which has a relatively high SNR in the optical, to assess the impact of the uncertainty in the \vrad{} derived from the optical spectrum. Changing $v_{\rm rad, optical}$ from its preferred value of \SI{300}{\kilo\meter\per\second} to \SI{352}{\kilo\meter\per\second} caused only minor changes in the fitted stellar parameters, within their uncertainties. For instance, the derived mass loss rate increased by 0.1\,dex. 

The UV normalization procedure and radial velocity determination works well in most of the cases. However, for the stars in Sextans\,A and Leo\,P, the results are suboptimal, while for WLM A11 and IC\,1613 B11 it is only the normalisation that is an issue. The main problem regarding the normalisation arises in the \civ\,$\lambda 1550$ feature, where the Fe forest is prominent. For LP26 and S3, because the model grids have a higher $Z$ than these stars, the grid search favours a high \vrot{} so as to make the Fe lines as shallow as possible. This then results in a relatively featureless pseudo-continuum that is a few percent lower than the actual Fe continuum of a low $Z$ star. Dividing through by the polynomial fit will then bring down the Fe forest regions in the observed spectrum, and therefore the \civ\,$\lambda 1550$ line, a few percent below the true value. This high \vrot{} consequently affects the \vrad{} determination, as there are fewer features in the spectrum available for accurate measurement.

While the grid search does not necessarily favour an abnormally high \vrot{} for A11 and B11, there is still an issue with the normalisation at the \civ\,$\lambda 1550$ feature in these stars. We therefore recognise this as an inherent uncertainty in our work\footnote{We note that a sub-SMC metallicity PoWR grid was released after the analysis was performed for this work, available at the PoWR website.}. In conclusion, our best-fitting stellar atmosphere models (described in the following section) are generally consistent with the observed spectra, save for two sources, A11 and B11. Possibly, A11 is a binary (see Sect.~\ref{subsec:results_anomalous}). 

\section{Methods and assumptions}
\label{sec:methods}

In this work, we use the stellar atmosphere code \fw{} to produce synthetic spectra and the genetic algorithm \kiwi{} to constrain stellar and wind parameters and their uncertainties. We make a handful of simplifying assumptions in our modelling approach given the quality of some of the spectra of the stars in our sample. We describe \fw{} and outline the assumptions we make in Sect. \ref{subsec:fw} which is followed by a description of \kiwi{} in Sect. \ref{subsec:kiwi}.

\subsection{Fastwind}
\label{subsec:fw}

\fw{} \citep[v10.6; ][]{santolaya1997atmospheric, puls2005atmospheric, rivero2012nitrogen, carneiro2016atmospheric, sundqvist2018atmospheric} is a one-dimensional NLTE stellar atmosphere code tailored for modelling massive stars with winds. It solves the equation of hydrostatic equilibrium in spherical symmetry to determine the velocity field in the photosphere, which is then connected to the wind outflow. The velocity profile of the wind is described by a $\beta$-type velocity law of the form 
\begin{equation}
    v(r) = v_\infty\left(1-\frac{b}{r}\right)^\beta,
    \label{eqn:vr}
\end{equation}
where \vinf{} is the terminal velocity of the wind, $\beta$ describes the shape of the velocity profile, and $b$ is a distance close to the stellar radius ($R$) at Rosseland optical depth $\tau_{\rm Ross} = 2/3$ \citep{santolaya1997atmospheric}. The effective temperature (\teff) is also defined at $\tau_{\rm Ross} = 2/3$, and the emergent spectrum is calculated at an outer boundary of $120\,R$. The mass loss rate ($\dot{M}$) is an input parameter which, along with $v(r)$, determines the density structure in the wind, $\rho(r)$, through the equation of mass conservation,
\begin{equation}
    \dot{M}=4\pi r^2 \rho(r) v(r).
    \label{eqn:mass_conservation}
\end{equation}
Turbulent motion high up in the wind is accounted for by the wind turbulent parameter ($v_{\rm windturb}$), which we fix to $0.14$\vinf{} -- the average value for LMC O-stars found by \cite{brands2022r136}.

\fw{} accounts for inhomogeneities in the wind using either an optically thin or optically thick formalism. In the optically thin formalism the ensemble of `clumps' in the outflow is assumed to be optically thin for line (and continuum) radiation; in the optically thick or macro-clumping formalism, the ensemble of clumps span an optical depth range from almost transparent to fully opaque \citep[for a description of optically thick clumping, see, e.g.][]{sundqvist2018atmospheric, brands2022r136}. In this work, we adopt the optically thin formalism primarily due to the limited ability to constrain clumping parameters given the data quality. In this case, the wind is described by a medium consisting of over-dense clumps of material with density $\rho_{\rm cl}$ within a void inter-clump medium (i.e. with an inter-clump density $\rho_{\rm ic} = 0$), where the density of the clumps and inter-clump medium is related to the mean wind density $\langle \rho \rangle$ through the clumping factor ($f_{\rm cl}$) and inter-clump density contrast ($f_{\rm ic}$):
\begin{eqnarray}
    \rho_{\rm cl} &=& f_{\rm cl}\,\langle\rho\rangle   \\
    \rho_{\rm ic} &=& f_{\rm ic}\,\langle\rho\rangle = 0,  \hspace{4mm} \textrm{i.e.}\,f_{\rm ic} = 0.
\end{eqnarray}

It has been suggested that these clumps can form as a result of the combination of sub-surface convection due to the Fe opacity bump \citep{davies2007modelling, cantiello2009sub} and the line-deshadowing instability \citep[LDI;][]{owocki1984ldi, owocki1985ldi}, the onset of which begins deep in the wind close to the stellar surface \citep{sundqvist2018clumping}. \fw{} assumes that the wind is smooth at its base (i.e. at $r\simeq R$), where its velocity structure until this point is determined by microturbulence in the photosphere (\microturb{})\footnote{We neglect effects of macroturbulent velocities in this work as the data quality inhibits us from disentangling the effects of both macroturbulence and rotational velocity.}, while the onset of clumping begins at some fraction of \vinf{} from the stellar surface, \vclstart{}, and reaches a maximum value of \fcl{} higher up at \vclmax{}. In this work, we assume $v_{\rm cl,start} = 0.15$\vinf{} and $v_{\rm cl,max} = 2v_{\rm cl,start} = 0.3$\vinf{}, which we estimate based on typical values found in previous analyses \citep{hawcroft2021empirical, brands2022r136}.

One of the benefits of \fw{} is its computation speed: one model is calculated in $\sim 30 - 45$ minutes on one CPU core. This is achieved in how it calculates the equations of statistical equilibrium and its treatment of line blocking and line blanketing (see \citealp{puls2005atmospheric} for a detailed description of this). In short, it treats the `explicit' elements with detail while accounting for the `background' elements in an approximate way. In this work, we leave the C, N, O, and Si abundances as free parameters (see Sect. \ref{subsec:abundances} for a discussion of this). We scale the Mg and Fe abundances from the SMC values determined by \cite{brott2011rotating} to the metallicity of the host galaxies, while we scale the other abundances from the solar values of \citet[][following \citealp{brott2011rotating}]{asplund2005abundances}. 

\begin{table}[]
\centering
{\caption{The shock velocity parameters of each star.}\label{tab:uinf}}\vspace{-2pt}
\begin{tabular}
{lccl}
\midrule\midrule
Star & $u_\infty$ [$\si{\kilo\meter\per\second}$] & \vinf{} [$\si{\kilo\meter\per\second}$] & \vinf{} Reference\\
\midrule
64066 & 600 & 2000 & \nv\,$\lambda 1240$ edge \\
A13 & 654 & 2180 & (a)\\
62024 & 375 & 1250 & (b)\\
B2 & 450 & 1500 & (b)\\
B11 & 390 & 1300 & (a)\\
A15 & 411 & 1370 & (c), blue edge of \civ\\
A11 & 420 & 1400 & (a)\\
N20 & 650 & 2166 & (d)\\
N34 & 407 & 1357 & (d)\\
S3 & 400 & 1333 & (e), \teff{} from (c) \\
LP26 & 334 & 1114 & (e), \teff{} from (c) \\
\midrule
\end{tabular}
\tablefoot{$u_\infty$ is the maximum shock velocity which assumes the corresponding $v_\infty$ value in its calculation.}
\tablebib{
(a) \cite{bouret2015spectro}, (b) \cite{garcia2014winds}, (c) \cite{telford2021far}, (d) \cite{tramper2014properties}, (e) \cite{hawcroft2024vinf}.
}
\end{table}

Finally, \fw{} accounts for X-ray emission in the wind that results from wind-embedded shocks \citep{carneiro2016atmospheric}. This can affect the population levels of highly ionised species, like \nv{}, and therefore potentially may affect the determination of important stellar parameters, such as \teff{} and nitrogen abundance, if not properly accounted for \citep[see][]{backs2024smc}. However, given no X-ray data exists for these stars, we have to assume values based on previous empirical studies. In doing so, we follow \cite{brands2022r136}. The X-ray emission is described by a number of input parameters in \fw, a number of which we fix following \cite{brands2022r136}\footnote{For a description of the parameters, see \cite{carneiro2016atmospheric} or Appendix G of \cite{brands2022r136}.}: $m_X=25$, $\gamma_X = 0.75$, and $R_{\rm min}^{\rm input} = 1.5$. We assume, for the maximum jump velocity of the shocks, $u_\infty = 0.3$\vinf{}, where we take \vinf{} values from previous literature or, in the event that these are not available, a by-eye estimate. Table \ref{tab:uinf} shows the $u_\infty$ values adopted in this work, for which we used \vinf{} values from the literature. Finally, the X-ray volume filling factor, $f_{\rm X}$, is calculated in \kiwi{} using \mdot{} and \vinf{} following \cite{1996rftu.proc....9K} such that an X-ray luminosity ($L_{\rm X}$) is obtained that gives a value, as close as possible, to $L_{\rm X}/L=10^{-7}$. This is the canonical value typically observed in Galactic O stars \citep[e.g. ][]{long1980survey,chlebowski1989einstein} that has also been observed in the Tarantula Nebula in the LMC \citep{crowther2022x}. 

As it is unclear whether this parametrisation represents well the situation in these low $Z$ stars, we choose not to fit the \nv\,$\lambda 1240$ line in our analysis due to its sensitivity. While it is a prominent feature in many of the spectra, it was revealed through test runs that including this line can lead to troublesome fits (for example, by forcing a higher \teff{} to produce enough \nv{} in the wind such that the profile at $\SI{1240}{\angstrom}$ is reproduced).

\subsection{Kiwi-GA}
\label{subsec:kiwi}

Another advantage of the fast computation time of \fw{} is that many models can be calculated within a reasonable time frame to explore a given parameter space and fit model spectra to observations. This capability aligns perfectly with the function of the genetic algorithm (GA) \kiwi{}\footnote{\url{https://github.com/sarahbrands/Kiwi-GA/}}, which we use in this work to determine stellar and wind parameters. 

For an in-depth description of \kiwi{}, see \cite{brands2022r136}. This algorithm uses the concepts of evolution, reproduction, and mutation to efficiently explore a large parameter space and find best-fit parameters with uncertainties. It consists of a `genome', the members (genes) of which are the set of free parameters of individual \fw{} models. An input to the GA is the parameter space to be explored for each of the fitting parameters. For the first generation, the parameters of the \fw{} models are randomly sampled from the parameter space and these models are then computed. The resulting spectra of each of the members are compared to the data and the chi-squared, $\chi^2$, values for each are calculated (see below for details). In this case, the data are the wavelength regions surrounding and including the spectral lines to be modelled.

\begin{table}[]
\centering
{\caption{The free parameters we fit in this work using \kiwi{}.}\label{tab:free_params}}
\begin{tabular}
{lll}
\midrule\midrule
Parameter & Description & Unit\\
\midrule
\teff{} & Effective temperature & K\\
$g$ & Surface gravity & $\si{\centi\meter\per\square\second}$\\
\mdot{} & Mass loss rate & $\si{\msun\per\year}$\\
$Y_{\rm He}$ & Helium abundance & $N_{\rm He} / N_{\rm H}$\\
$v_\infty$ & Wind terminal velocity & $\si{\kilo\meter\per\second}$\\
\vrot{} & Rotational velocity& $\si{\kilo\meter\per\second}$\\
$\xi$ & Microturbulent velocity& $\si{\kilo\meter\per\second}$\\
$\beta$ & Velocity law exponent & -- \\
$f_{\rm cl}$ & Clumping factor & -- \\
$\epsilon_{\rm C,N,O,Si}$ & C, N, O, Si abundances & $\log(N_i/N_{\rm H}) + 12$\\
\midrule	 
\end{tabular}
\end{table}

The parameters of the two best-fitting models of all previous generations are then randomly combined to produce the offspring that populate the next generation. Mutation may happen at this point; that is, a random change in the model parameters may occur. There is a large probability that a small mutation will occur, where the parameters of the offspring differ by only a few percent from the optimal parameters of the previous generation. Conversely, there is a smaller probability that a large mutation will occur, which sets the parameter value to a random point within the input parameter space. By including mutation, the entire parameter space is explored and regions around (local) minima are densely sampled, allowing for uncertainty determinations. This recombination process is then repeated for a specified number of generations and a best-fit spectrum and uncertainties are determined. For each star, we fit a total of 13 free parameters, which are outlined in Table \ref{tab:free_params}, and compute 60 generations of 128 models. A list of the spectral lines we model in this work is given in Table \ref{tab:linelist}.

In \kiwi{}, the radius, $R$, is calculated following \cite{mokiem2005ga}. For a given model with effective temperature $T_{\rm eff,mod}$, its SED is estimated as a blackbody with $0.9\,T_{\rm eff,mod}$ which is then scaled to the anchor absolute magnitude, $M_\lambda$, given in Table \ref{tab:phot}, to estimate $R$. Once the GA has finished, the SED of the best-fitting model is computed and scaled to the anchor magnitude to get the final value for $R$. We find that the difference between the estimated and final values of $R$ for the best-fitting models are $\sim 2\% - 5\%$. The mass loss rates are then scaled using the invariant wind-strength parameter $Q = \dot{M} \sqrt{f_{\rm cl}} / (v_\infty R/R_\odot)^{3/2}$ \citep{puls1996dmom, puls2008mass} using the newly determined radius. Not only \mdot{} is scaled, but also all other parameters that depend on $R$ such as spectroscopic mass and luminosity.

The uncertainties on the fit parameters are determined using either a standard $\chi^2$ test or the root-mean-square-error of approximation statistic \citep[RMSEA;][]{steiger1998rmsea}. For the former case, all $\chi^2$ values are first scaled such that the best-fitting model has a reduced $\chi^2$ of 1. Then the survival function, $P$, of the incomplete gamma function (i.e. the cumulative distribution function of the $\chi^2$ distribution) is calculated for $\nu = n_{\rm data} - n_{\rm free}$ degrees of freedom, where $n_{\rm data}$ and $n_{\rm free}$ are the number of data points in the spectrum that are considered in the fit and the number of free parameters, respectively. We consider the 1$\sigma$ uncertainty region to be the models where $P> 32$\%.

However, for most of the stars, a by-eye inspection of the final \kiwi{} fit revealed that the uncertainties on some parameters determined by this method were clearly underestimated. For stars with a reduced $\chi^2 > 1$, we therefore use the RMSEA to estimate the uncertainties, as this was designed to produce reasonable uncertainty estimates when models do not perfectly match data. In this case, the best-fit model has the lowest RMSEA, and models within the 1- and 2$\sigma$ uncertainties are those with an RMSEA less than 1.04 and 1.09 times the minimum RMSEA, respectively. This method in combination with Kiwi-GA was introduced by \cite{brands2025xshootingullysesmassivestars}. They also calibrated the 1- and 2$\sigma$ cutoff values using the sample of LMC O-type stars of \cite{brands2022r136}.

\section{Results}
\label{sec:results}
\begin{figure*}
   \centering
   \includegraphics[width=0.95\hsize]{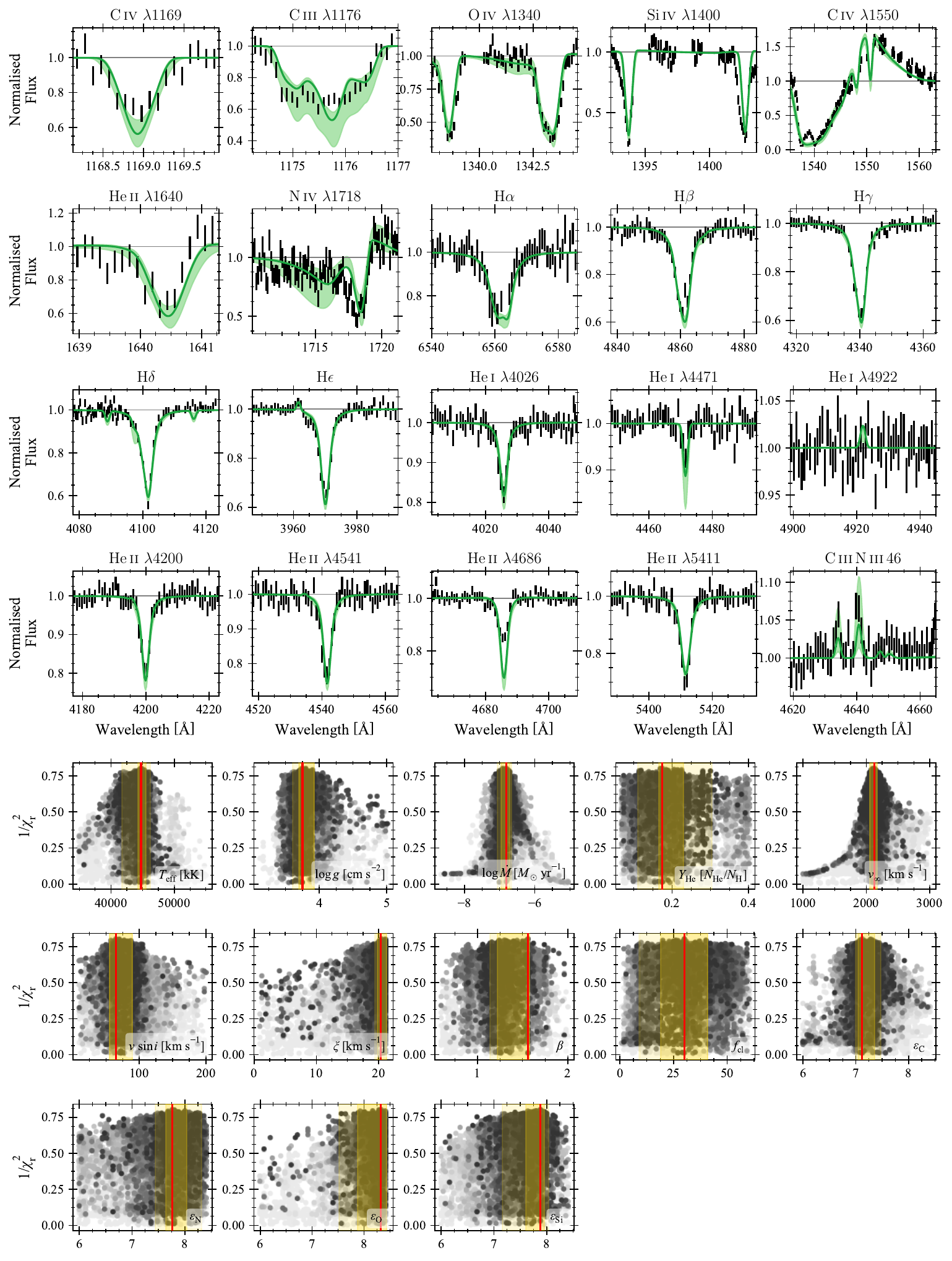}
      \caption{The \kiwi{} fit of A13. \textit{Top:} best-fit \fw{} model (solid green line) and 1$\sigma$ uncertainties (shaded green region) plotted on top of the observed spectrum (black vertical bars representing uncertainties; rebinned for clarity). \textit{Bottom:} Fitness plots for each parameter where darker coloured points represents models of later generations. The solid red line is the best-fit value, and darker and lighter yellow shaded regions represent 1- and 2$\sigma$ uncertainty regions, respectively.}
         \label{fig:example_fit}
\end{figure*}

\begin{table*}
\centering
\caption{Best-fit stellar parameters of the stars in the sample and corresponding upper and lower 1$\sigma$ uncertainties.}
\label{tab:best_fit_params}
\begin{tabular}
{l r@{}l  r@{}l r@{}l r@{}l r@{}l r@{}l r@{}l r@{}l}
\midrule\midrule
Star  & \multicolumn{2}{c}{$T_{\rm eff}$} & \multicolumn{2}{c}{$\log g$} &  \multicolumn{2}{c}{$v\sin i$} & \multicolumn{2}{c}{$\xi$} & \multicolumn{2}{c}{$\log \dot{M}$} & \multicolumn{2}{c}{$v_{\infty}$} & \multicolumn{2}{c}{$\beta$} & \multicolumn{2}{c}{$f_{\rm cl}$} \\
  & \multicolumn{2}{c}{[kK]} & \multicolumn{2}{c}{$[{\rm cm\,s}^{-2}]$} &  \multicolumn{2}{c}{$[{\rm km\,s}^{-1}]$} & \multicolumn{2}{c}{$[{\rm km\,s}^{-1}]$} & \multicolumn{2}{c}{[$M_\odot\,{\rm yr}^{-1}$]} & \multicolumn{2}{c}{$[{\rm km\,s}^{-1}]$} & \multicolumn{2}{c}{} & \multicolumn{2}{c}{} \\
\midrule
64066 & $38.0$ & $^{+4.0}_{-2.5}$ & $3.32$ & $^{+0.26}_{-0.12}$ &  $178$ & $^{+45}_{-38}$ & $21.0$ & $^{+0.3}_{-4.4}$ & $-7.9$ & $^{+1.4}_{\downarrow}$ & $2000$ & $^{+200}_{-200}$ & \multicolumn{2}{c}{0.95} & $56$ & $^{+5}_{-55}$  \\[3pt]
A13 & $44.75$ & $^{+0.75}_{-0.5}$ & $3.75$ & $^{+0.17}_{-0.05}$ &  $65$ & $^{+25}_{-10}$ & $20.4$ & $^{+0.9}_{-0.6}$ & $-6.81$ & $^{+0.10}_{-0.15}$ & $2125$ & $^{+50}_{-75}$ & $1.56$ & $^{+0.02}_{-0.34}$ & $30$ & $^{+11}_{-11}$  \\[3pt]
62024 & $38.75$ & $^{+1.5}_{-2.5}$ & $3.84$ & $^{+0.16}_{-0.4}$ &  $160$ & $^{+55}_{-55}$ & $19.1$ & $^{+2.2}_{-8.1}$ & $-6.84$ & $^{+0.15}_{-0.25}$ & $1050$ & $^{+120}_{-200}$ & $2.12$ & $^{+0.10}_{-0.47}$ & $45$ & $^{+16}_{-23}$  \\[3pt]
B2 & $37.5$ & $^{+2.2}_{-1.5}$ & $3.72$ & $^{+0.18}_{-0.16}$ &  $103$ & $^{+28}_{-35}$ & $1.9$ & $^{+4.1}_{-1.2}$ & $-8.65$ & $^{+0.30}_{-0.7}$ & $1025$ & $^{+650}_{-400}$ & $1.04$ & $^{+0.62}_{-0.46}$ & $13$ & $^{+48}_{-11}$  \\[3pt]
B11 & $33.25$ & $^{+0.25}_{-0.75}$ & $3.54$ & $^{+0.06}_{-0.08}$ &  $130$ & $^{+18}_{-28}$ & $5.1$ & $^{+3.1}_{-1.6}$ & $-8.58$ & $^{+0.55}_{-0.2}$ & $1350$ & $^{+25}_{-120}$ & $0.92$ & $^{+0.02}_{-0.14}$ & $4$ & $^{+2}_{-3}$  \\[3pt]
A15 & $37.5$ & $^{+0.19}_{-0.56}$ & $3.82$ & $^{+0.02}_{-0.14}$ &  $60$ & $^{+18}_{-5}$ & $12.2$ & $^{+3.1}_{-0.9}$ & $-8.89$ & $^{+0.15}_{-0.2}$ & $1200$ & $^{+75}_{-120}$ & $0.76$ & $^{+0.04}_{-0.18}$ & $21$ & $^{+19}_{-12}$  \\[3pt]
A11 & $33.25$ & $^{+1.5}_{-1.0}$ & $3.46$ & $^{+0.46}_{-0.14}$ &  $150$ & $^{+40}_{-30}$ & $20.4$ & $^{+0.9}_{-5.9}$ & $-9.39$ & $^{+0.30}_{-0.1}$ & $1325$ & $^{+400}_{-200}$ & $0.68$ & $^{+0.10}_{-0.1}$ & $41$ & $^{+20}_{-34}$  \\[3pt]
N20 & $30.75$ & $^{+1.5}_{-0.25}$ & $3.5$ & $^{+0.28}_{-0.02}$ &  $180$ & $^{+8}_{-42}$ & $16.6$ & $^{+4.4}_{-5.6}$ & $-7.48$ & $^{+0.35}_{-0.05}$ & $1600$ & $^{+150}_{-75}$ & $0.9$ & $^{+0.48}_{-0.16}$ & $46$ & $^{+9}_{-21}$  \\[3pt]
N34 & $33.75$ & $^{+2.2}_{-2.8}$ & $3.36$ & $^{+0.32}_{-0.32}$ &  $103$ & $^{+25}_{-52}$ & $3.2$ & $^{+15.0}_{-2.5}$ & $-7.3$ & $^{+1.2}_{-0.2}$ & $1725$ & $^{+25}_{-300}$ & $1.9$ & $^{+0.0}_{-1.1}$ & $18$ & $^{+28}_{-17}$  \\[3pt]
S3 & $33.25$ & $^{+1.5}_{-0.75}$ & $4.18$ & $^{+0.14}_{-0.08}$ &  $278$ & $^{+16}_{-44}$ & $4.8$ & $^{+5.0}_{-0.9}$ & $-9.21$ & $^{+0.3}_{\downarrow}$ & $250$ & $^{+180}_{-75}$ & $0.62$ & $^{+0.18}_{-0.04}$ & $57$ & $^{+4}_{-51}$  \\[3pt]
LP26 & $33.0$ & $^{+1.8}_{-1.8}$ & $4.22$ & $^{+0.18}_{-0.16}$ &  $145$ & $^{+65}_{-40}$ & $1.0$ & $^{+5.0}_{-0.3}$ & $-9.8$ & $^{+1.4}_{\downarrow}$ & $800$ & $^{+1400}_{-720}$ & $0.66$ & $^{+0.96}_{-0.08}$ & $23$ & $^{+38}_{-22}$  \\[3pt]
\midrule
\end{tabular}
\tablefoot{Down arrows indicate upper limits.}
\end{table*}

\begin{table*}
\centering
\caption{Parameters of the stars in the sample that have been derived from those in Table \ref{tab:best_fit_params} and the corresponding 1$\sigma$ uncertainties.}
\label{tab:derived_params}
\begin{tabular}
{l r@{}l r@{}l r@{}l r@{}l r@{}l r@{}l r@{}l c c}
\midrule\midrule
Star  & \multicolumn{2}{c}{$\log L$} & \multicolumn{2}{c}{$R$} & \multicolumn{2}{c}{$M_{\rm spec}$} & \multicolumn{2}{c}{$\Gamma_{\rm Edd}$} & \multicolumn{2}{c}{$v_{\rm esc}$} & \multicolumn{2}{c}{$v_{\infty} / v_{\rm esc}$} & \multicolumn{2}{c}{$D_{\rm mom}$} & $\log Q_0$ & $\log Q_1$ \\
  & \multicolumn{2}{c}{$[L_\odot]$} & \multicolumn{2}{c}{[$R_\odot$]} & \multicolumn{2}{c}{[$M_{\odot}$]} & \multicolumn{2}{c}{} & \multicolumn{2}{c}{[km\,s$^{-1}$]} && & \multicolumn{2}{c}{$[{\rm g}\,{\rm cm}\,{\rm s}^{-2}]$} & [s$^{-1}$] & [s$^{-1}$] \\
\midrule
64066 & $5.51$ & $^{+0.11}_{-0.07}$ & $13.2$ & $^{+0.6}_{-0.9}$ & $13.3$ & $^{+8.2}_{-2.0}$ & $0.65$ & $^{+0.11}_{-0.23}$ & $367$ & $^{+240}_{-75}$ & $5.4$ & $^{+3.5}_{-1.2}$ & $26.8$ & $^{+1.5}_{\downarrow}$ & $49.3$ & $48.4$  \\[3pt]
A13 & $5.65$ & $^{+0.02}_{-0.01}$ & $11.2$ & $^{+0.2}_{-0.2}$ & $25.6$ & $^{+10.0}_{-1.7}$ & $0.46$ & $^{+0.04}_{-0.12}$ & $685$ & $^{+210}_{-45}$ & $3.1$ & $^{+0.2}_{-0.7}$ & $27.8$ & $^{+0.1}_{-0.2}$ & $49.4$ & $48.8$ \\[3pt]
62024 & $5.39$ & $^{+0.04}_{-0.07}$ & $11.1$ & $^{+0.5}_{-0.3}$ & $30.9$ & $^{+11.0}_{-17.0}$ & $0.21$ & $^{+0.22}_{-0.05}$ & $917$ & $^{+190}_{-410}$ & $1.15$ & $^{+0.9}_{-0.2}$ & $27.5$ & $^{+0.2}_{-0.3}$ & $49.0$ & $48.2$  \\[3pt]
B2 & $5.29$ & $^{+0.06}_{-0.04}$ & $10.6$ & $^{+0.3}_{-0.4}$ & $21.5$ & $^{+8.4}_{-5.9}$ & $0.25$ & $^{+0.09}_{-0.06}$ & $764$ & $^{+180}_{-160}$ & $1.3$ & $^{+0.8}_{-0.4}$ & $25.7$ & $^{+0.7}_{-0.5}$ & $48.9$ & $48.1$  \\[3pt]
B11 & $5.50$ & $^{+0.01}_{-0.02}$ & $17.1$ & $^{+0.3}_{-0.2}$ & $37.1$ & $^{+3.7}_{-4.9}$ & $0.23$ & $^{+0.03}_{-0.02}$ & $798$ & $^{+49}_{-70}$ & $1.69$ & $^{+0.06}_{-0.19}$ & $26.0$ & $^{+1.1}_{-0.2}$ & $48.9$ & $47.2$ \\[3pt]
A15 & $5.17$ & $^{+0.03}_{-0.04}$ & $9.2$ & $^{+0.4}_{-0.4}$ & $20.2$ & $^{+1.6}_{-4.9}$ & $0.19$ & $^{+0.05}_{-0.00}$ & $823$ & $^{+17}_{-130}$ & $1.46$ & $^{+0.27}_{-0.04}$ & $25.5$ & $^{+0.2}_{-0.2}$  & $48.7$ & $47.8$\\[3pt]
A11 & $5.82$ & $^{+0.05}_{-0.05}$ & $24.7$ & $^{+1.1}_{-1.2}$ & $64.2$ & $^{+100.0}_{-16.0}$ & $0.28$ & $^{+0.09}_{-0.16}$ & $848$ & $^{+690}_{-160}$ & $1.6$ & $^{+0.6}_{-0.8}$ & $25.2$ & $^{+0.6}_{-0.2}$  & $49.2$ & $47.6$ \\[3pt]
N20 & $5.63$ & $^{+0.05}_{-0.01}$ & $23.3$ & $^{+0.4}_{-0.7}$ & $62.8$ & $^{+49.0}_{-2.8}$ & $0.18$ & $^{+0.02}_{-0.08}$ & $915$ & $^{+370}_{-19}$ & $1.75$ & $^{+0.08}_{-0.44}$ & $27.21$ & $^{+0.58}_{-0.07}$ & $48.7$ & $46.5$ \\[3pt]
N34 & $5.64$ & $^{+0.07}_{-0.09}$ & $19.5$ & $^{+1.1}_{-0.8}$ & $31.7$ & $^{+30.0}_{-14.0}$ & $0.37$ & $^{+0.17}_{-0.17}$ & $626$ & $^{+360}_{-240}$ & $2.8$ & $^{+1.2}_{-1.2}$ & $27.4$ & $^{+6.5}_{-0.2}$  & $49.2$ & $48.0$ \\[3pt]
S3 & $5.18$ & $^{+0.05}_{-0.04}$ & $11.9$ & $^{+0.4}_{-0.5}$ & $77.8$ & $^{+25.0}_{-12.0}$ & $0.05$ & $^{+0.01}_{-0.01}$ & $1539$ & $^{+240}_{-110}$ & $0.16$ & $^{+0.09}_{-0.05}$ & $24.5$ & $^{+0.5}_{\downarrow}$ & $48.3$ & $46.6$  \\[3pt]
LP26 & $4.94$ & $^{+0.09}_{-0.11}$ & $9.1$ & $^{+0.9}_{-0.9}$ & $49.8$ & $^{+23.0}_{-17.0}$ & $0.05$ & $^{+0.02}_{-0.01}$ & $1413$ & $^{+300}_{-230}$ & $0.6$ & $^{+1.2}_{-0.5}$ & $24.4$ & $^{+1.8}_{-\downarrow}$ & $48.0$ & $46.3$ \\[3pt]
\midrule
\end{tabular}
\tablefoot{Down arrows indicate upper limits. The quantities $\log Q_i$ are the ionizing photon production rate of H and He\,{\sc i}, for $i=0$ and 1, respectively, and were calculated by integrating the SED of the best-fitting \fw{} model. }
\end{table*}

In this section we present the results of the \kiwi{} fits of the stars in our sample. We provide a general overview of the results here, and discuss the individual targets in Appendix \ref{a_sec:comments}. In Sects. \ref{subsec:results_stellarparams} and \ref{subsec:results_windparams}, we discuss the stellar and wind parameters obtained, respectively. We mention the troublesome fits in Sect. \ref{subsec:results_anomalous}. We present abundance determinations in Sect. \ref{subsec:abundances} and conclude with a comment on the evolutionary status of the stars in Sect. \ref{subsec:results_evol}.

\subsection{Stellar parameters}
\label{subsec:results_stellarparams}

In Table \ref{tab:best_fit_params} we show the best-fit parameters determined by the \kiwi{} fits, and we report additional parameters derived from these in Table \ref{tab:derived_params}. We show an example of a best-fit spectrum in Fig. \ref{fig:example_fit}, while we show the rest of the fits in Appendix \href{https://zenodo.org/records/15078070}{H}\footnote{Available online at \url{https://zenodo.org/records/15078070}.}. 

In general, the temperatures and gravities of the stars are well constrained. The stars have effective temperatures in the range \SIrange{30}{40}{\kilo\kelvin}, with the only exception being the dwarf star A13 in IC1613, which is hotter at \SI{45}{\kilo\kelvin}. Values for \logg{} are typically between 3 an 4, except for the especially low $Z$ dwarf stars S3 and LP26 with values of 4.2 to 4.3. For LP26, we find \vrot{} significantly lower than that obtained by \cite{telford2021far}, while for S3, we find a value almost identical to that obtained in their study. The rest of the stars in the sample generally rotate with velocities $< \SI{200}{\kilo\meter\per\second}$. As for the stars that overlap with the sample of \cite{tramper2014properties}, we find \vrot{} consistent for all stars but A11, where we find that it rotates faster. We find that \vrot{} of B2 and 62024 is higher than that determined by \cite{garcia2014winds}, however, their determination included macroturbulent broadening and their value was determined to best match the continuum around the spectral lines that they analysed. We find that, for most stars, a microturbulence $> \SI{15}{\kilo\meter\per\second}$ is favoured. We note that the quoted uncertainties of $L$ in Table \ref{tab:derived_params} do not capture the degeneracy between $E(B-V)$ and $L$, so these are most likely underestimated.

\subsection{Wind parameters}
\label{subsec:results_windparams}
As for the wind parameters, values for \mdot{} are typically reasonably well constrained to within a factor of $\sim 2$ (or $\sim 0.3$\,dex in log space) for our SNR $ \sim 4-8$ at \civ\,$\lambda 1550$ spectra (see Table\,\ref{tab:uv_spec}). For 64066, S3, and LP26, we quote upper limits for \mdot{} due to the lack of wind signatures in their spectra. The mass loss rates of the stars in this sample are quite low, as expected in this metallicity regime, with typical values $<\SI{e-7}{\msun\per\year}$ apart from A13 and 62024 in IC\,1613 that have larger values than this. This highlights the importance of including the UV spectra when analysing the wind properties of low $Z$ massive stars: recombination lines in the optical spectrum, such as \halpha{}, are only sensitive when $\dot{M} \gtrsim \SI{e-7}{\msun\per\year}$, however, as seen here and in previous analyses (\citealp[e.g.][]{backs2024smc, bouret2015spectro, telford2024observations}), O-stars in this $Z$ regime typically have values that are lower than this, so the UV spectrum is required as this is where the more sensitive resonance lines lie.

We also measure the degree of clumpiness in the winds of massive stars at sub-SMC metallicity. For roughly half of the sample, we find that the winds of these stars appear clumped within uncertainties. For 64066, B2, S3, N34, and LP26, we do not find convincing constraints as the error bars cover almost the entire parameter space. For B11, the clumping appears modest with $f_{\rm cl} = 4^{+2}_{-3}$. The large spread in the values for \fcl{} and large uncertainties have been found before \citep[in e.g.][]{brands2022r136, backs2024smc}. We test the significance of the values we obtain for \fcl{} in Sect. \ref{subsubsec:thick_clumping}. We discuss the determination of \vinf{} in Sect. \ref{subsec:vinf}.  

\subsection{Anomalous fits}
\label{subsec:results_anomalous}
While, for most of the stars, \kiwi{} produced a satisfactory fit, it did have issues with some of them, which we discuss here.

\paragraph{B11 and A11} These stars share similar issues. Their fits can be seen in the Appendix in Figs. \href{https://zenodo.org/records/15078070}{H.4} and \href{https://zenodo.org/records/15078070}{H.6} for B11 and A11 respectively. The best-fitting \fw{} model for both of these stars produces a \ciii\,$\lambda 1176$ feature that is too strong, and \heii\,$\lambda 1640$ and \niv\,$\lambda 1718$ features that are too weak. The fits of the \oiv{} and \siiv{} wind features at 1340 and 1400\,\AA, respectively, are well fit within uncertainties but both best-fits fail to reproduce the red emission in the \siiv\,$\lambda 1400$ feature. As for the optical spectrum, both produce a \halpha{} feature that is too strong in absorption. Other than that, the fit to the rest of the lines are fine, apart from A11, whose best-fit \fw{} model produces too much absorption in the \heii\,$\lambda 4686$ feature and in the \ciii{} region in the \ciii\,\niii{} complex at $\sim\SI{4600}{\angstrom}$. We therefore decide to exclude these stars from the various fits performed in later sections, and mark them with red borders in any plots produced. Given its large $L$ ($10^{5.82}\,L_\odot$) and $M$ ($\SI{64.2}{\msun}$) for its spectral type, A11 could be part of a binary system, as was hinted by \cite{bouret2015spectro}. We return to these stars in Sect. \ref{subsubsec:thick_clumping}.

\subsection{Abundances}
\label{subsec:abundances}

\begin{table}
\centering
\caption{Abundances of our sample stars.}
\label{tab:abun_number}
\begin{small}
\begin{tabular}
{l r@{}l r@{}l r@{}l r@{}l r@{}l r@{}l}
\midrule
\midrule
Star  & \multicolumn{2}{c}{$Y_{\rm He}$} & \multicolumn{2}{c}{$\epsilon_{\rm C}$} & \multicolumn{2}{c}{$\epsilon_{\rm N}$} & \multicolumn{2}{c}{$\epsilon_{\rm O}$} & \multicolumn{2}{c}{$\epsilon_{\rm Si}$} & \multicolumn{2}{c}{$\epsilon_{\rm C+N+O}$} \\
  & \multicolumn{2}{c}{[$N_{\rm He}/N_{\rm H}$]} & \multicolumn{2}{c}{} & \multicolumn{2}{c}{} & \multicolumn{2}{c}{} & \multicolumn{2}{c}{} & \multicolumn{2}{c}{} \\
\midrule
64066 & $0.12$ & $^{+0.12}_{-0.06}$ & $7.1$ & $^{+0.9}_{-0.7}$ & $7.9$ & $^{+0.4}_{-1.8}$ & $8.4$ & $^{+0.0}_{-0.9}$ & $7.6$ & $^{+0.8}_{-1.2}$ & $8.5$ & $^{+0.2}_{-1.0}$ \\[3pt]
A13 & $0.18$ & $^{+0.06}_{-0.01}$ & $7.1$ & $^{+0.2}_{-0.1}$ & $7.8$ & $^{+0.3}_{-0.1}$ & $8.3$ & $^{+0.1}_{-0.4}$ & $7.9$ & $^{+0.1}_{-0.3}$ & $8.4$ & $^{+0.2}_{-0.3}$ \\[3pt]
62024 & $0.15$ & $^{+0.08}_{-0.06}$ & $6.0$ & $^{+0.3}_{-0.1}$ & $8.3$ & $^{+0.2}_{-0.9}$ & $7.1$ & $^{+1.1}_{-1.1}$ & $6.2$ & $^{+0.6}_{-0.2}$ & $8.3$ & $^{+0.3}_{-0.9}$ \\[3pt]
B2 & $0.1$ & $^{+0.08}_{-0.04}$ & $7.8$ & $^{+0.5}_{-0.5}$ & $7.3$ & $^{+1.0}_{-1.3}$ & $8.4$ & $^{+0.0}_{-0.9}$ & $8.2$ & $^{+0.3}_{-0.7}$ & $8.5$ & $^{+0.3}_{-0.8}$ \\[3pt]
B11 & $0.14$ & $^{+0.06}_{-0.05}$ & $8.0$ & $^{+0.2}_{-0.2}$ & $8.4$ & $^{+0.0}_{-0.4}$ & $8.2$ & $^{+0.0}_{-0.6}$ & $7.2$ & $^{+0.5}_{-0.3}$ & $8.7$ & $^{+0.1}_{-0.4}$ \\[3pt]
A15 & $0.14$ & $^{+0.04}_{-0.02}$ & $7.8$ & $^{+0.1}_{-0.2}$ & $8.0$ & $^{+0.2}_{-0.2}$ & $8.2$ & $^{+0.0}_{-1.0}$ & $6.6$ & $^{+0.3}_{-0.3}$ & $8.5$ & $^{+0.1}_{-0.4}$ \\[3pt]
A11 & $0.13$ & $^{+0.14}_{-0.07}$ & $8.4$ & $^{+0.1}_{-0.2}$ & $6.8$ & $^{+1.3}_{-0.4}$ & $6.9$ & $^{+0.8}_{-0.8}$ & $7.8$ & $^{+0.3}_{-0.4}$ & $8.4$ & $^{+0.3}_{-0.2}$ \\[3pt]
N20 & $0.08$ & $^{+0.1}_{-0.02}$ & $7.9$ & $^{+0.0}_{-0.4}$ & $6.4$ & $^{+1.9}_{-0.4}$ & $8.2$ & $^{+0.2}_{-0.2}$ & $6.7$ & $^{+0.1}_{-0.3}$ & $8.4$ & $^{+0.3}_{-0.3}$ \\[3pt]
N34 & $0.17$ & $^{+0.06}_{-0.07}$ & $7.4$ & $^{+0.6}_{-1.2}$ & $8.3$ & $^{+0.1}_{-0.8}$ & $8.2$ & $^{+0.2}_{-1.4}$ & $8.4$ & $^{+0.1}_{-1.6}$ & $8.6$ & $^{+0.2}_{-1.0}$ \\[3pt]
S3 & $0.064$ & $^{+0.016}_{-0.008}$ & $7.5$ & $^{+0.3}_{-0.3}$ & $7.4$ & $^{+0.7}_{-0.4}$ & $7.2$ & $^{+0.4}_{-1.2}$ & $7.0$ & $^{+0.4}_{-0.4}$ & $7.9$ & $^{+0.5}_{-0.4}$ \\[3pt]
LP26 & $0.06$ & $^{+0.04}_{-0.004}$ & $6.2$ & $^{+0.6}_{-0.2}$ & $6.9$ & $^{+0.8}_{-0.9}$ & $7.0$ & $^{+1.4}_{-1.1}$ & $5.6$ & $^{+0.8}_{-0.1}$ & $7.3$ & $^{+1.2}_{-0.9}$ \\[3pt]
\midrule
\end{tabular}
\end{small}
\tablefoot{For element $i$, the number fraction is $\epsilon_i = \log(N_i/N_{\rm H}) + 12$.}
\end{table}

\begin{figure}
   \centering
   \includegraphics[width=\hsize]{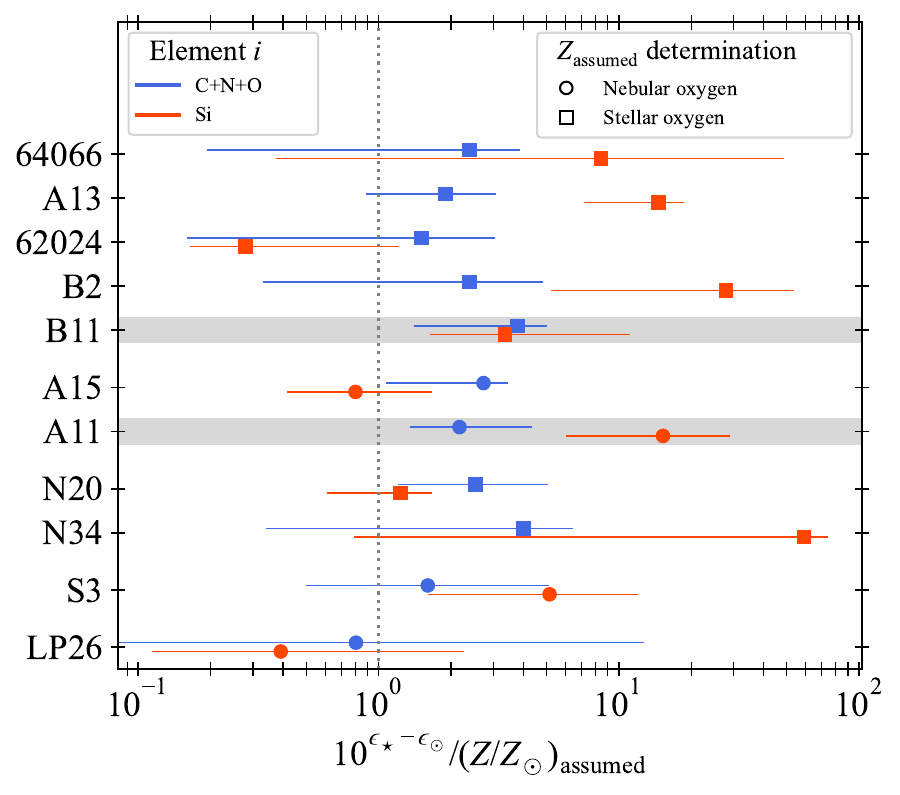}
      \caption{Photospheric abundance determinations of C+N+O (blue) and Si (red) for each star as fraction of solar compared to the adopted host galaxy metallicity, $Z$, from Table \ref{tab:gals}. Data points indicate whether the host galaxy metallicity $Z$ was determined from nebular emission  (circles) or through spectroscopic analysis of supergiants (squares). The vertical dotted line represents equality between our abundance determination and the solar value scaled to the adopted metallicity of the host galaxy. The troublesome runs of A11 and B11 are highlighted in grey.}
         \label{fig:abundances}
\end{figure}

We determine the abundances of C, N, O, and Si of the stars in the sample and show them, as number fractions, in Table \ref{tab:abun_number} (for completeness, we provide mass fractions in Table \ref{tab:abun_mass} in Appendix \ref{a_sec:abundance_mass}). In Fig. \ref{fig:abundances}, we show the Si and C+N+O abundances determined for each star, as a fraction of solar, and compare them to the metallicity of the host galaxies that we adopt in this study, $Z$. We also show the means by which $Z$ was determined, either through nebular O emission or through stellar spectroscopy. In this figure, we want to see if the abundances we determine are comparable to the solar value scaled to the adopted host galaxy metallicities (although we note, again, the uncertainties associated with scaling solar abundance values, as discussed in Sect. \ref{subsec:sample}, in that solar scaling does not always hold).

Our abundance determinations have large uncertainties, which is unsurprising given the modest SNR of the spectra and the limited number of lines available due to the low metallicity of the targets being studied. However, degeneracies between abundances and \mdot{} necessitate treating abundances as free parameters to determine accurate wind parameters. Nonetheless, we find that our C+N+O abundances are roughly equal within a factor of $2-3$ to the scaled-solar values, while Si abundances are significantly larger for five of the eleven stars.

We remark that the formal uncertainties on the abundances may underestimate the intrinsic uncertainties due to the low SNR of the spectra, especially for Si. This is because the modest SNR is unable to break potential degeneracies between abundances and other parameters, such as the mass loss rate. For example, the \siiv\,$\lambda 1400$ wind feature may favour one value in accordance with a degeneracy with \mdot, while the photospheric Si\,{\sc iv} components in the wings of H$\delta$ could favour another. If, in this example, the \siiv\,$\lambda 1400$ line has a greater weight in the final $\chi_{\rm r}^2$ value of the \kiwi{} fit, then its value will be preferred. If these two preferred values are very different to each other, the RMSEA uncertainty estimate may not capture the value preferred by \hdelta. If abundances are to be more accurately constrained, higher resolution and SNR spectra are needed.

\subsection{Evolutionary status}
\label{subsec:results_evol}
\begin{figure}
   \centering
   \includegraphics[width=0.97\hsize]{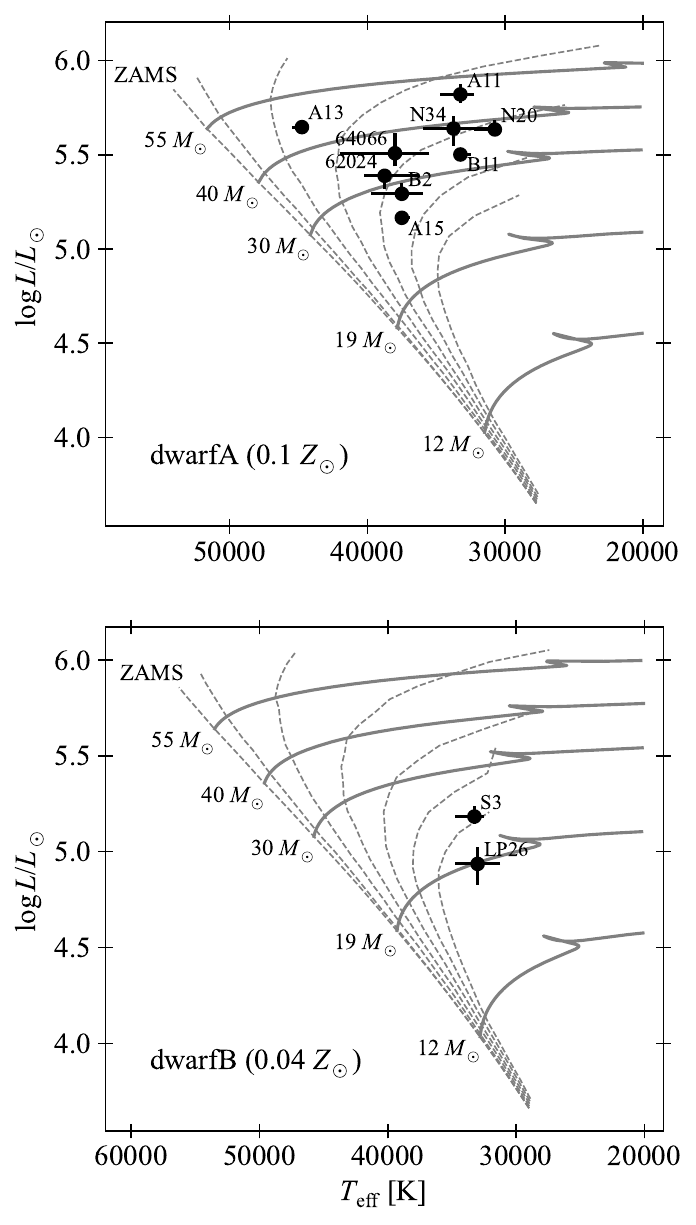}
      \caption{Positions of the stars on the HRD overplotted on the evolutionary models of \citet[][solid lines]{szecsi2022bonn}. The zero-age main sequence (ZAMS) and isochrones of $1-6$\,Myr (spaced 1\,Myr apart) are indicated with dashed lines. Top: the stars in IC\,1613, NGC\,3109, and WLM overplotted on the dwarfA model grid (0.1\,$Z_\odot$). Bottom: the stars in Sextans\,A and Leo\,P overplotted on the dwarfB model grid (0.04\,$Z_\odot$). }
         \label{fig:hrd}
\end{figure}
\begin{table}
\centering
\caption{Evolutionary parameters of our sample stars.\label{tab:evol_table}}
\begin{tabular}
{l r@{}l r@{}l r@{}l}
\midrule\midrule
Star & \multicolumn{2}{c}{$M_{\rm init}$} & \multicolumn{2}{c}{$M_{\rm evol}$} & \multicolumn{2}{c}{Age} \\
     & \multicolumn{2}{c}{[$M_{\odot}$]} &  \multicolumn{2}{c}{[$M_{\odot}$]} & \multicolumn{2}{c}{[Myr]} \\
\midrule
64066 & $35.4$ & $^{+6.4}_{-3.4}$ & $35.0$ & $^{+6.3}_{-3.4}$ & $3.7$ & $^{+0.5}_{-0.8}$\\[3pt]
A13 & $45.6$ & $^{+1.3}_{-1.0}$ & $45.1$ & $^{+1.3}_{-1.0}$ & $2.4$ & $^{+0.1}_{-0.1}$\\[3pt]
62024 & $31.6$ & $^{+2.0}_{-2.9}$ & $31.3$ & $^{+2.0}_{-2.8}$ & $3.9$ & $^{+0.6}_{-0.4}$\\[3pt]
B2 & $28.6$ & $^{+2.6}_{-1.6}$ & $28.4$ & $^{+2.6}_{-1.6}$ & $4.3$ & $^{+0.5}_{-0.6}$\\[3pt]
B11 & $33.1$ & $^{+0.4}_{-0.9}$ & $32.6$ & $^{+0.4}_{-0.9}$ & $4.4$ & $^{+0.1}_{-0.1}$\\[3pt]
A15 & $25.7$ & $^{+0.7}_{-0.9}$ & $25.6$ & $^{+0.7}_{-0.9}$ & $4.6$ & $^{+0.2}_{-0.1}$\\[3pt]
A11 & $48.6$ & $^{+4.3}_{-2.9}$ & $47.4$ & $^{+4.1}_{-2.7}$ & $3.4$ & $^{+0.2}_{-0.3}$\\[3pt]
N20 & $37.8$ & $^{+2.5}_{-0.5}$ & $37.1$ & $^{+2.4}_{-0.5}$ & $4.1$ & $^{+0.1}_{-0.3}$\\[3pt]
N34 & $38.9$ & $^{+4.3}_{-4.5}$ & $38.3$ & $^{+4.1}_{-4.4}$ & $3.9$ & $^{+0.5}_{-0.4}$\\[3pt]
S3 & $23.7$ & $^{+1.5}_{-1.0}$ & $23.6$ & $^{+1.5}_{-1.0}$ & $5.9$ & $^{+0.4}_{-0.5}$\\[3pt]
LP26 & $19.0$ & $^{+2.1}_{-1.7}$ & $18.9$ & $^{+2.1}_{-1.7}$ & $7.1$ & $^{+0.9}_{-0.8}$\\[3pt]
\midrule
\end{tabular}
\end{table}

In order to gauge the evolutionary status of the stars in this sample, we compare them to the Bonn Optimized Stellar Tracks \citep[\boost;][]{szecsi2022bonn}. Figure \ref{fig:hrd} shows the position of the stars on the Hertzsprung-Russell diagram (HRD). For the stars in IC\,1613, WLM, and NGC\,3109, we overplot stellar models from the dwarfA grid (\SI{0.1}{\zsun}) and for those in the lower $Z$ galaxies of Sextans\,A and Leo\,P, we overplot models from the dwarfB grid (\SI{0.04}{\zsun}). Also shown are isochrones in steps of \SI{1}{\mega\year}. We estimate evolutionary parameters by locating the position of each star in the interpolated grids within the uncertainties of $L$ and \teff, the results of which are shown in Table \ref{tab:evol_table}.   

All stars appear on the main sequence. In general, we see that, according to this set of single-star evolutionary models, the stars in Leo\,P and Sextans\,A are more evolved with an age $> \SI{5}{\mega\year}$, as found by \cite{telford2021far}, while the other stars are $\sim \SI{4}{\mega\year}$ in age, with A13 being an exception as it is quite young; around $\SI{2.5}{\mega\year}$ old. For 62024, B2, B11, A15, A11, and N34, we see a reasonable agreement between the current evolutionary mass, $M_{\rm evol}$ and the spectroscopic mass, $M_{\rm spec}$, determined by the \kiwi{} fits. However, there are large discrepancies between these quantities for the other stars. This is no surprise given the well-established `mass discrepancy problem' \citep{1992A&A...261..209H, weidner2010mass}. We do not comment further on this in this work.

The full main-sequence evolutionary timescale reported in the BoOST tracks for the stars in our sample range from $\sim 4-9$\,Myr. Given that the mass loss rates we derive are at most $10^{-7}$\,\msun\,yr$^{-1}$, they lose at most $0.1-1$\,\msun\, during this phase, namely for stars up to initially $\sim$\,50\,\msun\, in sub-SMC metallicity galaxies main-sequence mass loss, hence main-sequence angular momentum loss, is of minor importance for evolution. If the initial rotation speed of the star is large, this favors an efficient rotation induced mixing of CNO-products throughout the star and opens the possibility of chemically homogeneous evolution (CHE) for initially very fast spinning sources \citep[see e.g.][]{brott2011rotating,szecsi2022bonn}. The best candidate for CHE in our sample appears to be S3 in Sextans\,A for which we find $v\sin i = 278^{+16}_{-44}\,{\rm km\,s}^{-1}$. Its HRD position, non-enhanced helium abundance, and constraints for its CNO-abundance pattern derived here (see Table\,\ref{tab:abun_number}), though with considerable uncertainties, do not suggest a CHE pathway. 

\section{Discussion}
\label{sec:discussion}
In this section, we discuss the implications of the results we have obtained. In Sect. \ref{subsec:fits}, we quantify the dependence of \mdot{} on $L$ in this $Z$ regime by examining the modified wind momentum (\dmom{}). In Sect. \ref{subsec:prediction_comparison}, we compare our obtained values of \dmom{} and \mdot{} to the different theoretical predictions that exist for these quantities. We discuss implications of our findings in Sect. \ref{subsec:result_implications}. In Sects. \ref{subsec:comparing_to_prev} and \ref{subsec:lowz_talk}, we compare our results to previous analyses of SMC samples and samples of stars in lower $Z$ galaxies. In Sect. \ref{subsec:vinf}, we discuss the values for \vinf{} that we obtain and compare them to an empirical relation, and in Sect. \ref{subsec:assumptions}, we discuss potential consequences of the assumptions we have made in this work.

\subsection{Fits of $D_{\rm mom}$ vs. $L$}
\label{subsec:fits}

\begin{figure}
   \centering
   \includegraphics[width=\hsize]{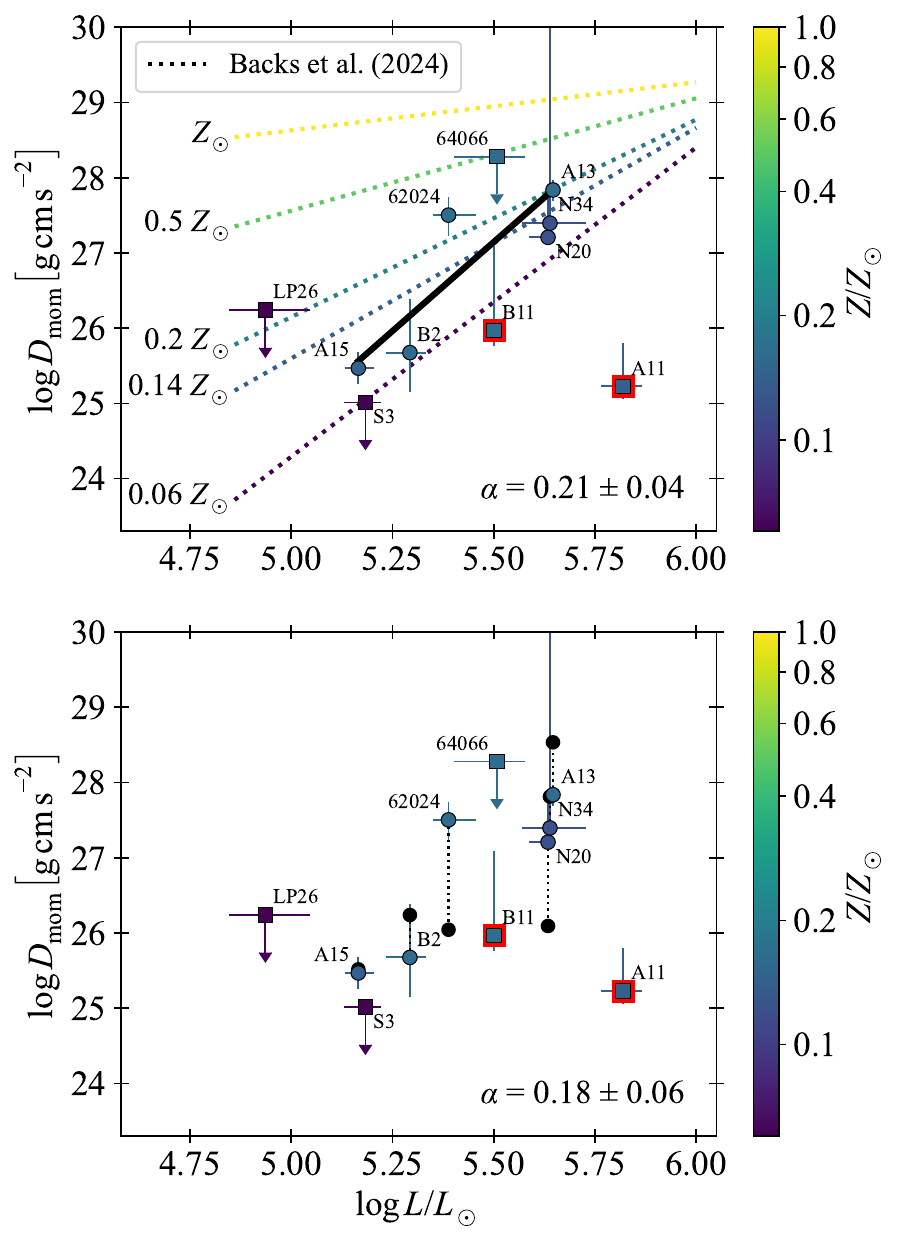}
      \caption{\textit{Top:} Fit of Eqn. \eqref{eqn:dmom_l} based on the six points denoted with circles. Overplotted with dotted lines is the empirical relation of $D_{\rm mom}(L,Z)$ determined by \cite{backs2024smc}. \textit{Bottom:} Projection of the two-dimensional fit to $D_{\rm mom}(L, M_{\rm eff})$ (Eqn. \ref{eqn:dmom_l_meff}) onto the luminosity axis, shown by black points (see text for further details). In both plots, the square points have been excluded from the fits, red borders indicate troublesome fits, the value for $\alpha$ determined from the fit is provided and the metallicity of both the tracks and the points are colour-coded. Uncertainty regions on the fit have been excluded for clarity. In the top panel they are large and span the entire vertical range of the panel.}
         \label{fig:dmom_fit}
\end{figure}

\begin{figure*}
   \centering
   \includegraphics[width=\hsize]{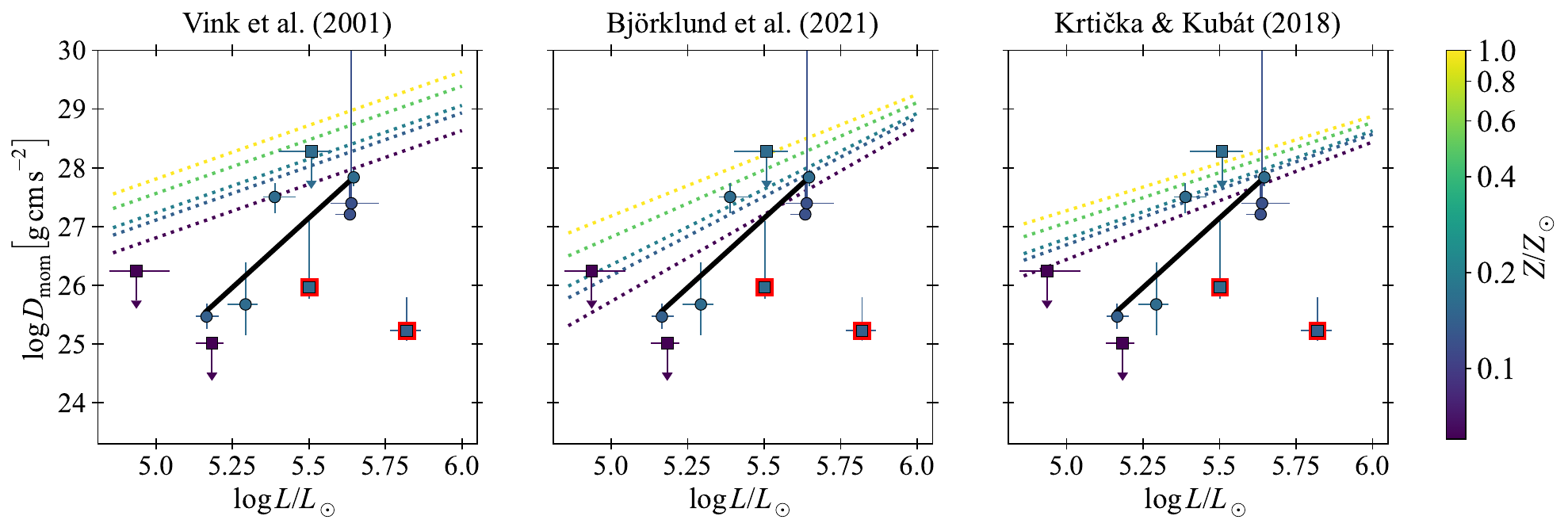}
      \caption{The fit of $D_{\rm mom}(L)$ obtained in this work compared to three different theoretical predictions (dotted coloured lines). In each plot, the points and the fit shown are the same as those shown in the top panel of Fig. \ref{fig:dmom_fit}, as are the $Z$ values of each theoretical track. The colour coding of the predictions is the same as in Fig.\,\ref{fig:dmom_fit}. Uncertainties in the fit span the entire vertical range of the panel.}
         \label{fig:comparing_dmom_theory}
\end{figure*}

First, we provide a brief overview of the relevant parts of radiation-driven wind theory (\citealp{lucy1970wind}; \citealp[hereafter \citetalias{castor1975radiation}]{castor1975radiation}; \citealp{abbott1982theory}) as we will be testing the scaling relations predicted by the theory and will compare our observations to the different predictions that exist.

In \citetalias{castor1975radiation} theory, the total line acceleration is expressed using a force-multiplier which is described by three parameters. One of these parameters is the ratio of the number of optically thick lines that contribute to the wind-driving force to the total number of lines: $\alpha$. This is important as it quantifies how the mass loss rate, $\dot{M}$, scales with $L$, the effective mass, $M_{\rm eff} = M(1-\Gamma_{\rm Edd})$, which is the mass of the star scaled by the Eddington parameter, and $Z$ in the following relation \citep{puls2008mass}:
\begin{equation}
    \dot{M}\propto Z^\frac{1-\alpha}{\alpha}L^{\frac{1}{\alpha}} M_{\rm eff}^{1 - \frac{1}{\alpha}},
\end{equation}
where the effect of a second parameter, the ionization parameter $\delta$, is neglected here as it is typically small ($\sim 0.05 - 0.1$; \citealp{abbott1982theory}).  One quantity derived from \citetalias{castor1975radiation} theory that is commonly used to examine the $\dot{M}(Z)$ dependence and to quantify $\alpha$ is the modified wind momentum, \dmom{} \citep{puls1996dmom, lamers1999introduction}. For a fixed $Z$, this quantity is given by
\begin{equation}
    D_{\rm mom} = \dot{M}v_\infty\sqrt{R} \propto L^{\frac{1}{\alpha}} M_{\rm eff}^{1 - \frac{1}{\alpha}+\frac{1}{2}}. \label{eqn:dmom}
\end{equation}
This is a useful relation because $\alpha$ is predicted to be $\sim 2/3$ for O-type stars at $\sim Z_{\odot}$ \citep{puls2000radiation}, meaning the dependency on $M_{\rm eff}$, a notoriously uncertain quantity if determined spectroscopically \citep{sander2015consistent}, effectively drops out. In this case, one may fit the relation,
\begin{equation}
    \log D_{\rm mom} = \log D_0 + \frac{1}{\alpha}\log L \,, \label{eqn:dmom_l}
\end{equation}
where $\log D_0$ is an offset term that captures the $Z$ dependence. 

However, recent analyses (discussed in Sect. \ref{subsec:comparing_to_prev}) suggest $\alpha$ may actually decrease with $Z$. In this case, the $M_{\rm eff}$ dependency becomes non-negligible, meaning it does not make sense to use Eqn. \eqref{eqn:dmom_l} and the entire relation must be fit:
\begin{equation}
    \log D_{\rm mom} = \log D_0 +  \frac{1}{\alpha}\log L + \left(1-\frac{1}{\alpha}+\frac{1}{2}\right)\log M_{\rm eff} \,,   \label{eqn:dmom_l_meff}
\end{equation}
for a fixed $Z$ and if $\alpha \neq 2/3$.

The top panel of Fig. \ref{fig:dmom_fit} shows the position of the stars in our sample on the \dmom{} vs. $L$ diagram. Also shown is the empirical $D_{\rm mom}(L,Z)$ relation of \cite{backs2024smc}. This was obtained from a sample of Milky Way, LMC, and SMC stars in the luminosity range $\log L/L_\odot = 5-6$; we extrapolate this relation to lower $Z$ and plot \dmom$(L)$ at $Z= 0.14$ and $\SI{0.06}{\zsun}$.

Present in the top panel of the figure is the fit to Eqn. \eqref{eqn:dmom_l}. To account for the $Z$ dependence, we chose to fit only the stars in IC\,1613, WLM, and NGC\,3109 and assume that these are of the same average $Z\sim \SI{0.14}{\zsun}$. This is a reasonable approximation to make because the adopted metallicities of these galaxies are similar. We excluded A11 and B11 (squares with red borders) due to the issues faced in the fitting procedure, as well as 64066 for which only an upper limit for \mdot{} was determined. This means that a total of six stars are considered in the fit (the circles in the figure). To account for the uncertainties on both $L$ and \dmom{}, we performed the fit using orthogonal distance regression \citep[ODR;][]{boggs1990orthogonal}. From the slope, we obtain $\alpha = 0.21 \pm 0.04$.

We present the fit of Eqn. \eqref{eqn:dmom_l_meff}, which incorporates the $M_{\rm eff}$ dependency, to the same 6 stars in the bottom panel of Fig. \ref{fig:dmom_fit}. As this is a two dimensional fit as a function of both $L$ and $M_{\rm eff}$, the best-fit solution to this equation is a plane. The best-fit solution of each star is therefore represented as a black point in Fig. \ref{fig:dmom_fit} which depicts, for a given star, the point on the best-fit plane at its $(L,M_{\rm eff})$ values projected onto the $L$ axis. By incorporating $M_{\rm eff}$ into this, we obtain a value for $\alpha$ of $0.18 \pm 0.06$. This is consistent with that obtained from the fit of Eqn. \eqref{eqn:dmom_l} and both values are significantly lower than the canonical value of 2/3.

 \begin{figure*}
   \centering
   \includegraphics[width=\hsize]{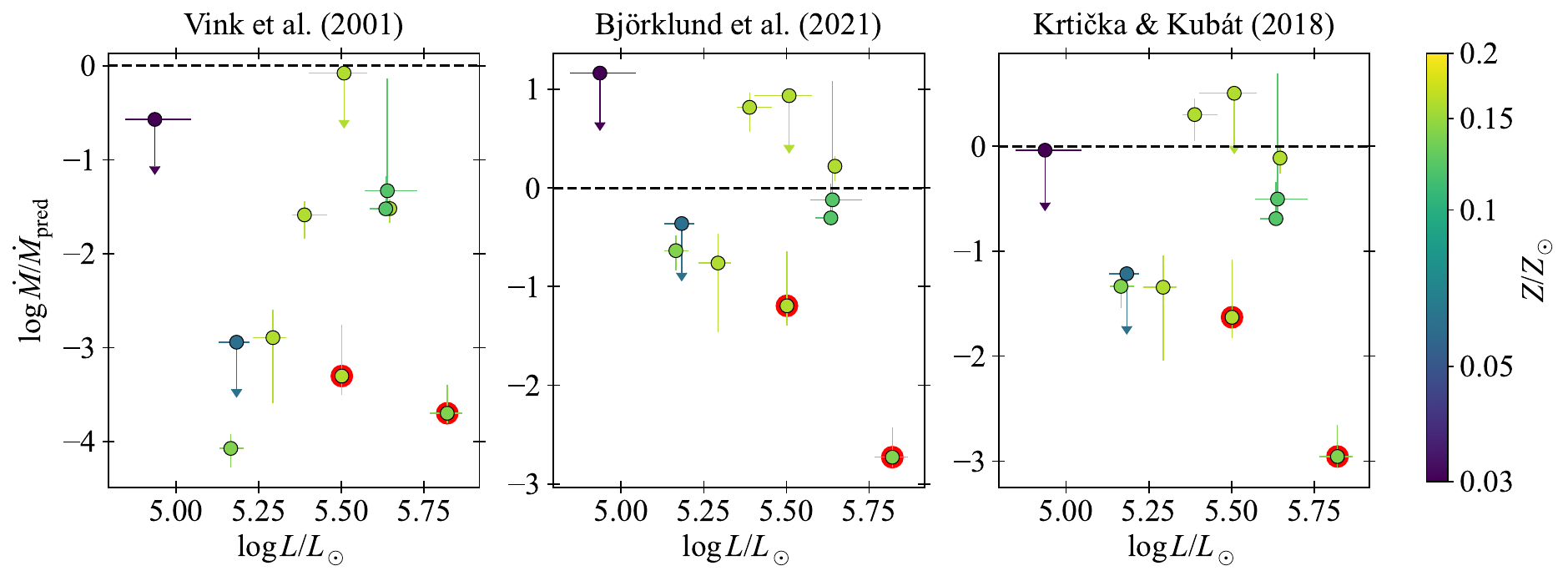}
      \caption{The ratio of the mass loss rate of the stars in the sample determined by the \kiwi{} fits ($\dot{M}$) and the prediction of this value ($\dot{M}_{\rm pred}$).}
         \label{fig:comparing_mdot}
\end{figure*}

We also perform this analysis using $\dot{M}$ and find similar results. This can be seen in Appendix \ref{a_sec:mdot}. That $\alpha < 2/3$ in this low $Z$ regime means that care should be taken when analysing \dmom{} of metal-poor massive stars, as the $M_{\rm eff}$ dependence becomes non-negligible. From its definition, a lower $\alpha$ means that less optically thick lines are contributing to the radiation force driving the wind, thus decreasing the overall radiation force and resulting in weaker winds, as expected in lower metallicity environments.  Future theoretical work should explore how $\alpha$ (in the \citetalias{castor1975radiation} formalism) is expected to scale with $Z$.

\subsection{Comparing $D_{\rm mom}(L)$ at $Z = 0.14\,Z_\odot$ and $\dot{M}$ to theory}
\label{subsec:prediction_comparison}
In this section, we compare the values we have obtained for \dmom{} and \mdot{} to three theoretical predictions: those of \cite{vink2001mass}, \cite{bjorklund2021new}, and \cite{krtivcka2018global}.

\subsubsection{Modified wind momentum}

Figure \ref{fig:comparing_dmom_theory} shows how our fit of Eqn. \eqref{eqn:dmom_l} compares to three sets of theoretical predictions of $D_{\rm mom}(L)$. To make this comparison, we extrapolated the results of \cite{bjorklund2021new} and \cite{krtivcka2018global} to metallicities below that of the SMC. The first thing to notice is that all prescriptions yield a shallower slope (i.e. a higher \citetalias{castor1975radiation} $\alpha$) at $Z \sim 0.14$\,$Z_{\odot}$ than our empirical result. Of the three sets, the \cite{bjorklund2021new} models have the steepest slope (i.e. have the lowest \citetalias{castor1975radiation} $\alpha$). Also, all predictions match best at the high $L$ end of stars studied here. The prescription of \cite{bjorklund2021new} is consistent with the values of the individual stars at high $L$, disregarding A11 and B11. It really only overpredicts for B2, A15, and S3 -- the stars in the low $L$ end of the sample. The same can be said for the predictions of \cite{krtivcka2018global}, however it is less consistent at low $L$. The \cite{vink2001mass} prescription predicts mass loss rates that are typically a factor of $\sim 2-3$ higher than those by the other two works \citep[e.g.][]{vink2022theory}. The reasons for this are possibly connected to the use of the energy equation rather than the momentum equation in constraining the wind properties (see \citealp{2008A&A...492..493M} and \citealp{muijres2012predictions} for a discussion of this), and to the use of the Sobolev approximation in computing the state of the gas. These authors overpredict \dmom{} of all stars except 64066, whose upper limit is consistent with the prescription. 

In conclusion, available theoretical works predict a shallower $D_{\rm mom}(L)$ relation causing discrepancies with empirical rates at $L \lesssim 10^{5.2}\,L_{\odot}$. Works requiring knowledge on the metallicity dependence of stellar winds may therefore best rely on empirical prescriptions \citep[e.g. those of][]{backs2024smc}. If one wants to rely on theory, the prediction of \cite{bjorklund2021new} best represents our findings. This is consistent with previous studies \citep[e.g.][]{backs2024smc, rickard2022stellar, telford2024observations}, which we discuss in Sect. \ref{subsec:comparing_to_prev}.

\subsubsection{Mass loss rates}

For completeness, we compare our determinations of \mdot{} to the same three theoretical predictions, which can be seen in Fig. \ref{fig:comparing_mdot}. Again, the prediction of \cite{bjorklund2021new} is closest to the values we obtain, with a spread of $\sim \pm 1$\,dex around a ratio of unity. That of \cite{krtivcka2018global} slightly overpredicts \mdot{} of our sample, while that of \cite{vink2001mass} highly overpredicts the mass loss rates of most of the stars by more than a factor of ten. This finding is in line with the conclusion formulated at the end of the previous sub-section.

\subsection{Implications of our results}
\label{subsec:result_implications}

In this section, we discuss the implications of our findings from Sects. \ref{subsec:fits} and \ref{subsec:prediction_comparison}. Our results indicate a fit of $D_{\rm mom}(L)$ at $Z=\SI{0.14}{\zsun}$ for six of the stars in our sample that is in fair agreement with an extrapolation of the empirical $D_{\rm mom}(L,Z)$ relation of \cite{backs2024smc} to this $Z$. Notably, this relationship, calibrated in the metallicity range from solar to 1/5th solar metallicity,  predicts a steeper dependence on $Z$ of \dmom{} at relatively low luminosity ($L\lesssim 10^{5.2}\,L_\odot$) than what is predicted by theory (compare Figs.~\ref{fig:dmom_fit} and \ref{fig:comparing_dmom_theory}). 

In this low $L$ regime at $Z\sim \SI{0.14}{\zsun}$, our findings suggest O-stars experience minimal mass loss during their main sequence evolution. For a $\SI{40}{\msun}$ star, representative of the average luminosity of $\sim10^{5.4}\,L_\odot$ of our sample, a main sequence mass loss rate of $\sim 10^{-7}-\SI{e-8}{\msun\per\year}$ would remove only $\sim 0.05-0.5$\,\SI{}{\msun} over the roughly 5 million years that this stage of evolution lasts \citep[e.g.][]{brott2011rotating,2015A&A...581A..15S}. Consequently, the effect of main-sequence mass loss for mass stripping and angular momentum loss is much less significant for these stars compared to equally bright O-stars in higher $Z$ regions. 

Whether this also holds for their post-main sequence evolution remains to be investigated. If later on these sub-SMC metallicity massive stars experience a Luminous Blue Variable phase (LBV; see \citealp{herrero2010nature} for a discussion of LBV candidate V39 in IC\,1613) during which mass is shedded efficiently, similar to what is hypothesised for Galactic stars \citep{2014ARA&A..52..487S}, the situation may be different, especially if the (unknown) mechanism for giant LBV eruptions is metallicity independent. 

Additionally, we note that the `weak-wind' problem has been observed in this low $L$ regime (at $L\la 10^{5.2}\,L_\odot$; \citealp{vink2022theory}) at solar \citep{martins2005stars, marcolino2009analysis} and SMC metallicities \citep{martins2004puzzling}. We find a steeper $D_{\rm mom}(L,Z)$ relation than what is predicted by theory, so this may be linked to it. Furthermore, it has also been shown that, in OB stars with these weak winds, material in the wind can be shocked to higher ionisation stages, meaning that the traditional wind signatures such as \civ{} or \siiv{} in the UV disappear, and instead the wind is detectable in the form of X-ray emission \citep[e.g.][]{huenemoerder2012xray}. This is a possible solution to the weak-wind problem, as a shocked wind scenario could result in lower \mdot{} and/or \vinf{} inferred from the UV than the true values across all phases of the wind. Whether such X-ray emission is present in our sample stars is uncertain and would require dedicated observations to confirm. Given we see these traditional features form in the wind in all but one of our sample stars, this would suggest that this ionisation due to shocks has not taken place, at least not completely.

At the high luminosity end, at $L \sim \SI{e+6}{\lsun}$, the situation is different. Here, the empirical relation of \cite{backs2024smc} and the theoretical relations by \citet{bjorklund2021new} and \citet{krtivcka2018global} show a particularly weak metallicity dependence. Therefore, in this regime, massive stars in galaxies similar to those studied here may be capable of shedding (almost) as much mass as their counterparts in the Magellanic Clouds or even in the Milky Way. In fact, this further suggests that low $Z$ stars at even higher $L$, in the regime of very massive stars, may be subject to the same upturn (or `kink') in \mdot{} as a function of $\Gamma_{\rm Edd}$ as experienced by stars at higher $Z$ \citep[e.g. ][]{bestenlehner2014vlt}. This occurs when the flux-weighted mean optical depth, $\tau_F$, of the wind transitions from below to above 1 \citep{vink2011wind, vink2012transition}. When the winds are optically thin ($\tau_F$ < 1), $\dot{M} \propto \Gamma_{\rm Edd}^2$ until they grow more dense and become optically thick ($\tau_F$ > 1) and this dependence sharply increases to $\dot{M}\propto\Gamma_{\rm Edd}^5$. To test this would require observations of more low $Z$ stars with masses high enough to place them in this $L$ regime.

We find that the stars in our sample only reach masses up to $\sim 30-50$\,\msun{} (Table \ref{tab:derived_params}), likely because they are found in regions of modest star-formation. If this is indicative of the general massive star population in these galaxies, mass loss is not as important in their evolution compared to similar $L$ stars in higher $Z$ regions. Vigorous star-formation, similar to the star-forming activity in the Large Magellanic Cloud \citep[where the maximum stellar mass is $\sim 250-300\,M_{\odot}$; e.g.][]{brands2022r136}, is required to produce stars of at least about $90\,\SI{}{\msun}$ or $\sim 10^6\,L_\odot$ that would suffer strongly from mass loss \citep[e.g.][]{2015A&A...573A..71K,2015A&A...581A..15S,2023MNRAS.524.1529S}. 

\subsection{Comparing $D_{\rm mom}(L,Z)$ to previous empirical studies}
\label{subsec:comparing_to_prev}
One of the first investigations of the $\dot{M}(Z)$ dependence was performed by \cite{mokiem2007empirical}. In this study, they gathered samples of Milky Way, LMC, and SMC stars that had been studied with either \fw{} or \textsc{Cmfgen} \citep{hillier1998treatment}. They found, at $\log (L/L_\odot) = 5.75$, $\dot{M}\propto Z^{0.83 \pm 0.16}$ for smooth winds, or $\dot{M}\propto Z^{0.72 \pm 0.15}$ when clumping was accounted for. Furthermore, this derived dependence incorporated the terminal velocity scaling $v_\infty \propto Z^{0.13}$ of \cite{leitherer1992deposition}.

\cite{backs2024smc} analysed a sample of 13 SMC O-stars using \fw{} and \kiwi{}. These authors found that the predictions of \cite{bjorklund2021new} best match the mass loss rates of the stars in their sample. Further, using their results, along with the LMC samples of \cite{brands2025xshootingullysesmassivestars} and \cite{hawcroft2024empirical}, and the Milky Way sample of \cite{hawcroft2021empirical}, they derived an empirical fit for $D_{\rm mom}(L,Z)$ to which we compare our findings in previous sections. They found a stronger $Z$ dependent $L$ dependence on \dmom{} than what is suggested by theoretical predictions of this quantity. Our fit of six stars is relatively in line with their relation for $Z = \SI{0.14}{\zsun}$ -- the average metal content of the host galaxies of the stars considered in the fit. 

\cite{rickard2022stellar} analysed the O-star population in the NGC 346 cluster in the SMC. Through analysis of \dmom, they found a steeper $Z$ dependent $L$ dependence of \mdot{} than what is currently predicted; instead of a fixed scaling, say $\dot{M} \propto Z^{x}$, they find $x \propto L^{-1.2}$, much steeper\footnote{At $L = 10^{5}\,L_{\odot}$ they find $x = 4.3$; at $L = 10^{6}\,L_{\odot}$ it is $x = 3.1$.} than the $L$ dependence of $L^{-0.32}$ predicted by \cite{bjorklund2021new}.

A steep $Z$ dependence was also found by \cite{ramachandran2019testing}. In this work, they  fit \mdot{} as a function of $L$ for nine OB stars in the wing of the SMC. When considered along with the $\dot{M}(L)$ relation found by \cite{ramachandran2018stellar} for LMC stars, where $Z = 0.5\,Z_\odot$, they suggested a value for $x \sim 2$. This is steeper than an exponent 0.69, as predicted by \cite{vink2001mass}, or 0.83, determined empirically by \cite{mokiem2007empirical}.

On the other hand, \cite{marcolino2022wind} found contrasting behaviour to the previous studies mentioned. They found, for their sample of Milky Way and SMC stars, that their fits of $D_{\rm mom}(L)$ for the two galaxies diverge towards higher $L$, as opposed to lower $L$ as is seen in the aforementioned studies. They suggest this could be due to the large number of low $L$ targets in their sample, where various factors can become important in this $L$ regime, such as additional dependencies on $\alpha$, like \teff{} for example, becoming non-negligible \citep{puls2000radiation}. In fact, when the low $L$ targets are excluded from fits, they find good agreement with the empirical study of \cite{mokiem2007empirical}.

What does all of this suggest? Apart from the study of \cite{marcolino2022wind}, all recent analyses of SMC samples have found a steeper metallicity dependence than what is currently predicted by theory. Our results are in line with this finding (c.f. Fig. \ref{fig:dmom_fit}). To avoid uncertainties by comparing different studies that use different methods and assumptions, a full homogeneous analysis of Milky Way, LMC, SMC, and low $Z$ stars would be extremely beneficial to gain further insight into the $Z$ scaling of the winds of massive stars. 

\subsection{Comparing to other quantitative spectroscopic analyses of our sample stars}
\label{subsec:lowz_talk}

\begin{figure}
   \centering
   \includegraphics[width=\hsize]{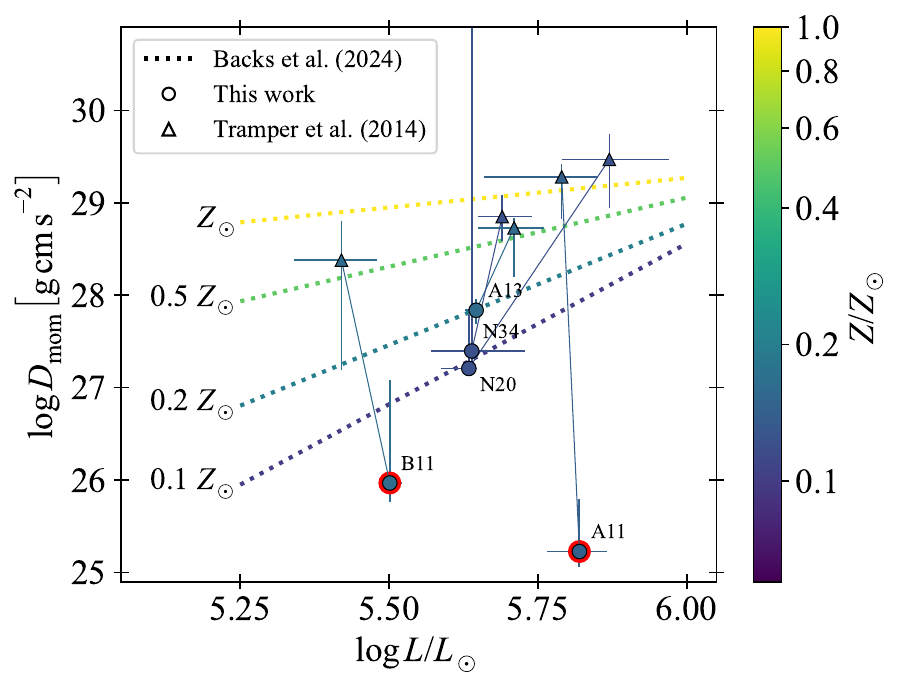}
      \caption{Comparing the values obtained for $D_{\rm mom}$ in this work (circles) to those obtained by \cite{tramper2014properties} (triangles). Solid lines join  the $(L,D_{\rm mom})$ pairs obtained for the same star but in the two different studies.}
         \label{fig:dmom_redetermine}
\end{figure}

Here we discuss previous analyses that included our sample stars. In Appendix \href{https://zenodo.org/records/15078070}{G}\footnote{Available online at \url{https://zenodo.org/records/15078070}.}, we detail the methods used in each in Table \href{https://zenodo.org/records/15078070}{G.1} and tabulate our results along with those obtained in these analyses in Tables \href{https://zenodo.org/records/15078070}{G.2} and \href{https://zenodo.org/records/15078070}{G.3}.

One of the first analyses of low $Z$ (i.e. $Z<Z_{\rm SMC}$) environments was carried out by \cite{tramper2014properties}. In this study, the optical spectra of a sample of 10 stars from IC\,1613, WLM, and NGC\,3109 were analysed. They found a significant discrepancy between their results and theoretical predictions and concluded that these low $Z$ stars produce winds that are stronger than predicted. However, these authors did not have access to the crucial resonance lines in the UV. These are much more sensitive to lower values of \mdot{} compared to optical wind lines, notably H$\alpha$ and, moreover, encode vital information about the acceleration behaviour of the wind outflow and the terminal velocity it can achieve. Therefore, without the UV spectrum in this low $Z$ regime, mass loss rates $\lesssim \SI{e-7}{\msun\per\year}$ are uncertain, and \vinf{} and $\beta$ become degenerate with \mdot{}. 

\cite{tramper2014properties} made assumptions in order to navigate the problems regarding \vinf{} and $\beta$. For \vinf, it was first scaled with the surface escape velocity, $v_{\rm esc}$ where $v_\infty = 2.65v_{\rm esc}$ \citep{lamers1995terminal} and then scaled with $Z$ as $v_\infty \propto Z^{0.13}$. \cite{garcia2014winds} found a large scatter around the canonical relation of $v_\infty / v_{\rm esc} = 2.65$, meaning this scaling may not hold in general, and recent results have found somewhat higher exponents in the $v_\infty(Z)$ scaling of $\sim 0.2$ \citep{hawcroft2021empirical, vink2021metallicity}. As for $\beta$, a value of 0.95 was adopted -- a value predicted for O5 supergiants \citep{muijres2012predictions} -- but this could be an uncertain assumption, as, in the case of a rapid rotator, for example, it may depend on factors such as the angle of inclination at which the star is being observed \citep{herrero2012peculiar}. 

We demonstrate the problems of obtaining \mdot{} from mere optical diagnostics in Fig. \ref{fig:dmom_redetermine} where we show updated \dmom{} and $L$ determinations from our optical+UV analysis compared to the optical only analysis of \cite{tramper2014properties} for the stars in both of our samples.  By including the UV spectra, we obtain lower \dmom{} values for all stars and, excluding the troublesome fits of A11 and B11, we find values for \dmom{} that are more in line with the empirical relation of \cite{backs2024smc}. We also find different values for $L$ for some stars, most notably N20. This is because the wind lines in the UV, and of course the other lines in this wavelength range, are sensitive to \teff, which impacts the luminosity determination. This highlights the necessity to include both the optical and UV spectra in the analysis of low $Z$ massive stars.

\cite{bouret2015spectro} also highlight the importance of including the UV spectra in such analyses. They analysed the optical and UV spectra of A13, A11, and B11 and also found lower values for \mdot{} than \cite{tramper2014properties} (see Table \href{https://zenodo.org/records/15078070}{G.3} for the values obtained in each study). For A13, we find a value for \mdot{} consistent with theirs, however, we find lower values for A11 and B11. Regarding these latter two stars, \cite{bouret2015spectro} speculate that, for such weak winds, porosity and vorosity effects on wind structure and the presence of hot gas may severely bias the estimated mass loss rate. We revisit this in Sect. \ref{subsubsec:thick_clumping}.

Interestingly, both \cite{bouret2015spectro} and \cite{garcia2014winds} found evidence that the Fe abundance of the stars they studied may be higher than the typically adopted literature values for their host galaxies (IC\,1613 for both authors, and WLM for \citealp{bouret2015spectro}) and are instead closer to an SMC-like abundance of $\SI{0.2}{\zsun}$. \cite{bouret2015spectro} found that models with Fe/Fe$_\odot$ of both 0.14 and 0.20 were compatible with the Fe lines in the observed spectra, although there was some degeneracy with the micro-turbulent velocity, \microturb. However, they motivated that a value of $\SI{0.2}{\zsun}$ is a more realistic metallicity value for both galaxies. For the IC\,1613 stars, their measured value of 0.2\,Fe$_\odot$ is consistent with iron abundances of late type stars \citep{tautvaivsiene2007first} and with the star formation history of the galaxy \citep{skillman2003deep}. \cite{garcia2014winds} concur with this, as they find that model spectra with $Z=\SI{0.2}{\zsun}$ were required for their analysis and that the spectra of the stars in their sample closely resemble stars in the SMC with the same spectral type. We return to the metallicity of IC\,1613 in Sect. \ref{subsubsec:new_z}.

To conclude this section, we show that the UV spectra are necessary when analysing the wind properties of low $Z$ massive stars, as previously shown by \cite{bouret2015spectro} and \cite{garcia2014winds}. 

\subsection{Terminal velocities}
\label{subsec:vinf}

\begin{figure}
   \centering
   \includegraphics[width=\hsize]{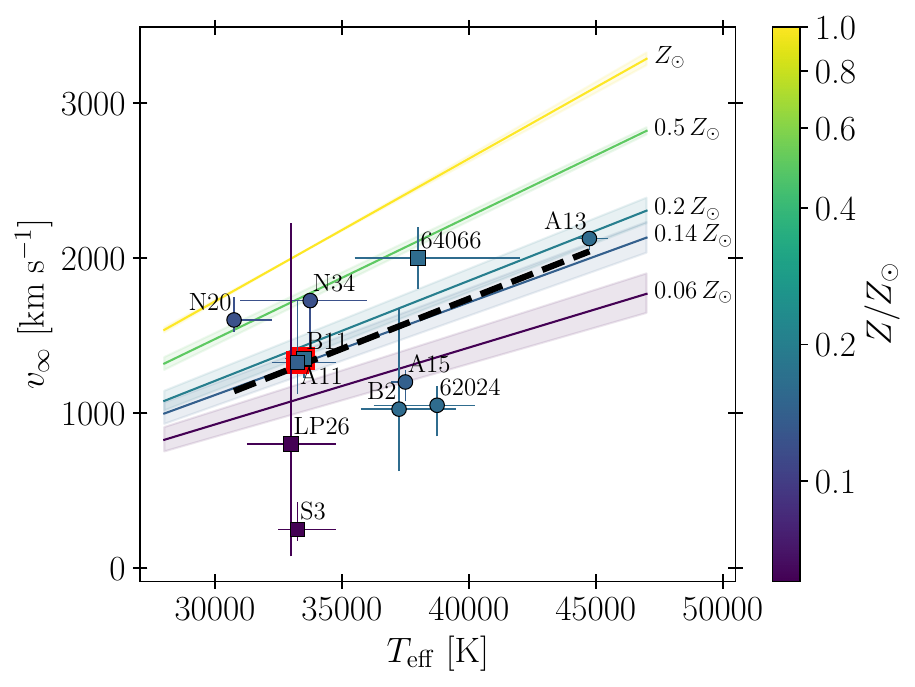}
      \caption{Terminal velocities overplotted on the empirical relation of \cite{hawcroft2024vinf}. The black dashed line represents an ODR fit through the circle points. As B2 and A15 have the same \teff{}, that of the former has been offset by $-\SI{250}{\kelvin}$ for clarity. For similar reasons, that of B11 has been offset by $+\SI{250}{\kelvin}$.}
         \label{fig:vinf_plot}
\end{figure}

The determination of the terminal velocity, \vinf, proved to be quite challenging for many of the stars in this sample.  Because of the weak winds of these stars, many do not have saturated P Cygni profiles, with some barely showing any wind signatures. In these cases, both \vinf{} and $\beta$ are hard to constrain. There are, however, stars with \civ\,$\lambda 1550$ features that show a well defined P Cygni profile (A13, N20, N34), so the parameters obtained for these stars can be considered robust. Furthermore, for the very low metallicity stars, LP26 and S3, we find very low values of $800^{+1400}_{-720}$\,km\,s$^{-1}$ and $250^{+180}_{-75}$\,km\,s$^{-1}$, respectively, consistent with \cite{telford2024observations}.

We compare our findings to the empirical relation of \cite{hawcroft2024vinf}. They analysed the \civ{}\,$\lambda 1550$ profiles of samples of Milky Way, LMC, and SMC stars. If the profile was saturated, \vinf{} was determined from the bluest edge of the P Cygni absorption trough, while the SEI method \citep{1987ApJ...314..726L} was employed if the profile was not saturated. The following relation was determined after fitting \vinf{} as a function of \teff{} and $Z$ for the entire sample:
\begin{equation}
    v_\infty (T_{\rm eff}, Z)=( \num{9.2e-2}T_{\rm eff} - 1040)\times Z^{0.22}.
    \label{eqn:hawcroftvinf}
\end{equation}
\begin{figure*}
   \centering
   \includegraphics[width=\hsize]{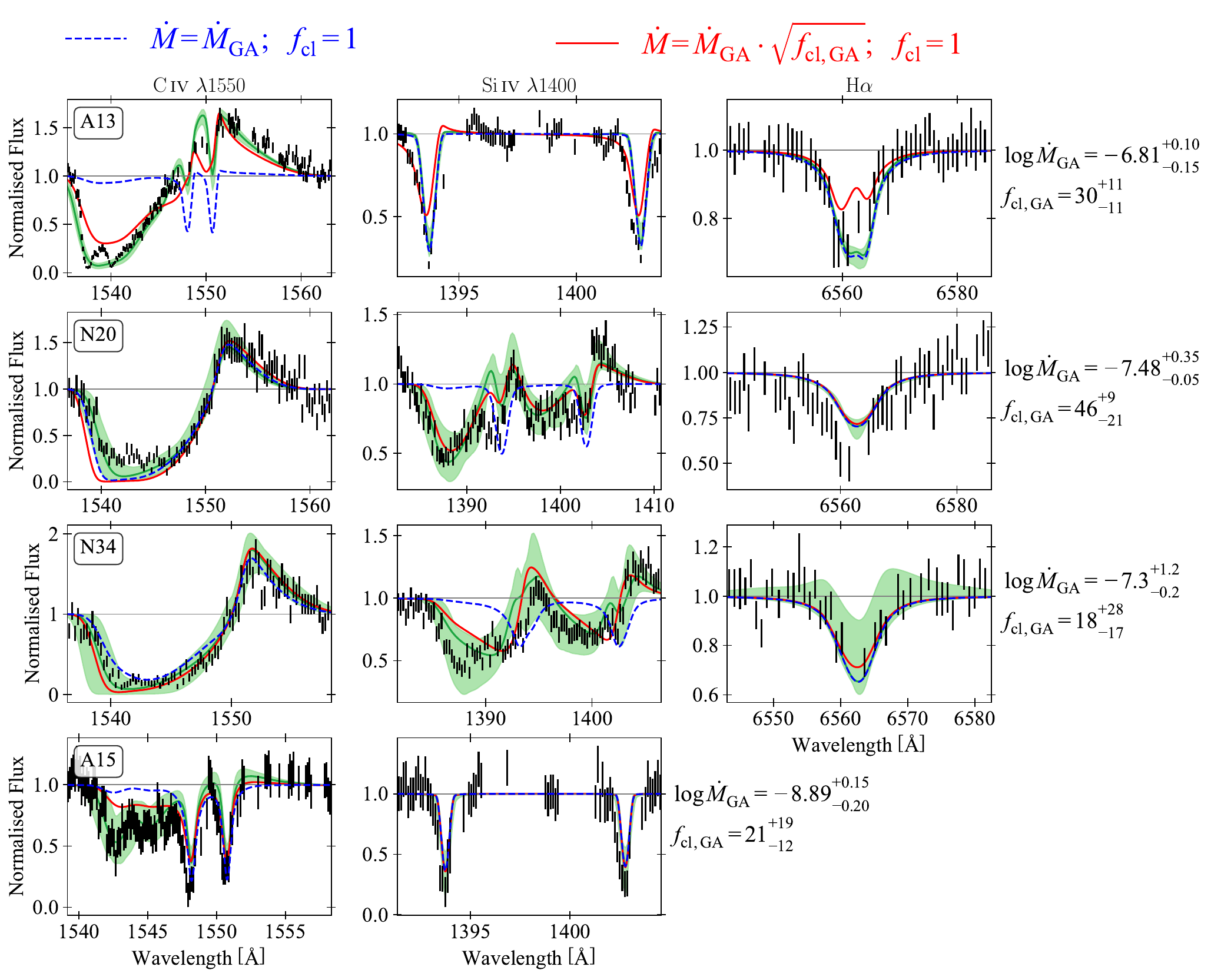}
      \caption{A comparison between the best-fitting \fw{} models determined by \kiwi{} where optically thin clumping was assumed (green), a \fw{} model with the same stellar and wind parameters but with a smooth wind (i.e. $f_{\rm cl}=1$;  blue dashed line), and a smooth \fw{} model, again with the same stellar parameters as the \kiwi{} fit, but with a mass loss rate scaled by a factor of $\sqrt{f_{\rm cl,GA}}$. }
         \label{fig:clumping_comparing}
\end{figure*}
As this relation was only determined from stars with metal contents as low as SMC metallicity, we extrapolate it here. Figure \ref{fig:vinf_plot} shows \vinf{} obtained for the stars studied in this work overplotted on Eqn. \eqref{eqn:hawcroftvinf} at Milky Way, LMC, and SMC metallicity, and extrapolated to lower metallicities of $0.14\,Z_\odot$ and $0.06\,Z_\odot$. We perform an ODR fit through our data, for which we limit ourselves to the same six points as used to fit $D_{\rm mom}$ in Fig.\,\ref{fig:comparing_dmom_theory}; that is, we excluded 64066, S3 and LP26, and those for which the spectral fits were of poor quality (B11 and A11). We find that the fit is consistent with the empirical relation at 0.14\,$Z_{\odot}$, however, there is a lot of scatter; the uncertainty region has been excluded for clarity, but it covers all other empirical tracks. This is a small sample whose \vinf{} values were determined from low resolution and low SNR spectra. Care must therefore be taken with this fit; a larger sample of higher resolution spectra is needed to infer more about this. Furthermore, as discussed in Sect. \ref{subsec:result_implications}, if X-rays are shocking material in the winds of these stars, the values for \vinf{} may potentially be underestimated here, especially in S3 and LP26. Finally, we remark that the terminal velocities of the three fitted sources in IC\,1613 (A13, B2, and 62024) are not remarkably high relative to the other three fitted sources.

\begin{table*}
\centering
\caption{Best-fit parameters for A13 and A15 where optically thick clumping has been assumed.}
\label{tab:thick_table}
\begin{small}
\begin{tabular}
{l r@{}l r@{}l r@{}l r@{}l r@{}l r@{}l r@{}l r@{}l r@{}l r@{}l r@{}l}
\midrule\midrule
Star  & \multicolumn{2}{c}{$T_{\rm eff}$} & \multicolumn{2}{c}{$\log g$} & \multicolumn{2}{c}{$Y_{\rm He}$} & \multicolumn{2}{c}{$\xi$} & \multicolumn{2}{c}{$\log \dot{M}$} & \multicolumn{2}{c}{$v_{\infty}$} & \multicolumn{2}{c}{$\beta$} & \multicolumn{2}{c}{$f_{\rm cl}$} & \multicolumn{2}{c}{$f_{\rm ic}$} & \multicolumn{2}{c}{$f_{\rm vel}$} & \multicolumn{2}{c}{$v_{\rm turb}/v_{\infty}$} \\
  & \multicolumn{2}{c}{[kK]} & \multicolumn{2}{c}{$[{\rm cm\,s}^{-2}]$} & \multicolumn{2}{c}{[$N_{\rm He}/N_{\rm H}$]} & \multicolumn{2}{c}{$[{\rm km\,s}^{-1}]$} & \multicolumn{2}{c}{[$M_\odot\,{\rm yr}^{-1}$]} & \multicolumn{2}{c}{$[{\rm km\,s}^{-1}]$} & \multicolumn{2}{c}{} & \multicolumn{2}{c}{} & \multicolumn{2}{c}{} & \multicolumn{2}{c}{} & \multicolumn{2}{c}{} \\
\midrule
A13 & $47.0$ & $^{+2.0}_{-0.3}$ & $3.85$ & $^{+0.15}_{-0.03}$ & $0.2$ & $^{+0.03}_{-0.08}$ & $21.0$ & $^{+0.3}_{-0.9}$ & $-6.6$ & $^{+0.1}_{-0.1}$ & $2218$ & $^{+31}_{-62}$ & $1.10$ & $^{+0.24}_{-0.06}$ & $23$ & $^{+1}_{-9}$ & $0.21$ & $^{+0.06}_{-0.01}$ & $0.64$ & $^{+0.01}_{-0.32}$ & $0.02$ & $^{+0.04}_{-0.01}$  \\[3pt]
A15 & $37.5$ & $^{+0.2}_{-0.2}$ & $3.78$ & $^{+0.04}_{-0.02}$ & $0.12$ & $^{+0.044}_{-0.004}$ & $11.9$ & $^{+4.1}_{-0.3}$ & $-8.14$ & $^{+0.06}_{-0.10}$ & $1225$ & $^{+25}_{-75}$ & $0.68$ & $^{+0.02}_{-0.04}$ & $2$ & $^{+3}_{-1}$ & $0.18$ & $^{+0.17}_{-0.01}$ & $0.16$ & $^{+0.01}_{-0.12}$ & \multicolumn{2}{c}{$-$}  \\[3pt]
\midrule
\end{tabular}
\end{small}
\end{table*}
\begin{figure*}
    \centering\includegraphics[width=\hsize]{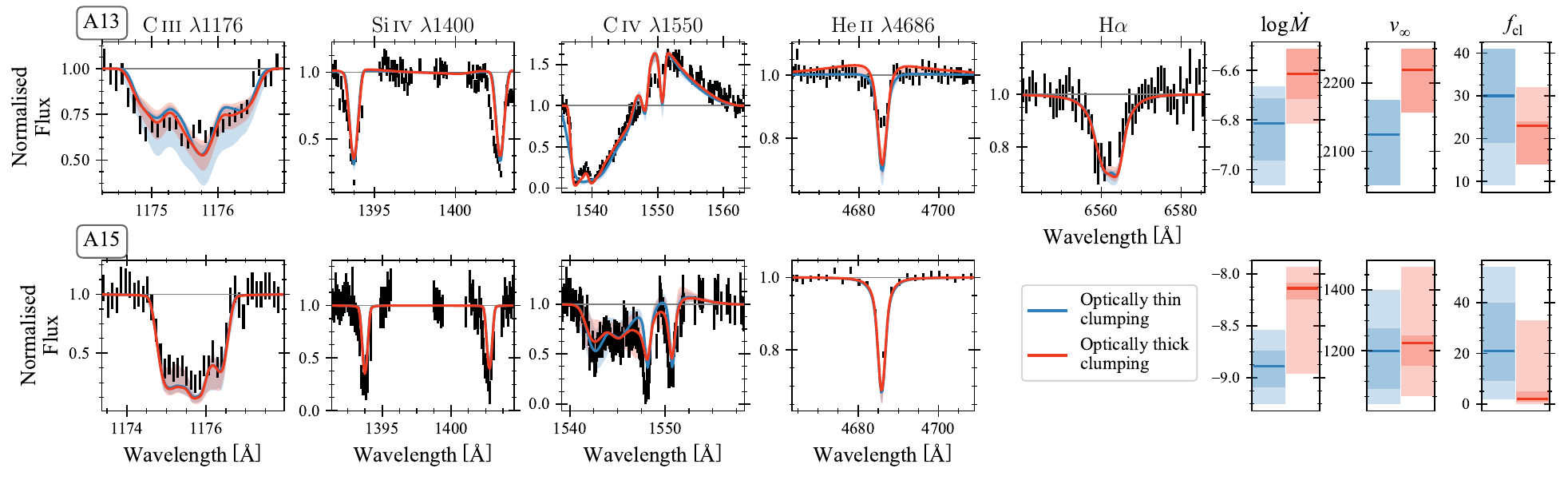}
    \caption{Comparing the optically thin formalism used in this work (blue) to the optically thick formalism (red) for A13 in IC\,1613 (for which $v_{\rm windturb}$ is also a free parameter; top) and A15 in WLM (bottom). Also shown in the three rightmost plots are the best-fit (solid line), 1$\sigma$ (dark shaded region), and 2$\sigma$ (light shaded region) uncertainties obtained for \mdot{}, \vinf{}, and \fcl{} for both runs.}
        \label{fig:thickthincompare}
\end{figure*}

\subsection{Impact of our assumptions}
\label{subsec:assumptions}
As in all modelling, for practical purposes, we have made a number of simplifying assumptions to achieve the goals of this paper. In this section, we examine the potential consequences of making such assumptions which will inform future work once larger samples become available. First, in Sect. \ref{subsubsec:thick_clumping}, we discuss the assumptions we have made regarding the wind structure, particularly in the context of A11 and B11. Then, in Sect. \ref{subsubsec:new_z}, we examine our assumption that the metal content of the stars in our sample are equal to those of their host galaxy, given in Table \ref{tab:gals}.

\subsubsection{Wind structure}
\label{subsubsec:thick_clumping}

Here, we first comment on the robustness of the values we obtain for the clumping factors in the optically thin clumping formalism. We then discuss possible effects of optically thick clumping, and conclude this section with a discussion of A11 and B11 in this context.

\paragraph{Optically thin clumping.}

As mentioned in Sect. \ref{subsec:results_windparams}, many of the values we obtain for \fcl{} are largely unconstrained as they have large uncertainties -- some spanning the entire parameter space. This raises the question whether these values hold physical significance or are a result of overfitting. We tested this on a number of stars in the sample: A13, N20, and A15, with reasonably well constrained posteriors of \fcl, and N34, whose values are essentially unconstrained.

We demonstrate this in Fig. \ref{fig:clumping_comparing}. The green fit represents the best-fitting \fw{} models and 1$\sigma$ uncertainties determined by \kiwi{} with optically thin clumping implemented. To see if the UV resonance lines are clumping sensitive in this $Z$ regime, we first calculated \fw{} models with identical parameters as the best fitting stellar and wind parameters determined by \kiwi{} but with a smooth wind, shown as the dashed blue lines in the figure. We find that these features are indeed sensitive to variations in \fcl{}, provided they are formed in the wind, as wind features become photospheric (\civ\,$\lambda 1550$ for A13 and A15, and \siiv\,$\lambda 1400$ for N20 and N34). For A13, the reason why the \civ\,$\lambda 1550$ profile becomes photospheric in the model with a smooth wind is because C\,{\sc v} is the dominant ionisation stage of carbon in the wind here, resulting from its high \teff{} of $\sim\SI{45}{\kilo\kelvin}$, and therefore no \civ{} is seen in the wind. This is because the recombination rate increases quadratically with density, while the ionisation rate increases only linearly. Therefore, in the model with a clumped wind, \civ{} is the dominant ionisation stage due to the increased recombination from C\,{\sc v} in the overdense regions, resulting in a well-developed P Cygni profile.

To compensate for the lack of clumping, we then calculated smooth \fw{} models, again with the same best-fitting stellar parameters determined by \kiwi{}, but with a larger mass loss rate scaled by $\sqrt{f_{\rm cl,GA}}$; where $f_{\rm cl,GA}$ is the best-fitting clumping factor determined by \kiwi{}. For A13, we see that, while \civ\,$\lambda 1550$ is formed in the wind at this increased \mdot, there is also a wind signature in \halpha{} as the line core is filled in, which is not observed in the spectrum. For N20, this test model reproduces the \siiv{} feature reasonably well but saturates the \civ{} profile, which is not observed in this star. For N34, the test model reproduces the \civ{} profile, but is unable to reproduce the absorption observed in the blue feature of the \siiv{} doublet. This minimal change in the shape of the \civ{} profile can be expected in these two stars as it is saturated in both models, meaning it is less affected by clumping \citep[e.g. ][]{bouret2005lower}. Finally, for A15, we can see that clumping is required to reproduce the blue absorption in the \civ\,$\lambda 1550$ line. These discrepancies suggest that wind inhomogeneities should be accounted for in stars in this $Z$ regime.

We conclude that, provided they are formed in the wind, the UV resonance lines in the spectra of stars in this $Z$ regime are sensitive to variations in \fcl{} in the optically thin clumping formalism implemented in \fw. As such, these features can provide somewhat meaningful constraints on the clumping factor. This suggests that our obtained values of \fcl{} for our sample stars with non-saturated wind signatures in their spectra are not entirely a result of overfitting.

\paragraph{Optically thick clumping.}
\begin{figure*}
   \centering
   \includegraphics[width=\hsize]{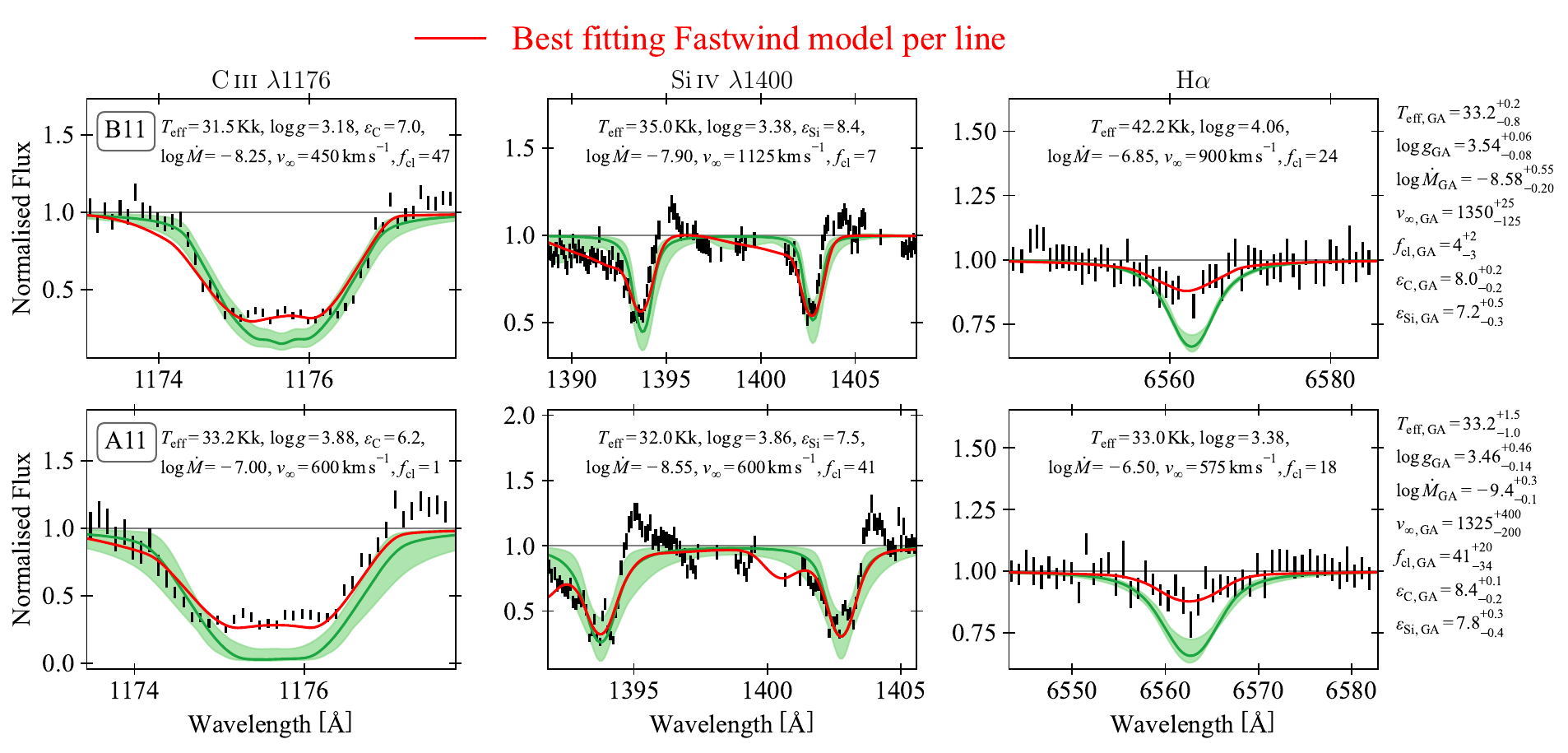}
      \caption{Best-fitting \fw{} models per line determined by \kiwi{} (red) compared to the overall best-fitting \fw{} model and 1$\sigma$ uncertainties (green) in the optically thin formalism. The top and bottom rows are B11 and A11, respectively. On each axis, the main parameters of the best-fitting model of the line shown on the axis are given. For comparison, we give the parameters of the overall best-fitting model determined by \kiwi{} on the side of the figure.}
         \label{fig:fw_mode_per_line_a11b11}
\end{figure*}

In Fig. \ref{fig:thickthincompare} we present test calculations assessing the possible effects of the assumption of optically thin clumping, as well as others we have made, on two stars: A13 and A15; the most and least luminous stars, respectively, considered in the $D_{\rm mom}(L)$ fits in previous sections.

To this end, we test the possibility that the wind contains an ensemble of clumps that may range from being optically thin to optically thick (see Sect.~\ref{subsec:fw}). For a detailed explanation of this so-called macro clumping assumption and the way in which it is implemented in {\sc Fastwind}, see \cite{sundqvist2018clumping,brands2022r136}.

In the figure, we compare our adopted results, shown in blue, to a \kiwi{} fit adopting different assumptions, shown in red. For A13, optically thick clumping is assumed in this additional \kiwi{} run and $v_{\rm windturb}$ is left as a free parameter, while for A15, only optically thick clumping is assumed. We also show the effect that this has on the determination of the wind parameters \mdot{}, \vinf{}, and \fcl{} in panels on the right side of the figure. We decide to leave $v_{\rm windturb}$ free for A13 only in this test because it is clear, from the mismatch of our best-fit model and the blue edge of the \civ\,$\lambda 1550$ profile, that our assumed value of $0.14\,v_{\infty}$ is too large. Such a conclusion cannot be drawn from the fit of A15, as its wind features are not as prominent. 

This optically thick clumping formalism in \fw{} allows for the inter-clump density contrast, $f_{\rm ic}$, to be left as a free parameter, as well as for the velocity-porosity in the wind to be accounted for. Velocity-porosity is a measure of the velocity dispersion within the clumps and is described by a factor $f_{\rm vel}$ \citep[see, e.g.][]{sundqvist2018atmospheric, brands2022r136} which we make a free parameter in this exercise. To allow for a more thorough exploration of the parameter space, we fix \vrot{} in each run to the value determined from the original fit where optically thin clumping was assumed. We also show the results obtained from the fits in Table \ref{tab:thick_table}.

For A13 (top row of Fig. \ref{fig:thickthincompare}), we find that the assumption of optically thick clumping and leaving $v_{\rm windturb}$ free does not significantly affect the determinations of \mdot{}, \vinf{}, and \fcl{}, as they are equal within 1$\sigma$ uncertainties. We note that the fit to the \civ\,$\lambda 1550$ profile is better if we assume optically thick clumping and fit $v_{\rm windturb}$, however, this is probably more affected by the latter. If $v_{\rm windturb}$ is lower than what we assume, the blue edge of the P Cygni profile will be steeper, as there is less excess velocity at \vinf{} in the wind. Furthermore, the smaller $v_{\rm windturb}$ allows for the best-fitting \fw{} model to reproduce the blueshifted doublet absorption feature at $\sim \SI{1540}{\angstrom}$. In this new fit, a lower value of $\beta = 1.10^{+0.24}_{-0.06}$ is found that is closer to the value of 0.8 obtained by \cite{garcia2014winds} and \cite{bouret2015spectro}, though still equal within uncertainties to the higher value obtained in the original, optically thin fit. Other than this, we find no significant differences between the two assumptions for the rest of the line profiles.

Here we find that $v_{\rm windturb} = \left(0.02 ^{+0.04}_{-0.01}\right)\,v_\infty$ for A13 -- significantly lower than the value of $0.14\,v_\infty$ that we assume, and the value of $(0.14 \pm 0.06)\,v_\infty$ determined from the LMC sample of \cite{brands2022r136}. This could suggest that lower $Z$ stars exhibit lower turbulence in their winds, however, to be further investigated, this must be explored with high-resolution and high SNR spectra of more stars that exhibit P Cygni profiles.

\begin{figure*}
    \centering\includegraphics[width=\hsize]{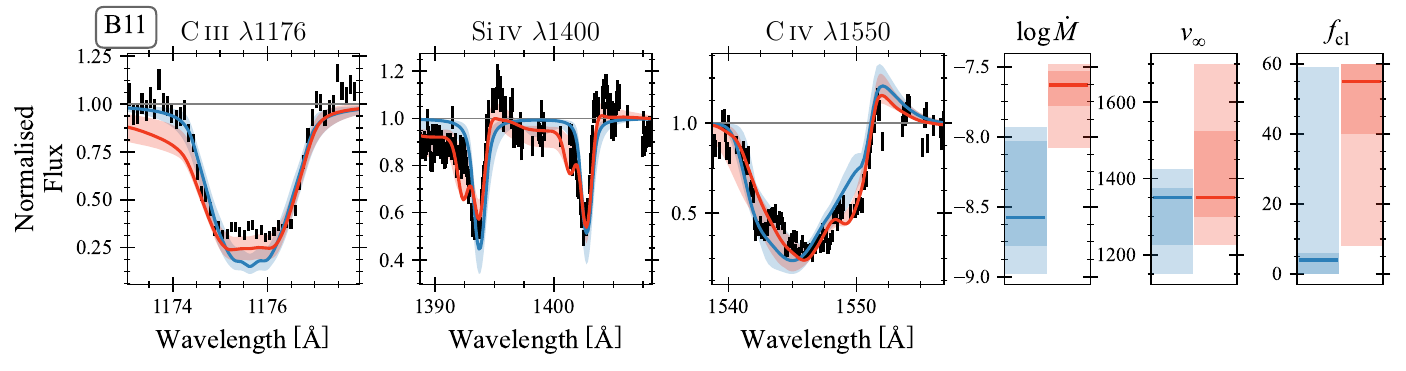}
    \caption{\kiwi{} fits of B11 where optically thin (blue) and thick (red) clumping has been assumed. As in Fig. \ref{fig:thickthincompare}, we show the best-fitting values and uncertainties for \mdot, \vinf, and \fcl{} obtained in each run.}
        \label{fig:thickthincompare_b11}
\end{figure*}

For the case of A15 (bottom row), if the clumps are allowed to become optically thick, we find an \mdot{} that is a factor of $\sim 6$ higher than that obtained in the optically thin case while also finding a value for \vinf{} that is equal within uncertainties to that obtained when optically thin clumping was assumed. As \mdot{} can be affected, this test shows that the assumption of optically thick clumping may influence the slope and offset of the $D_{\rm mom}(L)$ relation.

That being said, we do find that a smooth wind cannot be ruled out in the optically thick \kiwi{} fit of A15, as $f_{\rm cl} = 2^{+3}_{-1}$ here. This value is significantly lower than that obtained from the fit under the assumption of optically thin clumping, where $f_{\rm cl} = 21^{+19}_{-12}$. Such a change in \fcl{} can change the shape of the \civ\,$\lambda 1550$ profile, and consequently, \mdot. This introduces a degeneracy between \mdot{} and \fcl{}. The modest SNR of the data, combined with the limited diagnostic features present in the spectra of these low $Z$ targets, makes it difficult to break this degeneracy at the present moment. Furthermore, due to the large uncertainties on the data for this star, the fits from both runs are consistent within uncertainties. Future work should investigate the effects of optically thick clumping in low $Z$ stars and their potential impact on the $D_{\rm mom}(L)$ relation.

\paragraph{B11 and A11.}
As mentioned in Sect. \ref{subsec:results_anomalous}, our \kiwi{} fits to these stars were not satisfactory, therefore we did not include them in the $D_{\rm mom}(L)$ fits in Sect. \ref{subsec:fits}. To test whether this was an issue of an insufficiently large parameter space, we show for both stars, in Fig. \ref{fig:fw_mode_per_line_a11b11}, the best-fitting \fw{} models of the common features between the two stars that are poorly fit, \ciii\,$\lambda1176$, \siiv\,$\lambda1400$, and \halpha, and compare them to the overall best-fitting \fw{} model determined by \kiwi. For both stars, the best-fitting models for each line provide improved fits compared to the overall best-fitting model. The core of the \ciii\,$\lambda1176$ profile is desaturated for both stars and the red wing is better reproduced for A11. The blue absorption of both features of the \siiv{} doublet is captured by the best-fitting model of this line complex for both stars (not so much for the blue component of A11), while the red emission is still not present in these models. The \halpha{} profile of the best-fitting model of this line is better fit for both stars than the best-fitting overall model. For these stars, no unique set of fitting parameters can simultaneously provide a satisfactory fit for all lines in the spectra of these stars, while models with different combinations of parameters can provide good fits individually. This suggests that there may exist a more optimal solution using some combination of parameters from outside the parameter space. This could also mean, however, that there are some other effects at play in these stars that are not accounted for by our assumptions made in this work, which we explore next.

These stars were also analysed by \cite{bouret2015spectro} who similarly reported low $\dot{M}$ values. They suggest that their values may be underestimated due to a more complex wind structure than simply optically thin clumping, with spatial- and velocity-porosity effects possibly becoming important.

They investigate this hypothesis by computing a model for A11 with a wind structure that mimics a wind with hot gas that does not contribute to the UV spectrum. Focusing on the \ciii$\,\lambda 1176$ and \siiv$\,\lambda 1400$ lines only, they found that this new model better reproduced the observed features by desaturating the centre of the \ciii$\,\lambda 1176$ line and exhibiting red emission and blueshifted absorption in the \siiv$\,\lambda 1400$ line. 

We also find that our best-fitting \fw{} model cannot reproduce these features in these lines for B11 and A11 (whose fits are shown in Figs. \href{https://zenodo.org/records/15078070}{H.4} and \href{https://zenodo.org/records/15078070}{H.6}, respectively, and as shown above). Therefore, to test the statement of \cite{bouret2015spectro}, specifically regarding the impact of porosity in the winds, we performed \kiwi{} fits for these stars assuming optically thick clumping, where this effect is accounted for. We found a similar unsatisfactory fit for A11, so we only discuss B11 here. Figure \ref{fig:thickthincompare_b11} shows, for B11, the \ciii$\,\lambda 1176$, \siiv$\,\lambda 1400$, and \civ$\,\lambda 1550$ lines and the best-fitting \fw{} models determined by \kiwi{} where optically thin (blue) and optically thick (red) clumping has been assumed. We find similar results to \cite{bouret2015spectro}, namely that, when optically thick clumping is assumed, the centre of the \ciii{} line desaturates (albeit at the expense of excess blue absorption), the blue absorption and red emission in the \siiv{} line is reproduced within uncertainty, and that \mdot{} increases by a factor of $\sim 10$.

This shows that the assumption of optically thick clumping in \fw{} can help to reproduce features in observed line profiles that optically thin clumping cannot account for, suggesting that these effects may indeed become important in the winds of these stars as hypothesized by \cite{bouret2015spectro}. Furthermore, because a unique solution was found here using stellar and wind parameters from the original parameter space explored (albeit with the introduction of new free parameters like porosity), this suggests that the parameter space explored in the optically thin runs are sufficient and the original troublesome fits are more likely due to effects not accounted for by our original assumptions. However, as noted above, the optically thick clumping formalism in \fw{} introduces a number of extra free parameters that are difficult to constrain with data of this quality. Therefore, care should be taken when interpreting this.

\subsubsection{Adopted metal contents}
\label{subsubsec:new_z}
As mentioned in Sect.\,\ref{subsec:sample} and discussed in Sect. \ref{subsec:lowz_talk}, it has been suggested that IC\,1613 and WLM may have higher metal contents than have been assumed in this work, and are probably closer to an SMC-like value \citep{garcia2014winds, bouret2015spectro}. This would therefore affect the $D_{\rm mom}(L)$ fits presented in previous sections as $\SI{0.2}{\zsun}$ is considerably different to the other lower metal contents. Here, we examine this for IC\,1613 only, as indications for a higher metallicity appear somewhat more substantial for this dwarf irregular.

\begin{figure}
   \centering
   \includegraphics[width=\hsize]{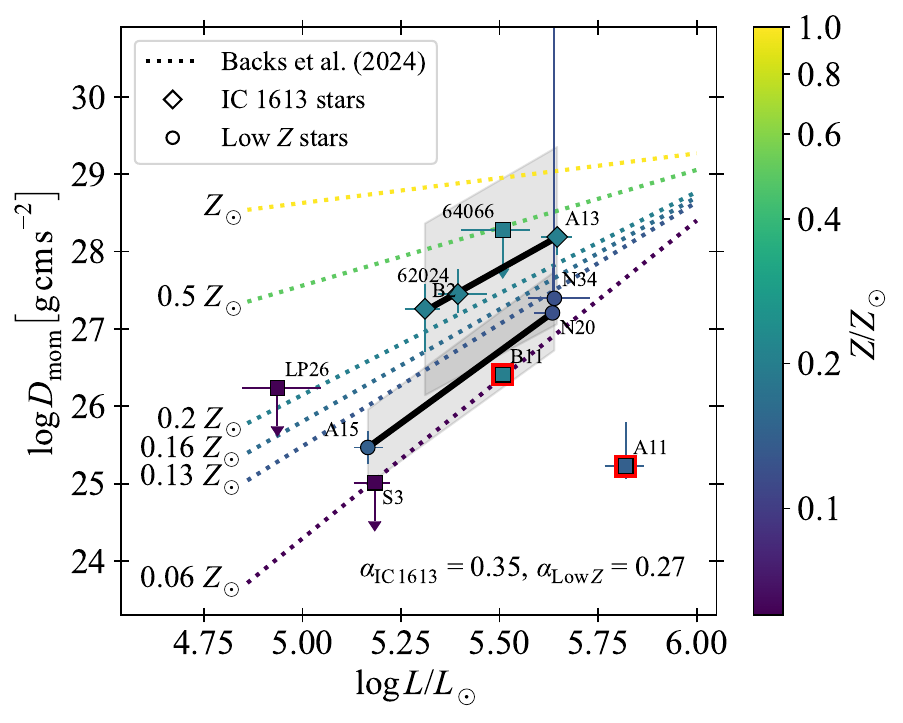}
      \caption{Separate fits of $D_{\rm mom}(L)$ for the stars in WLM and NGC\,3109 (circles) and IC\,1613 (diamonds) where the points representing the latter stars are from \kiwi{} runs where $Z=\SI{0.2}{\zsun}$ was assumed. Grey shaded regions represent 1$\sigma$ uncertainties on the lines of best-fit. Square points have not been included in the fit and red borders indicate troublesome runs. The colour scheme is the same as that of Fig. \ref{fig:dmom_fit}.} 
         \label{fig:dmom_twofits}
\end{figure}

We investigate potential consequences of our assumed metallicity of $\SI{0.16}{\zsun}$ for this galaxy by performing \kiwi{} fits for these stars with metallicities of $\SI{0.2}{\zsun}$ instead, and then fitting $D_{\rm mom}(L)$ both for these stars alone and for the stars in the lower $Z$ galaxies. Figures where, for the same source, fits for different $Z$ are compared, as well as comments on these new fits, are given in Appendix \ref{a_sec:ic1613_fits}. The fit of B11 was, again, troublesome as it had similar issues with that of the lower $Z$ case, while the rest of the fits were satisfactory. We therefore only fit $D_{\rm mom}(L)$ for B2, 62024, and A13 in this case, and perform a separate fit for A15, N34, and N20, assuming an average $Z$ of $\sim \SI{0.13}{\zsun}$. The resulting $D_{\rm mom}(L)$ fits are shown in Fig. \ref{fig:dmom_twofits} where, unless otherwise stated in the legend, colours and symbols have the same meanings as in Fig. \ref{fig:dmom_fit}. We also show the uncertainty region as it is clearer than for the case where all six stars were fitted together. 

Both fits produce a different slope and therefore a different $\alpha$, the two values of which are quoted in the bottom right of the figure. The higher $Z$ IC\,1613 sample resulted in a higher $\alpha$ than that of the lower $Z$ sample, which is to be expected from \citetalias{castor1975radiation} theory. The higher $Z$ fit is now in line with the empirical relation of \cite{backs2024smc} at $\SI{0.2}{\zsun}$, albeit rather uncertain as it is consistent with the LMC line but also the $\SI{0.13}{\zsun}$ line within 1$\sigma$ uncertainties. The fit of the lower $Z$ sample is positioned lower than the SMC line and is consistent with $\SI{0.13}{\zsun}$ line. The slopes of both fits are in good agreement with the empirical relation of \cite{backs2024smc} and the higher positioning of the IC\,1613 stars is, at face value, in line with the suggestion that perhaps the metallicity of this galaxy is more SMC-like.

However, the sample size (three stars per fit) is very small. At the present moment, therefore, the trade off that has to be made is one between reducing further an already small sample size, or having a larger sample (of six stars) and assuming a same average $Z$. Future work incorporating larger samples will be able to better account for different metallicities.

\section{Summary and conclusions}
\label{sec:summary}
In this work, we analysed the combined optical and UV spectra of a sample of 11 O-stars in nearby dwarf galaxies with $Z < 0.2\,Z_\odot$. These eleven stars in the irregular dwarf galaxies IC\,1613, WLM, NGC\,3109, Sextans\,A and Leo\,P constitute the full sample for which both spectral ranges are currently available in the literature. 
We used \kiwi{} and \fw{} to determine stellar and wind parameters for each of the stars. Given the large distances to these galaxies the quality of the spectra (with SNR $\sim 2-8$ in the UV and $\sim 10-60$ in the optical at modest spectral resolution), we opted to assume that clumps in the outflows are optically thin.
Our main findings can be summarised as follows: 

\begin{itemize}
    \item{We corroborate the important finding by \cite{garcia2014winds} and \cite{bouret2015spectro} that constraining mass loss rates of O-type stars in galaxies of metallicities below that of the SMC (with $Z = 0.2\,Z_{\odot}$) requires ultraviolet spectra of resonance line ions such as \civ\,$\lambda 1550$. If only optical spectra are available, with H$\alpha$ as the primary wind diagnostic, mass loss rates may be overestimated by as much as an order of magnitude.}

    \vspace{0.7mm}

    \item{We compare our fit of the modified wind momentum $D_{\rm mom}$, a measure of wind strength that most easily allows for a comparison of stars with different $Z$ and for a confrontation with theory, with the most sophisticated empirical relation currently available for $D_{\rm mom}(L,Z)$, established so far for the metallicity range $Z = 0.2-1\,Z_{\odot}$ \citep[see][]{backs2024smc}. Utilizing the six stars for which the wind strength is most reliable, we find that at (an average) 0.14\,$Z_{\odot}$ these are in line with this empirical relation (Fig. \ref{fig:dmom_fit}). This supports the conclusion discussed and reviewed in \cite{backs2024smc} that at lower luminosity mass loss shows a steeper $Z$-dependence than at higher luminosity.}

    \vspace{0.7mm}

    \item{The slope of $D_{\rm mom}(L)$ is a measure of the line-force parameter $\alpha$ in \citetalias{castor1975radiation} theory. For the six stars with $Z \sim 0.14\,Z_{\odot}$ mentioned above, in IC\,1613, WLM, and NGC\,3109, we find $\alpha = 0.21\pm0.04$ ignoring mass effects. If a mass dependence is considered in \citetalias{castor1975radiation} theory, $\alpha = 0.18\pm0.06$. The steepness of the slope, mentioned in the previous finding, linked to the low $\alpha$ values at $Z \sim 0.14\,Z_{\odot}$ mentioned here is not reproduced by hydrodynamical simulations of wind driving. These predict a shallower $D_{\rm mom}(L)$ dependence at the metallicity of our sample stars.
    }

    \vspace{0.7mm}

    \item{Empirically, in the metallicity range $Z = 0.2-1.0\,Z_{\odot}$, it is found that the terminal flow velocity \vinf{} is less at lower metallicity \citep{hawcroft2024vinf}. The sample is currently too small to make meaningful statements regarding a \vinf $(Z)$-behaviour at lower metallicity.}
    
\end{itemize}

Local Group irregular dwarf galaxies with metallicities below that of the SMC (at $Z = 0.2\,Z_{\odot}$) are probes of the environmental conditions in massive galaxies {\em before} the peak of cosmic star formation at redshift $2-3$ \citep{2014ARA&A..52..415M,2024MNRAS.532.3102S}. Establishing properties of populations of massive stars in these nearby galaxies are therefore key to calibrate our theories of atmospheres and evolution of these objects, providing us with predictions of statistical properties of their end-of-life products, including supernova types and black hole masses. For both these evolution outcomes an understanding of mass loss properties of O-type stars is crucial. To this end, future analyses using stellar atmosphere codes that can be used to infer the iron abundances of these metal-poor stars, such as \powr{}, as was used by \cite{telford2024observations}, would be useful. This most comprehensive and complete study of these properties at $Z<Z_{\rm SMC}$, to date, reveals (and strengthens earlier indications) that theoretical expectations and empirical findings do not match in full. Larger samples of such objects are needed. First steps to enlarge the sample of UV spectra are underway with ongoing low- and moderate-resolution HST observations (Program ID: 17491, PI: Telford). A giant leap forward in securing optical spectra can be expected once the second-generation multi-object spectrograph MOSAIC is mounted on the Extremely Large Telescope.

\section*{Data availability}
Appendix \href{https://zenodo.org/records/15078070}{G}, which contains a comparison with literature values of the sample stars, and Appendix \href{https://zenodo.org/records/15078070}{H}, presenting our \kiwi{} fits to our sample stars, are available online at Zenodo (\url{https://zenodo.org/records/15078070}). Other data may be made available upon request to the lead author.

\begin{acknowledgements}
We thank the anonymous referee for the insightful comments that have helped to improve the quality of this manuscript. We thank SURF (\url{https://www.surf.nl/}) for the support in using the Dutch National supercomputer, Snellius. OGT acknowledges support from a Carnegie-Princeton Fellowship, through Princeton University and the Carnegie Observatories. FB acknowledges the support of the European Research Council (ERC) Horizon Europe grant under grant agreement number 101044048. F. Tramper and M. Garcia gratefully acknowledge support by grant PID2022-137779OB-C41, funded by the Spanish Ministry of Science, Innovation and Universities/State Agency of Research MICIU/AEI/10.13039/501100011033, and M. Garcia further acknowledges grant PID2022-140483NB-C22. J. Gomez-Mantecon acknowledgements CSIC funding by the JAE Intro program JAEINT\_24\_01658. Based on observations with the NASA/ESA Hubble Space Telescope obtained at the Space Telescope Science Institute, which is operated by the Association of Universities for Research in Astronomy, Incorporated, under NASA contract NAS5-26555. Support for Program number GO-15967 was provided through a grant from the STScI under NASA contract NAS5-26555. This work was supported by a NASA Keck PI Data Award, administered by the NASA Exoplanet Science Institute. Data presented herein were obtained at the W. M. Keck Observatory from telescope time allocated to the National Aeronautics and Space Administration through the agency's scientific partnership with the California Institute of Technology and the University of California. The Observatory was made possible by the generous financial support of the W. M. Keck Foundation.
\end{acknowledgements}

\bibliographystyle{aa}
\bibliography{references}

\begin{thebibliography}{115}
\expandafter\ifx\csname natexlab\endcsname\relax\def\natexlab#1{#1}\fi

\bibitem[{Abbott(1982)}]{abbott1982theory}
Abbott, D.~C. 1982, ApJ, 259, 282

\bibitem[{{Asplund} {et~al.}(2005){Asplund}, {Grevesse}, \& {Sauval}}]{asplund2005abundances}
{Asplund}, M., {Grevesse}, N., \& {Sauval}, A.~J. 2005, in Astronomical Society of the Pacific Conference Series, Vol. 336, Cosmic Abundances as Records of Stellar Evolution and Nucleosynthesis, ed. I.~{Barnes}, Thomas~G. \& F.~N. {Bash}, 25

\bibitem[{Backs {et~al.}(2024{\natexlab{a}})Backs, {Brands}, {de Koter}, {et~al.}}]{backs2024smc}
Backs, F., {Brands}, S.~A., {de Koter}, A., {et~al.} 2024{\natexlab{a}}, \aap, 692, A88

\bibitem[{Backs {et~al.}(2024{\natexlab{b}})Backs, Brands, Ram{\'\i}rez-Tannus, {et~al.}}]{backs2024properties}
Backs, F., Brands, S.~A., Ram{\'\i}rez-Tannus, M., {et~al.} 2024{\natexlab{b}}, A\&{A}, 690, A113

\bibitem[{Berger {et~al.}(2018)Berger, Kudritzki, Urbaneja, {et~al.}}]{berger2018quantitative}
Berger, T.~A., Kudritzki, R.-P., Urbaneja, M.~A., {et~al.} 2018, ApJ, 860, 130

\bibitem[{Bestenlehner {et~al.}(2014)Bestenlehner, Gr{\"a}fener, Vink, {et~al.}}]{bestenlehner2014vlt}
Bestenlehner, J.~M., Gr{\"a}fener, G., Vink, J.~S., {et~al.} 2014, A\&{A}, 570, A38

\bibitem[{Bj{\"o}rklund {et~al.}(2021)Bj{\"o}rklund, Sundqvist, Puls, \& Najarro}]{bjorklund2021new}
Bj{\"o}rklund, R., Sundqvist, J., Puls, J., \& Najarro, F. 2021, A\&{A}, 648, A36

\bibitem[{Boggs \& Rogers(1990)}]{boggs1990orthogonal}
Boggs, P.~T. \& Rogers, J.~E. 1990, Contemporary mathematics, 112, 183

\bibitem[{Bouret {et~al.}(2005)Bouret, Lanz, \& Hillier}]{bouret2005lower}
Bouret, J.-C., Lanz, T., \& Hillier, D. 2005, A\&{A}, 438, 301

\bibitem[{{Bouret} {et~al.}(2015){Bouret}, {Lanz}, {Hillier}, {et~al.}}]{bouret2015spectro}
{Bouret}, J.~C., {Lanz}, T., {Hillier}, D.~J., {et~al.} 2015, MNRAS, 449, 1545

\bibitem[{Brands {et~al.}(2025)Brands, Backs, de~Koter, {et~al.}}]{brands2025xshootingullysesmassivestars}
Brands, S.~A., Backs, F., de~Koter, A., {et~al.} 2025, arXiv e-prints, arXiv:2503.14687

\bibitem[{Brands {et~al.}(2022)Brands, de~Koter, Bestenlehner, {et~al.}}]{brands2022r136}
Brands, S.~A., de~Koter, A., Bestenlehner, J.~M., {et~al.} 2022, A\&{A}, 663, A36

\bibitem[{Bresolin {et~al.}(2006)Bresolin, Pietrzy{\'n}ski, Urbaneja, {et~al.}}]{bresolin2006araucaria}
Bresolin, F., Pietrzy{\'n}ski, G., Urbaneja, M.~A., {et~al.} 2006, ApJ, 648, 1007

\bibitem[{Bresolin {et~al.}(2007)Bresolin, Urbaneja, Gieren, {et~al.}}]{bresolin2007vlt}
Bresolin, F., Urbaneja, M.~A., Gieren, W., {et~al.} 2007, ApJ, 671, 2028

\bibitem[{Brott {et~al.}(2011)Brott, de~Mink, Cantiello, {et~al.}}]{brott2011rotating}
Brott, I., de~Mink, S.~E., Cantiello, M., {et~al.} 2011, {A}\&{A}, 530, A115

\bibitem[{Cantiello {et~al.}(2009)Cantiello, Langer, Brott, {et~al.}}]{cantiello2009sub}
Cantiello, M., Langer, N., Brott, I., {et~al.} 2009, A\&{A}, 499, 279

\bibitem[{Carneiro {et~al.}(2016)Carneiro, Puls, Sundqvist, \& Hoffmann}]{carneiro2016atmospheric}
Carneiro, L.~P., Puls, J., Sundqvist, J., \& Hoffmann, T. 2016, A\&{A}, 590, A88

\bibitem[{Castor {et~al.}(1975)Castor, Abbott, \& Klein}]{castor1975radiation}
Castor, J.~I., Abbott, D.~C., \& Klein, R.~I. 1975, ApJ, 195, 157

\bibitem[{Chlebowski {et~al.}(1989)Chlebowski, Harnden~Jr, \& Sciortino}]{chlebowski1989einstein}
Chlebowski, T., Harnden~Jr, F., \& Sciortino, S. 1989, \apj, 341, 427

\bibitem[{Crowther {et~al.}(2022)Crowther, Broos, Townsley, {et~al.}}]{crowther2022x}
Crowther, P.~A., Broos, P.~S., Townsley, L.~K., {et~al.} 2022, \mnras, 515, 4130

\bibitem[{Dalcanton {et~al.}(2009)Dalcanton, Williams, Seth, {et~al.}}]{dalcanton2009acs}
Dalcanton, J.~J., Williams, B.~F., Seth, A.~C., {et~al.} 2009, ApJS, 183, 67

\bibitem[{Davies {et~al.}(2007)Davies, Vink, \& Oudmaijer}]{davies2007modelling}
Davies, B., Vink, J.~S., \& Oudmaijer, R.~D. 2007, A\&{A}, 469, 1045

\bibitem[{Evans {et~al.}(2007)Evans, Bresolin, Urbaneja, {et~al.}}]{evans2007araucaria}
Evans, C., Bresolin, F., Urbaneja, M., {et~al.} 2007, ApJ, 659, 1198

\bibitem[{Fitzpatrick(1999)}]{fitzpatrick1999correcting}
Fitzpatrick, E.~L. 1999, PASP, 111, 63

\bibitem[{Fullerton {et~al.}(2006)Fullerton, Massa, \& Prinja}]{fullerton2006discordance}
Fullerton, A.~W., Massa, D., \& Prinja, R. 2006, ApJ, 637, 1025

\bibitem[{Garcia \& Herrero(2013)}]{garcia2013young}
Garcia, M. \& Herrero, A. 2013, A\&{A}, 551, A74

\bibitem[{Garcia {et~al.}(2014)Garcia, Herrero, Najarro, {et~al.}}]{garcia2014winds}
Garcia, M., Herrero, A., Najarro, F., {et~al.} 2014, ApJ, 788, 64

\bibitem[{Garcia {et~al.}(2009)Garcia, Herrero, Vicente, Castro, {et~al.}}]{garcia2009young}
Garcia, M., Herrero, A., Vicente, B., Castro, N., {et~al.} 2009, A\&{A}, 502, 1015

\bibitem[{Geen {et~al.}(2015)Geen, Rosdahl, Blaizot, Devriendt, \& Slyz}]{geen2015detailed}
Geen, S., Rosdahl, J., Blaizot, J., Devriendt, J., \& Slyz, A. 2015, MNRAS, 448, 3248

\bibitem[{{G{\"u}del} {et~al.}(2008){G{\"u}del}, {Briggs}, {Montmerle}, {Audard}, {Rebull}, \& {Skinner}}]{2008Sci...319..309G}
{G{\"u}del}, M., {Briggs}, K.~R., {Montmerle}, T., {et~al.} 2008, Science, 319, 309

\bibitem[{Hainich {et~al.}(2019)Hainich, Ramachandran, Shenar, {et~al.}}]{hainich2019powr}
Hainich, R., Ramachandran, V., Shenar, T., {et~al.} 2019, A\&{A}, 621, A85

\bibitem[{Hawcroft {et~al.}(2024{\natexlab{a}})Hawcroft, Mahy, Sana, {et~al.}}]{hawcroft2024empirical}
Hawcroft, C., Mahy, L., Sana, H., {et~al.} 2024{\natexlab{a}}, A\&{A}, 690, A126

\bibitem[{Hawcroft {et~al.}(2024{\natexlab{b}})Hawcroft, Sana, Mahy, Sundqvist, de~Koter, Crowther, Bestenlehner, Brands, David-Uraz, Decin, {et~al.}}]{hawcroft2024vinf}
Hawcroft, C., Sana, H., Mahy, L., {et~al.} 2024{\natexlab{b}}, A\&{A}, 688, A105

\bibitem[{Hawcroft {et~al.}(2021)Hawcroft, Sana, Mahy, {et~al.}}]{hawcroft2021empirical}
Hawcroft, C., Sana, H., Mahy, L., {et~al.} 2021, A\&{A}, 655, A67

\bibitem[{Heap {et~al.}(2006)Heap, Lanz, \& Hubeny}]{heap2006fundamental}
Heap, S.~R., Lanz, T., \& Hubeny, I. 2006, ApJ, 638, 409

\bibitem[{Herrero {et~al.}(2012)Herrero, Garcia, Puls, {et~al.}}]{herrero2012peculiar}
Herrero, A., Garcia, M., Puls, J., {et~al.} 2012, A\&{A}, 543, A85

\bibitem[{Herrero {et~al.}(2010)Herrero, Garcia, Uytterhoeven, {et~al.}}]{herrero2010nature}
Herrero, A., Garcia, M., Uytterhoeven, K., {et~al.} 2010, A\&{A}, 513, A70

\bibitem[{{Herrero} {et~al.}(1992){Herrero}, {Kudritzki}, {Vilchez}, {Kunze}, {Butler}, \& {Haser}}]{1992A&A...261..209H}
{Herrero}, A., {Kudritzki}, R.~P., {Vilchez}, J.~M., {et~al.} 1992, \aap, 261, 209

\bibitem[{Hillier \& Miller(1998)}]{hillier1998treatment}
Hillier, D.~J. \& Miller, D.~L. 1998, ApJ, 496, 407

\bibitem[{Hirano {et~al.}(2014)Hirano, Hosokawa, Yoshida, {et~al.}}]{hirano2014one}
Hirano, S., Hosokawa, T., Yoshida, N., {et~al.} 2014, ApJ, 781, 60

\bibitem[{Hosek {et~al.}(2014)Hosek, Kudritzki, Bresolin, {et~al.}}]{hosek2014quantitative}
Hosek, M.~W., Kudritzki, R.-P., Bresolin, F., {et~al.} 2014, ApJ, 785, 151

\bibitem[{{Huenemoerder} {et~al.}(2012){Huenemoerder}, {Oskinova}, {Ignace}, {Waldron}, {Todt}, {Hamaguchi}, \& {Kitamoto}}]{huenemoerder2012xray}
{Huenemoerder}, D.~P., {Oskinova}, L.~M., {Ignace}, R., {et~al.} 2012, \apjl, 756, L34

\bibitem[{Jacobs {et~al.}(2009)Jacobs, Rizzi, Tully, {et~al.}}]{jacobs2009extragalactic}
Jacobs, B.~A., Rizzi, L., Tully, R.~B., {et~al.} 2009, AJ, 138, 332

\bibitem[{{Kobayashi} {et~al.}(2020){Kobayashi}, {Karakas}, \& {Lugaro}}]{2020ApJ...900..179K}
{Kobayashi}, C., {Karakas}, A.~I., \& {Lugaro}, M. 2020, \apj, 900, 179

\bibitem[{{K{\"o}hler} {et~al.}(2015){K{\"o}hler}, {Langer}, {de Koter}, {de Mink}, {Crowther}, {Evans}, {Gr{\"a}fener}, {Sana}, {Sanyal}, {Schneider}, \& {Vink}}]{2015A&A...573A..71K}
{K{\"o}hler}, K., {Langer}, N., {de Koter}, A., {et~al.} 2015, \aap, 573, A71

\bibitem[{Krti{\v{c}}ka \& Kub{\'a}t(2018)}]{krtivcka2018global}
Krti{\v{c}}ka, J. \& Kub{\'a}t, J. 2018, A\&{A}, 612, A20

\bibitem[{{Krti{\v{c}}ka} \& {Kub{\'a}t}(2014)}]{2014A&A...567A..63K}
{Krti{\v{c}}ka}, J. \& {Kub{\'a}t}, J. 2014, \aap, 567, A63

\bibitem[{{Kudritzki} {et~al.}(1996){Kudritzki}, {Palsa}, {Feldmeier}, {Puls}, \& {Pauldrach}}]{1996rftu.proc....9K}
{Kudritzki}, R.~P., {Palsa}, R., {Feldmeier}, A., {Puls}, J., \& {Pauldrach}, A.~W.~A. 1996, in Roentgenstrahlung from the Universe, ed. H.~U. {Zimmermann}, J.~{Tr{\"u}mper}, \& H.~{Yorke}, 9--12

\bibitem[{Lamers \& Cassinelli(1999)}]{lamers1999introduction}
Lamers, H.~J. \& Cassinelli, J.~P. 1999, {Introduction to Stellar Winds} (Cambridge University Press)

\bibitem[{Lamers {et~al.}(1995)Lamers, Snow, \& Lindholm}]{lamers1995terminal}
Lamers, H.~J., Snow, T.~P., \& Lindholm, D.~M. 1995, ApJ, 455, 269

\bibitem[{{Lamers} {et~al.}(1987){Lamers}, {Cerruti-Sola}, \& {Perinotto}}]{1987ApJ...314..726L}
{Lamers}, H.~J.~G.~L.~M., {Cerruti-Sola}, M., \& {Perinotto}, M. 1987, \apj, 314, 726

\bibitem[{Lee {et~al.}(2005)Lee, Skillman, \& Venn}]{lee2005investigating}
Lee, H., Skillman, E.~D., \& Venn, K.~A. 2005, ApJ, 620, 223

\bibitem[{Leitherer {et~al.}(1992)Leitherer, Robert, \& Drissen}]{leitherer1992deposition}
Leitherer, C., Robert, C., \& Drissen, L. 1992, ApJ, 401, 596

\bibitem[{Long \& White(1980)}]{long1980survey}
Long, K.~S. \& White, R.~L. 1980, \apj, 239, L65

\bibitem[{{Lorenzo} {et~al.}(2022){Lorenzo}, {Garcia}, {Najarro}, {Herrero}, {Cervi{\~n}o}, \& {Castro}}]{2022MNRAS.516.4164L}
{Lorenzo}, M., {Garcia}, M., {Najarro}, F., {et~al.} 2022, \mnras, 516, 4164

\bibitem[{{Lucy} \& {Solomon}(1970)}]{lucy1970wind}
{Lucy}, L.~B. \& {Solomon}, P.~M. 1970, ApJ, 159, 879

\bibitem[{{Madau} \& {Dickinson}(2014)}]{2014ARA&A..52..415M}
{Madau}, P. \& {Dickinson}, M. 2014, \araa, 52, 415

\bibitem[{Marcolino {et~al.}(2009)Marcolino, Bouret, Martins, {et~al.}}]{marcolino2009analysis}
Marcolino, W., Bouret, J.-C., Martins, F., {et~al.} 2009, A\&{A}, 498, 837

\bibitem[{Marcolino {et~al.}(2022)Marcolino, Bouret, Rocha-Pinto, {et~al.}}]{marcolino2022wind}
Marcolino, W., Bouret, J.-C., Rocha-Pinto, H., {et~al.} 2022, MNRAS, 511, 5104

\bibitem[{Martins \& Plez(2006)}]{martins2006ubvjhk}
Martins, F. \& Plez, B. 2006, A\&{A}, 457, 637

\bibitem[{Martins {et~al.}(2004)Martins, Schaerer, Hillier, \& Heydari-Malayeri}]{martins2004puzzling}
Martins, F., Schaerer, D., Hillier, D.~J., \& Heydari-Malayeri, M. 2004, A\&{A}, 420, 1087

\bibitem[{Martins {et~al.}(2005)Martins, Schaerer, Hillier, {et~al.}}]{martins2005stars}
Martins, F., Schaerer, D., Hillier, D.~J., {et~al.} 2005, A\&{A}, 441, 735

\bibitem[{McQuinn {et~al.}(2015)McQuinn, Skillman, Dolphin, {et~al.}}]{mcquinn2015leo}
McQuinn, K.~B., Skillman, E.~D., Dolphin, A., {et~al.} 2015, ApJ, 812, 158

\bibitem[{Mokiem {et~al.}(2007)Mokiem, de~Koter, Vink, {et~al.}}]{mokiem2007empirical}
Mokiem, M., de~Koter, A., Vink, J., {et~al.} 2007, A\&{A}, 473, 603

\bibitem[{{Mokiem} {et~al.}(2005){Mokiem}, {de Koter}, {Puls}, \& other}]{mokiem2005ga}
{Mokiem}, M.~R., {de Koter}, A., {Puls}, J., \& other. 2005, \aap, 441, 711

\bibitem[{{Mokiem} {et~al.}(2007){Mokiem}, {de Koter}, {Vink}, {Puls}, {Evans}, {Smartt}, {Crowther}, {Herrero}, {Langer}, {Lennon}, {Najarro}, \& {Villamariz}}]{2007A&A...473..603M}
{Mokiem}, M.~R., {de Koter}, A., {Vink}, J.~S., {et~al.} 2007, \aap, 473, 603

\bibitem[{Muijres {et~al.}(2012)Muijres, Vink, de~Koter, {et~al.}}]{muijres2012predictions}
Muijres, L., Vink, J.~S., de~Koter, A., {et~al.} 2012, A\&{A}, 537, A37

\bibitem[{{M{\"u}ller} \& {Vink}(2008)}]{2008A&A...492..493M}
{M{\"u}ller}, P.~E. \& {Vink}, J.~S. 2008, \aap, 492, 493

\bibitem[{{Owocki} \& {Rybicki}(1984)}]{owocki1984ldi}
{Owocki}, S.~P. \& {Rybicki}, G.~B. 1984, \apj, 284, 337

\bibitem[{{Owocki} \& {Rybicki}(1985)}]{owocki1985ldi}
{Owocki}, S.~P. \& {Rybicki}, G.~B. 1985, \apj, 299, 265

\bibitem[{{Pauli} {et~al.}(2022){Pauli}, {Oskinova}, {Hamann}, {et~al.}}]{pauli2022binary}
{Pauli}, D., {Oskinova}, L.~M., {Hamann}, W.~R., {et~al.} 2022, \aap, 659, A9

\bibitem[{Pietrzy{\'n}ski {et~al.}(2006)Pietrzy{\'n}ski, Gieren, Soszy{\'n}ski, {et~al.}}]{pietrzynski2006araucaria}
Pietrzy{\'n}ski, G., Gieren, W., Soszy{\'n}ski, I., {et~al.} 2006, ApJ, 642, 216

\bibitem[{Poelarends {et~al.}(2008)Poelarends, Herwig, Langer, \& Heger}]{poelarends2008supernova}
Poelarends, A., Herwig, F., Langer, N., \& Heger, A. 2008, ApJ, 675, 614

\bibitem[{Puls {et~al.}(1996)Puls, {Kudritzki}, {Herrero}, {et~al.}}]{puls1996dmom}
Puls, J., {Kudritzki}, R.~P., {Herrero}, A., {et~al.} 1996, \aap, 305, 171

\bibitem[{Puls {et~al.}(2000)Puls, Springmann, \& Lennon}]{puls2000radiation}
Puls, J., Springmann, U., \& Lennon, M. 2000, A\&{A}S, 141, 23

\bibitem[{Puls {et~al.}(2005)Puls, Urbaneja, Venero, {et~al.}}]{puls2005atmospheric}
Puls, J., Urbaneja, M., Venero, R., {et~al.} 2005, A\&{A}, 435, 669

\bibitem[{Puls {et~al.}(2008)Puls, Vink, \& Najarro}]{puls2008mass}
Puls, J., Vink, J.~S., \& Najarro, F. 2008, A\&{A} Rev., 16, 209

\bibitem[{Ramachandran {et~al.}(2018)Ramachandran, Hamann, Hainich, {et~al.}}]{ramachandran2018stellar}
Ramachandran, V., Hamann, W.-R., Hainich, R., {et~al.} 2018, A\&{A}, 615, A40

\bibitem[{Ramachandran {et~al.}(2019)Ramachandran, Hamann, Oskinova, {et~al.}}]{ramachandran2019testing}
Ramachandran, V., Hamann, W.-R., Oskinova, L., {et~al.} 2019, A\&{A}, 625, A104

\bibitem[{{Ramachandran} {et~al.}(2024){Ramachandran}, {Sander}, {Pauli}, {et~al.}}]{ramachandran2024bebinaries}
{Ramachandran}, V., {Sander}, A.~A.~C., {Pauli}, D., {et~al.} 2024, \aap, 692, A90

\bibitem[{Renzo {et~al.}(2017)Renzo, Ott, Shore, \& de~Mink}]{renzo2017systematic}
Renzo, M., Ott, C.~D., Shore, S.~N., \& de~Mink, S.~E. 2017, A\&{A}, 603, A118

\bibitem[{Rickard {et~al.}(2022)Rickard, Hainich, Hamann, {et~al.}}]{rickard2022stellar}
Rickard, M., Hainich, R., Hamann, W.-R., {et~al.} 2022, A\&{A}, 666, A189

\bibitem[{Rivero~Gonz{\'a}lez {et~al.}(2012)Rivero~Gonz{\'a}lez, Puls, Najarro, \& Brott}]{rivero2012nitrogen}
Rivero~Gonz{\'a}lez, J., Puls, J., Najarro, F., \& Brott, I. 2012, A\&{A}, 537, A79

\bibitem[{Roman-Duval {et~al.}(2020)Roman-Duval, Proffitt, Taylor, {et~al.}}]{roman2020ultraviolet}
Roman-Duval, J., Proffitt, C.~R., Taylor, J.~M., {et~al.} 2020, Research Notes of The American Astronomical Society, 4, 205

\bibitem[{{Sabhahit} {et~al.}(2023){Sabhahit}, {Vink}, {Sander}, \& {Higgins}}]{2023MNRAS.524.1529S}
{Sabhahit}, G.~N., {Vink}, J.~S., {Sander}, A. A.~C., \& {Higgins}, E.~R. 2023, \mnras, 524, 1529

\bibitem[{{Sana} {et~al.}(2012){Sana}, {de Mink}, {de Koter}, {et~al.}}]{sana2012binary}
{Sana}, H., {de Mink}, S.~E., {de Koter}, A., {et~al.} 2012, Science, 337, 444

\bibitem[{{Sana} {et~al.}(2024){Sana}, {Tramper}, {Abdul-Masih}, {et~al.}}]{sana2024xshootu}
{Sana}, H., {Tramper}, F., {Abdul-Masih}, M., {et~al.} 2024, \aap, 688, A104

\bibitem[{Sander {et~al.}(2015)Sander, Shenar, Hainich, {et~al.}}]{sander2015consistent}
Sander, A., Shenar, T., Hainich, R., {et~al.} 2015, A\&{A}, 577, A13

\bibitem[{Santolaya-Rey {et~al.}(1997)Santolaya-Rey, Puls, \& Herrero}]{santolaya1997atmospheric}
Santolaya-Rey, A., Puls, J., \& Herrero, A. 1997, A\&{A}, 323, 488

\bibitem[{{Silva-Farf{\'a}n} {et~al.}(2024){Silva-Farf{\'a}n}, {F{\"o}rster}, {Moriya}, {Hern{\'a}ndez-Garc{\'\i}a}, {Mu{\~n}oz Arancibia}, {S{\'a}nchez-S{\'a}ez}, {Anderson}, {Tonry}, \& {Clocchiatti}}]{2024ApJ...969...57S}
{Silva-Farf{\'a}n}, J., {F{\"o}rster}, F., {Moriya}, T.~J., {et~al.} 2024, \apj, 969, 57

\bibitem[{Skillman {et~al.}(1989)Skillman, Kennicutt, \& Hodge}]{skillman1989oxygen}
Skillman, E.~D., Kennicutt, R., \& Hodge, P. 1989, ApJ, 347, 875

\bibitem[{Skillman {et~al.}(2013)Skillman, Salzer, Berg, Pogge, {et~al.}}]{skillman2013alfalfa}
Skillman, E.~D., Salzer, J.~J., Berg, D.~A., Pogge, R.~W., {et~al.} 2013, ApJ, 146, 3

\bibitem[{Skillman {et~al.}(2003)Skillman, Tolstoy, Cole, Dolphin, Saha, Gallagher, Dohm-Palmer, \& Mateo}]{skillman2003deep}
Skillman, E.~D., Tolstoy, E., Cole, A.~A., {et~al.} 2003, ApJ, 596, 253

\bibitem[{{Smith}(2014)}]{2014ARA&A..52..487S}
{Smith}, N. 2014, \araa, 52, 487

\bibitem[{Soszy{\'n}ski {et~al.}(2006)Soszy{\'n}ski, Gieren, Pietrzy{\'n}ski, {et~al.}}]{soszynski2006araucaria}
Soszy{\'n}ski, I., Gieren, W., Pietrzy{\'n}ski, G., {et~al.} 2006, ApJ, 648, 375

\bibitem[{{Stanton} {et~al.}(2024){Stanton}, {Cullen}, {McLure}, {Shapley}, {Arellano-C{\'o}rdova}, {Begley}, {Amor{\'\i}n}, {Barrufet}, {Calabr{\`o}}, {Carnall}, {Cirasuolo}, {Dunlop}, {Donnan}, {Hamadouche}, {Liu}, {McLeod}, {Pentericci}, {Pozzetti}, {Sanders}, {Scholte}, \& {Topping}}]{2024MNRAS.532.3102S}
{Stanton}, T.~M., {Cullen}, F., {McLure}, R.~J., {et~al.} 2024, \mnras, 532, 3102

\bibitem[{Steiger(1998)}]{steiger1998rmsea}
Steiger, J.~H. 1998, Structural Equation Modeling: A Multidisciplinary Journal, 5, 411

\bibitem[{Sundqvist \& Puls(2018)}]{sundqvist2018atmospheric}
Sundqvist, J. \& Puls, J. 2018, A\&{A}, 619, A59

\bibitem[{{Sundqvist} {et~al.}(2018){Sundqvist}, {Owocki}, \& {Puls}}]{sundqvist2018clumping}
{Sundqvist}, J.~O., {Owocki}, S.~P., \& {Puls}, J. 2018, \aap, 611, A17

\bibitem[{Sz{\'e}csi {et~al.}(2022)Sz{\'e}csi, Agrawal, W{\"u}nsch, \& Langer}]{szecsi2022bonn}
Sz{\'e}csi, D., Agrawal, P., W{\"u}nsch, R., \& Langer, N. 2022, A\&{A}, 658, A125

\bibitem[{{Sz{\'e}csi} {et~al.}(2015){Sz{\'e}csi}, {Langer}, {Yoon}, {Sanyal}, {de Mink}, {Evans}, \& {Dermine}}]{2015A&A...581A..15S}
{Sz{\'e}csi}, D., {Langer}, N., {Yoon}, S.-C., {et~al.} 2015, \aap, 581, A15

\bibitem[{Tautvai{\v{s}}ien{\.e} {et~al.}(2007)Tautvai{\v{s}}ien{\.e}, Geisler, Wallerstein, {et~al.}}]{tautvaivsiene2007first}
Tautvai{\v{s}}ien{\.e}, G., Geisler, D., Wallerstein, G., {et~al.} 2007, AJ, 134, 2318

\bibitem[{Telford {et~al.}(2021)Telford, Chisholm, McQuinn, \& Berg}]{telford2021far}
Telford, O.~G., Chisholm, J., McQuinn, K.~B., \& Berg, D.~A. 2021, ApJ, 922, 191

\bibitem[{Telford {et~al.}(2024)Telford, Chisholm, Sander, {et~al.}}]{telford2024observations}
Telford, O.~G., Chisholm, J., Sander, A.~A., {et~al.} 2024, ApJ, 974, 85

\bibitem[{{Telford} {et~al.}(2023){Telford}, McQuinn, Chisholm, \& Berg}]{telford2023ionizing}
{Telford}, O.~G., McQuinn, K.~B., Chisholm, J., \& Berg, D.~A. 2023, ApJ, 943, 65

\bibitem[{Tramper {et~al.}(2011)Tramper, Sana, de~Koter, \& Kaper}]{tramper2011mass}
Tramper, F., Sana, H., de~Koter, A., \& Kaper, L. 2011, ApJL, 741, L8

\bibitem[{Tramper {et~al.}(2014)Tramper, Sana, de~Koter, {et~al.}}]{tramper2014properties}
Tramper, F., Sana, H., de~Koter, A., {et~al.} 2014, A\&{A}, 572, A36

\bibitem[{Urbaneja {et~al.}(2023)Urbaneja, Bresolin, \& Kudritzki}]{urbaneja2023metallicity}
Urbaneja, M.~A., Bresolin, F., \& Kudritzki, R.-P. 2023, ApJ, 959, 52

\bibitem[{{Vernet} {et~al.}(2011){Vernet}, {Dekker}, {D'Odorico}, {et~al.}}]{vernet2011xshooter}
{Vernet}, J., {Dekker}, H., {D'Odorico}, S., {et~al.} 2011, \aap, 536, A105

\bibitem[{Vink(2022)}]{vink2022theory}
Vink, J.~S. 2022, ARA\&{A}, 60, 203

\bibitem[{Vink {et~al.}(2001)Vink, de~Koter, \& Lamers}]{vink2001mass}
Vink, J.~S., de~Koter, A., \& Lamers, H. 2001, A\&{A}, 369, 574

\bibitem[{Vink \& Gr{\"a}fener(2012)}]{vink2012transition}
Vink, J.~S. \& Gr{\"a}fener, G. 2012, ApJL, 751, L34

\bibitem[{Vink {et~al.}(2011)Vink, Muijres, Anthonisse, {et~al.}}]{vink2011wind}
Vink, J.~S., Muijres, L., Anthonisse, B., {et~al.} 2011, A\&{A}, 531, A132

\bibitem[{Vink \& Sander(2021)}]{vink2021metallicity}
Vink, J.~S. \& Sander, A.~A. 2021, MNRAS, 504, 2051

\bibitem[{Weidner \& Vink(2010)}]{weidner2010mass}
Weidner, C. \& Vink, J.~S. 2010, A\&{A}, 524, A98

\end{thebibliography}

\begin{appendix} 
\setcounter{secnumdepth}{1}
\section{SIMBAD-resolvable names}
\label{a_sec:simbad_names}

In this section, we provide, in Table \ref{tab:simbad_names}, identifiers of the sample stars that allows one to easily find them on SIMBAD\footnote{\url{https://simbad.u-strasbg.fr/simbad/sim-fid}}.

\begin{table}[h!]
\centering
\caption{SIMBAD-resolvable target names of the sample stars.\label{tab:simbad_names}}
\begin{tabular}
{l l}
\midrule\midrule
Star Name & SIMBAD Resolvable\\
(This work) & Identifier \\
\midrule
64066 & [GHV2009] Star 64066\\
A13 & [BUG2007] A 13\\
62024 & [GHV2009] Star 62024\\
B2 & [BUG2007] B 2\\
B11 & [BUG2007] B 11\\
A15 & [BPU2006] A 15\\
A11 & [BPU2006] A 11\\
N20 & [EBU2007] 20\\
N34 & [EBU2007] 34\\
S3 & [VPW98] 1744\\
LP26 & [ECG2019] LP 26\\
\midrule
\end{tabular}
\end{table}

\section{Abundances as mass fractions}
\label{a_sec:abundance_mass}

Here we provide the best-fitting abundances for our sample stars determined by \kiwi{} in the form of mass fractions in Table \ref{tab:abun_mass}.

\begin{table}[h!]
\centering
\caption{As in Table \ref{tab:abun_number}, but for mass fractions, $X_i$, for element $i$.}
\label{tab:abun_mass}
\begin{tiny}
\begin{tabular}
{l r@{}l r@{}l r@{}l r@{}l r@{}l }
\midrule
\midrule
Star  &\multicolumn{2}{c}{$X_{\rm He}$} & \multicolumn{2}{c}{$\log X_{\rm C}$} & \multicolumn{2}{c}{$\log X_{\rm N}$} & \multicolumn{2}{c}{$\log X_{\rm O}$} & \multicolumn{2}{c}{$\log X_{\rm Si}$} \\
\midrule
64066 & $0.32$ & $^{+0.17}_{-0.12}$ & $-4.0$ & $^{+1.0}_{-0.9}$ & $-3.1$ & $^{+0.5}_{-1.9}$ & $-2.6$ & $^{+0.1}_{-1.0}$ & $-3.1$ & $^{+0.8}_{-1.3}$   \\[3pt]
A13 & $0.41$ & $^{+0.07}_{-0.01}$ & $-4.0$ & $^{+0.3}_{-0.1}$ & $-3.3$ & $^{+0.3}_{-0.2}$ & $-2.7$ & $^{+0.1}_{-0.5}$ & $-2.9$ & $^{+0.1}_{-0.3}$   \\[3pt]
62024 & $0.38$ & $^{+0.10}_{-0.12}$ & $-5.1$ & $^{+0.4}_{-0.2}$ & $-2.8$ & $^{+0.2}_{-1.0}$ & $-3.9$ & $^{+1.2}_{-1.2}$ & $-4.6$ & $^{+0.7}_{-0.3}$   \\[3pt]
B2 & $0.29$ & $^{+0.12}_{-0.09}$ & $-3.3$ & $^{+0.5}_{-0.6}$ & $-3.7$ & $^{+1.1}_{-1.4}$ & $-2.6$ & $^{+0.1}_{-1.0}$ & $-2.5$ & $^{+0.3}_{-0.8}$   \\[3pt]
B11 & $0.36$ & $^{+0.09}_{-0.09}$ & $-3.1$ & $^{+0.2}_{-0.3}$ & $-2.7$ & $^{+0.1}_{-0.5}$ & $-2.8$ & $^{+0.1}_{-0.6}$ & $-3.5$ & $^{+0.6}_{-0.3}$   \\[3pt]
A15 & $0.37$ & $^{+0.05}_{-0.04}$ & $-3.3$ & $^{+0.2}_{-0.3}$ & $-3.0$ & $^{+0.2}_{-0.2}$ & $-2.8$ & $^{+0.1}_{-1.1}$ & $-4.2$ & $^{+0.4}_{-0.3}$   \\[3pt]
A11 & $0.34$ & $^{+0.17}_{-0.16}$ & $-2.7$ & $^{+0.2}_{-0.3}$ & $-4.2$ & $^{+1.4}_{-0.5}$ & $-4.1$ & $^{+0.9}_{-1.0}$ & $-2.9$ & $^{+0.4}_{-0.5}$   \\[3pt]
N20 & $0.24$ & $^{+0.17}_{-0.05}$ & $-3.1$ & $^{+0.1}_{-0.5}$ & $-4.6$ & $^{+1.9}_{-0.5}$ & $-2.7$ & $^{+0.2}_{-0.4}$ & $-4.0$ & $^{+0.2}_{-0.4}$   \\[3pt]
N34 & $0.40$ & $^{+0.08}_{-0.12}$ & $-3.7$ & $^{+0.6}_{-1.2}$ & $-2.8$ & $^{+0.2}_{-0.9}$ & $-2.9$ & $^{+0.2}_{-1.5}$ & $-2.4$ & $^{+0.2}_{-1.6}$   \\[3pt]
S3 & $0.2$ & $^{+0.04}_{-0.02}$ & $-3.5$ & $^{+0.3}_{-0.3}$ & $-3.6$ & $^{+0.7}_{-0.4}$ & $-3.7$ & $^{+0.4}_{-1.3}$ & $-3.7$ & $^{+0.4}_{-0.4}$   \\[3pt]
LP26 & $0.19$ & $^{+0.09}_{-0.01}$ & $-4.8$ & $^{+0.7}_{-0.3}$ & $-4.1$ & $^{+0.9}_{-1.0}$ & $-3.9$ & $^{+1.4}_{-1.1}$ & $-5.1$ & $^{+0.8}_{-0.2}$   \\[3pt]\midrule
\end{tabular}
\end{tiny}
\end{table}

\section{Diagnostic spectral line list}
\label{a_sec:linelist}
In this section, we provide, in Table \ref{tab:linelist}, a list of the spectral lines used in this work to determine the stellar parameters of our sample stars. In some stars, \ciii\,$\lambda 4070$, \civ\,$\lambda 5801$, and \niv\,$\lambda 6380$ were modelled, but these were not present in the observed spectra and best-fitting \fw{} models, and had no influence on parameter determination, so these profiles are not shown in the plots that show the best fits.
\begin{table}
\centering
\caption{The spectral lines examined in this work. Within each line complex, we show the lines of each ion considered.\label{tab:linelist}}
\begin{tabular}{lll}
\midrule
\midrule
Ion & Wavelength (\AA) & Complex \\
\midrule
C\,{\sc iv} & 1168.9, 1169.0 & C\,{\sc iv}\,$\lambda$1169$^*$ \\
C\,{\sc iii} & 1174.9, 1175.3, 1175.6, & C\,{\sc iii}\,$\lambda$1176$^*$ \\
 &  1175.7, 1176.0, 1176.4 & \\
O\,{\sc iv} & 1338.6, 1343.0, 1343.5 & O\,{\sc iv}\,$\lambda$1340 \\
Si\,{\sc iv} & 1393.8, 1402.8 & Si\,{\sc iv}\,$\lambda$1400 \\
C\,{\sc iv} & 1548.2, 1550.8 & C\,{\sc iv}\,$\lambda$1550 \\
He\,{\sc ii} & 1640.4 & He\,{\sc ii}\,$\lambda$1640 \\
N\,{\sc iv} & 1718.6 & N\,{\sc iv}\,$\lambda$1718 \\
\midrule
He\,{\sc ii} & 4025.4 & He\,{\sc i}\,$\lambda$4026 \\
He\,{\sc i} & 4026.2 & He\,{\sc i}\,$\lambda$4026 \\
N\,{\sc iii} & 4379.0, 4379.2 & He\,{\sc i}\,$\lambda$4387 \\
He\,{\sc i} & 4387.9 & He\,{\sc i}\,$\lambda$4387 \\
He\,{\sc i} & 4471.5 & He\,{\sc i}\,$\lambda$4471 \\
He\,{\sc i} & 4921.9 & He\,{\sc i}\,$\lambda$4922 \\
He\,{\sc i} & 5875.6 & He\,{\sc i}\,$\lambda$5875 \\
N\,{\sc iii} & 4097.4, 4103.4 & He\,{\sc ii}\,$\lambda$4200 \\
He\,{\sc ii} & 4199.6 & He\,{\sc ii}\,$\lambda$4200 \\
N\,{\sc iii} & 4534.6 & He\,{\sc ii}\,$\lambda$4541 \\
He\,{\sc ii} & 4541.4 & He\,{\sc ii}\,$\lambda$4541 \\
He\,{\sc ii} & 4685.6 & He\,{\sc ii}\,$\lambda$4686 \\
He\,{\sc ii} & 5411.3 & He\,{\sc ii}\,$\lambda$5411 \\
O\,{\sc iii} & 3961.6 & H$\epsilon^\dagger$ \\
He\,{\sc i} & 3964.7 & H$\epsilon^\dagger$ \\
H\,{\sc i} & 3970.1 & H$\epsilon^\dagger$ \\
Si\,{\sc iv} & 4088.9, 4116.1 & H$\delta$ \\
N\,{\sc iii} & 4097.4, 4103.4 & H$\delta$ \\
He\,{\sc ii} & 4099.9 & H$\delta$ \\
H\,{\sc i} & 4101.7 & H$\delta$ \\
He\,{\sc ii} & 4338.7 & H$\gamma$ \\
H\,{\sc i} & 4340.5 & H$\gamma$ \\
N\,{\sc iii} & 4858.7, 4859.0, 4861.1, & H$\beta$ \\
&  4867.1, 4867.2, 4873.6 & \\
He\,{\sc ii} & 4859.1 & H$\beta$ \\
H\,{\sc i} & 4861.4 & H$\beta$ \\
He\,{\sc ii} & 6559.8 & H$\alpha^\dagger$ \\
H\,{\sc i} & 6562.8 & H$\alpha^\dagger$ \\
N\,{\sc iii} & 4634.1, 4640.6, 4641.9 & C\,{\sc iii} N\,{\sc iii} 46 \\
C\,{\sc iii} & 4647.4, 4650.2, 4651.5 & C\,{\sc iii} N\,{\sc iii} 46 \\
C\,{\sc iii} & 4068.90, 4070.26 & C\,{\sc iii}\,$\lambda$4070$^\ddag$ \\
C\,{\sc iv} & 5801.3, 5812.0 & C\,{\sc iv}\,$\lambda$5801$^{**}$ \\
N\,{\sc iv} & 6380.8 & N\,{\sc iv}\,$\lambda$6380$^{\dagger\dagger}$ \\
\midrule
\end{tabular}
\tablefoot{$^*$ These profiles are blended in N20. $^\dagger$ Not available in the spectra of A15, S3, and LP26. $^\ddag$ Modelled in all but A15 and A11. $^{**}$ Modelled in A13, 62024, B11, N20, and N34 only. $^{\dagger\dagger}$ Modelled in A13 only.}
\end{table}

\section{Comments on individual targets}
\label{a_sec:comments}
\subsection{64066}
\label{a_subsec:64066}
64066 is the only star without a G160M spectrum, meaning the \civ\,$\lambda 1550$ line could not be modelled. Due to our modelling approach, in that we chose not to include the \nv\,$\lambda 1240$ line due to its sensitivity to X-rays, no strong wind signatures were present in the UV, so upper limits for \mdot{} were determined from optical recombination lines.

As there were no strong wind signatures present in the available UV resonance lines, we had to assume values for \vinf{} and $\beta$. For \vinf, which is constrained by the absorption troughs of P Cygni profiles, we took the bluemost edge of the \nv\,$\lambda 1240$ line as an estimate, which gave $\SI{2000}{\kilo\meter\per\second}$, and adopt conservative uncertainty estimates of $\pm \SI{200}{\kilo\meter\per\second}$. For $\beta$, which is constrained from the shape of a P Cygni profile, we assumed a value of 0.95. Overall, the fit is good as all lines are reproduced within uncertainties (Fig. \href{https://zenodo.org/records/15078070}{H.1}). 

The only time 64066 was analysed before was in \cite{herrero2012peculiar}. In this study, they only determine a lower limit on the effective temperature of $\SI{49}{\kilo\kelvin}$, and $\log g = 3.8$ from the optical spectrum alone. We find significantly lower \teff{} and \logg{} values than \cite{garcia2013young} of $38^{+4.0}_{-2.8}\,{\rm Kk}$ and $3.32^{+0.26}_{-0.12}$ respectively. Furthermore, they used a grid search approach, where the models of the grid had a fixed $\xi=\SI{10}{\kilo\meter\per\second}$. From our combined optical and UV analysis, we find a factor of two higher value of this quantity.

\subsection{A13}
\label{a_subsec:A13}

A13 has the highest quality spectrum of the entire sample in that it has a \civ\,$\lambda 1550$ feature in the form of a developed P Cygni profile, a \niv\,$\lambda 1718$ feature showing a wind signature, and strong \heii{} lines in the optical. We therefore used this star to converge on our modelling approach.

Overall, the fit is good (Fig. \ref{fig:example_fit}). As for the UV fit, one of the components of the \ciii\,$\lambda 1176$ complex at $\sim \SI{1175.75}{\angstrom}$ is not reproduced and the bottom of the absorption trough \civ\,$\lambda 1550$ is not reproduced by our best-fit models (however this is due to fixing $v_{\rm windturb}$; see Sect. \ref{subsec:assumptions} for a discussion of this). As for the optical spectrum, the \heii\,$\lambda 4686$ of our best-fit model is too deep. Given the rest of the He lines are well reproduced, this line is probably sensitive to some other assumptions we made, perhaps where the onset of clumping begins.

Our best-fit parameters are generally in line with those obtained by \cite{bouret2015spectro}, although we find it to be slightly hotter. Our \vinf{} values are consistent with both theirs and that determined by \cite{garcia2014winds}. 

\subsection{62024}
\label{a_subsec:62024}

The optical spectrum of this star was previously analysed by \cite{herrero2012peculiar}. Through the analysis of the \heii\,$\lambda 4686$ profile that was in the form of a P Cygni profile in their low resolution VIMOS spectrum, they found a high value of $\beta = 2.0$ for this star, as anything lower failed to reproduce this profile. From this, they conclude that 62024 may be a fast rotator seen pole on due to different wind conditions between the equator and the pole that are captured in this high $\beta$ value.

We also find a high $\beta$ value, consistent with a value of 2. We also find a relatively high $v\sin i = \SI{160(55)}{\kilo\meter\per\second}$. However, we have neglected macroturbulent broadening effects, which may become significant in stars with a sufficiently high $\Gamma_{\rm Edd}$. We therefore cannot disprove this hypothesis that it is a fast rotator.

Our values for $\beta$ and \vinf{} are also in line with those determined by \cite{garcia2014winds} within uncertainties. The fit is satisfactory (Fig. \href{https://zenodo.org/records/15078070}{H.2}).   

\subsection{B2}
\label{a_subsec:B2}

The fit to B2 is satisfactory (Fig. \href{https://zenodo.org/records/15078070}{H.3}). Only the \niv\,$\lambda 1718$ feature is not properly reproduced by our best-fit. Our $\beta$ and \vinf{} determinations are consistent with the values determined by \citet[their 65426]{garcia2014winds} within uncertainties, as are our \teff{} and \logg{} values with \cite{garcia2013young}.

\subsection{B11}
\label{a_subsec:B11}

We faced issues fitting this star which have been subject to discussion in Sect. \ref{subsec:results_anomalous}, meaning it was excluded from fitting. The fit to the optical spectrum is satisfactory apart from discrepancies with the \halpha{} feature. Potentially, macroclumping effects, such as spatial and velocity porosity, may be at play in the winds of this star, as we discuss in Sect. \ref{subsubsec:thick_clumping}. The fit to this star is shown in Fig. \href{https://zenodo.org/records/15078070}{H.4}

\subsection{A15}
\label{a_subsec:A15}

This fit is satisfactory overall (Fig. \href{https://zenodo.org/records/15078070}{H.5}). All lines in the UV and optical spectrum are well reproduced by our best-fit model. The only exception is the \civ\,$\lambda 1169$ line, the profile of which is slightly too deep in the best-fit model compared to the data. We find stellar and wind parameters generally consistent with those found by \cite{telford2024observations} for this star, only we find a lower \logg.

\subsection{A11}
\label{a_subsec:A11}

We faced similar problems as B11 in the fit for A11 (Fig. \href{https://zenodo.org/records/15078070}{H.6}). This star could potentially be a binary due to an abnormally large luminosity for its spectral type, as speculated by \cite{bouret2015spectro}.

\subsection{N20}
\label{a_subsec:N20}
The fit to this star is for the most part satisfactory (Fig. \href{https://zenodo.org/records/15078070}{H.7}). The fits to the wind features -- the \civ\,$\lambda 1169$ + \ciii\,$\lambda 1176$ blend, \siiv\,$\lambda 1400$, and \civ\,$\lambda 1550$ -- are good, as are those of the hydrogen Balmer and \hei{} lines. Also sufficiently fit within uncertainties are the \heii{} lines, apart from the one at $\SI{4686}{\angstrom}$. This is not immediately evident, since the spectra have been rebinned for clarity.

The \heii\,$\lambda 4686$ is not matched by our best-fit model. Interestingly, this feature in the spectrum of \cite{tramper2011mass} and \cite{tramper2014properties} is filled in, while ours is in absorption. This anomaly cannot be explained. The `Programme ID' in the \texttt{.fits} file containing the spectrum is the same as that given in \cite{tramper2011mass} where the observation of this star was carried out. There could perhaps be a difference in data processing between our work and theirs that is not immediately evident. We note that the \texttt{PROCSOFT} entry of the header of the \texttt{.fits} file says that the `xshoo/2.3.12dev1' version of the ESO pipeline was used to reduce the spectrum contained in the \texttt{.fits} file used in this work, while \cite{tramper2011mass} state the X-Shooter pipeline
v1.2.2 was used to reduce their data.

Given the rest of the wind lines are sufficiently fit, this star is not excluded from the \dmom{} analysis.

\subsection{N34}
\label{a_subsec:N34}

The fit for N34 is shown in Fig. \href{https://zenodo.org/records/15078070}{H.8}. Overall the fit is satisfactory, as all models reproduce the observed spectrum within uncertainties. What is notable are the large uncertainties on the parameters, especially the $>1$\,dex uncertainty on \mdot. That, and the large uncertainties on $\log g$, which translate to uncertainties on $R$ and therefore \dmom, which are evident in Fig. \ref{fig:dmom_fit}. 

This large uncertainty on \mdot{} is due to the low resolution and SNR of the G140L spectrum. Closer inspection of the fit revealed that only the \siiv\,$\lambda 1400$ and \civ\,$\lambda 1550$ lines favoured $\dot{M} = \SI{e-7}{\msun\per\year}$ with a large error margin, while all optical wind features as well as the \heii\,$\lambda 1640$ in the UV were insensitive for $\dot{M} < \SI{e-6}{\msun\per\year}$. Therefore, this combination of the many optical lines where only upper limits were determined and the two UV lines where a single \mdot{} was slightly more favoured than this upper limit, resulted in a final determination with large uncertainties.

\subsection{S3}
\label{a_subsec:S3}

Fig. \href{https://zenodo.org/records/15078070}{H.9} shows the fit for S3. Given its large distance, its UV spectrum is especially noisy. It shows negligible wind features, meaning only upper limits for \mdot{} can be obtained. This can be seen from the posterior of \mdot{} in Fig. \href{https://zenodo.org/records/15078070}{H.9}. Nonetheless, the fit is satisfactory. We find that it is slightly hotter and has slightly higher \logg{} than that determined by \cite{telford2024observations}, while the rest of the parameters are generally consistent.

\subsection{LP26}
\label{a_subsec:LP26}

LP 26 is the most distant and most metal-poor star in the sample. Similar to S3, only upper limits for \mdot{} could be determined for this star. The fit, shown in Fig. \href{https://zenodo.org/records/15078070}{H.10}, is satisfactory, however the terminal velocity could not be constrained. We obtain parameters consistent with those of \cite{telford2024observations}.

\section{Mass loss rates fits}
\label{a_sec:mdot}
Here we show a figure equivalent to Fig. \ref{fig:dmom_fit} for \mdot. We find low values of $\alpha$ consistent with that obtained from fitting \dmom, and consistent values depending on whether $M_{\rm eff}$ is included in the fit or not. We also find a trend of $\dot{M}(L)$ that is steeper than that predicted by \cite{bjorklund2021new} at $Z\sim 0.14\,Z_\odot$ at low $L$.

\begin{figure}
   \centering
   \includegraphics[width=\hsize]{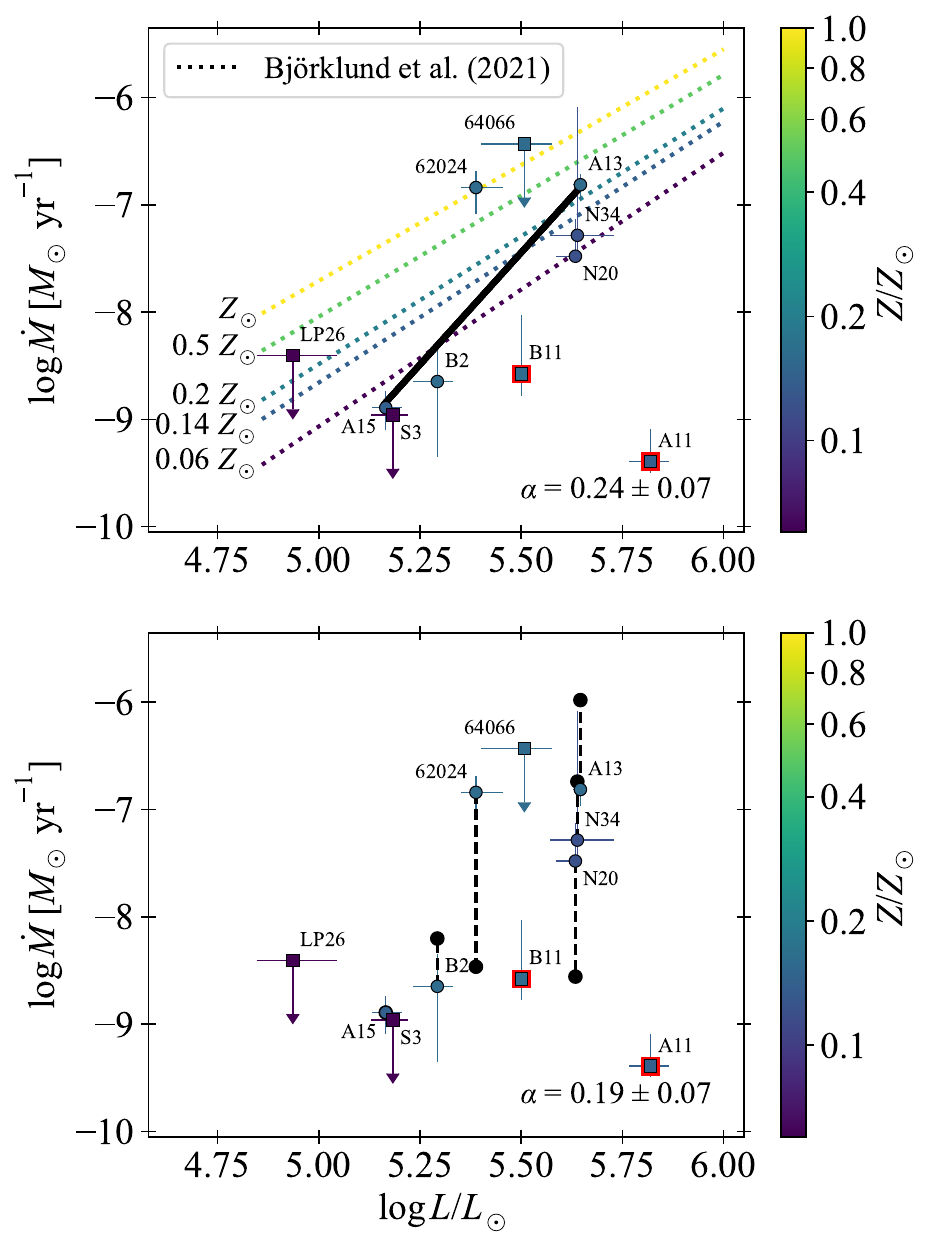}
      \caption{Fits of $\dot{M}(L)$ (top) and $\dot{M}(L,M_{\rm eff})$ (bottom) for the stars in the sample, where, as in Fig. \ref{fig:dmom_fit}, only stars with reliable \mdot{} determinations are considered. Overplotted in the top panel are the theoretical predictions of \cite{bjorklund2021new}.}
         \label{fig:mdot_preds}
\end{figure}

\section{Fits of IC\,1613 stars at different \textit{Z}}
\label{a_sec:ic1613_fits}
\begin{figure*}
   \centering
   \includegraphics[width=\hsize]{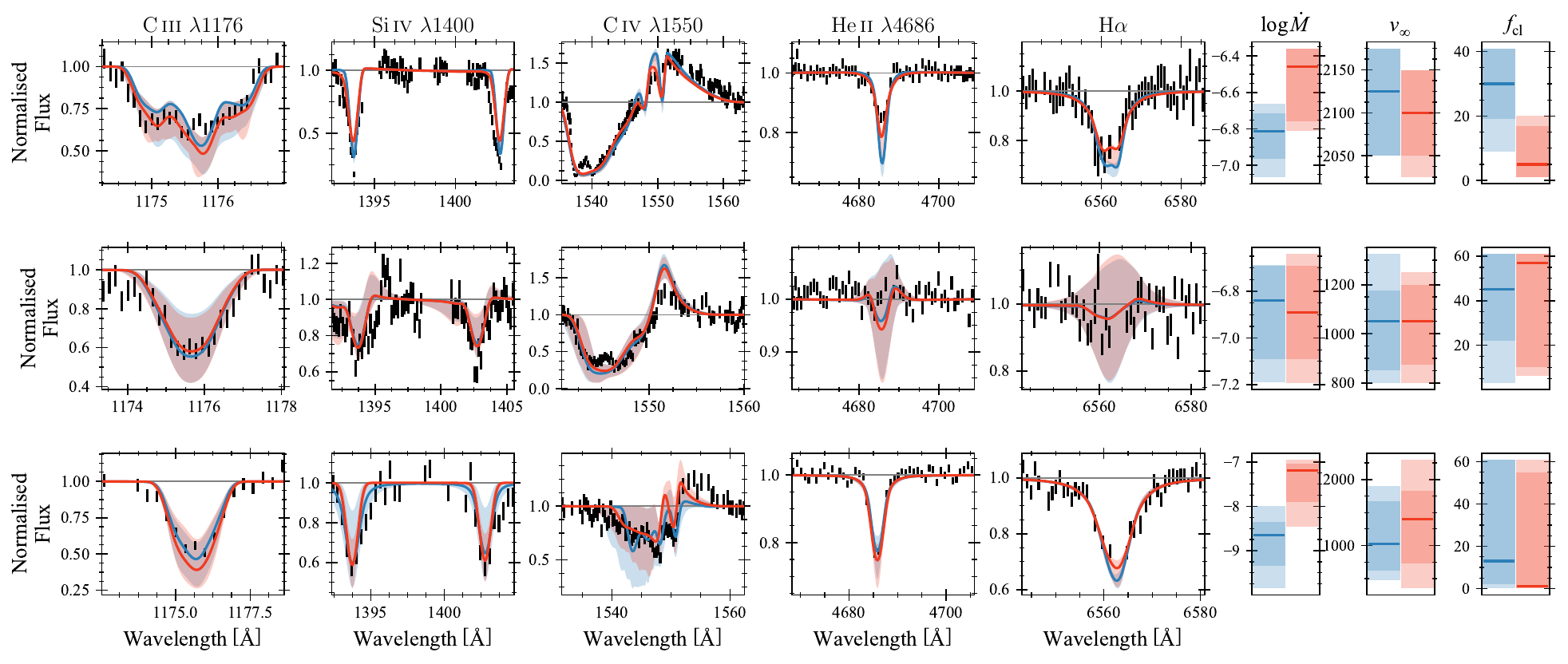}
      \caption{Comparison of \kiwi{} fits of the 3 stars in IC\,1613 considered in the fitting in Sect. \ref{subsubsec:new_z} at $\SI{0.16}{\zsun}$ (blue) and $\SI{0.2}{\zsun}$ (red) of A13 (top), 62024 (middle), and B2 (bottom). Similar to Fig. \ref{fig:thickthincompare}, we show the best-fit values and uncertainties of \mdot{} and \vinf.}
         \label{fig:lowzhighz}
\end{figure*}

Here we present a comparison of the fits of the IC\,1613 stars at the adopted metal content of the galaxy in this paper, $\SI{0.16}{\zsun}$, and \kiwi{} fits of these stars at $\SI{0.2}{\zsun}$. We show the best-fit parameters and uncertainties in Table \ref{tab:highz_ic1613_vals} and compare the best-fitting spectra in Fig. \ref{fig:lowzhighz}.

We would also like to note some of the differences between the \kiwi{} fits at higher $Z$ for these stars and those at lower $Z$ so as to explain the position of the former on the $D_{\rm mom}(L)$ vs. $L$ diagram (Fig. \ref{fig:dmom_twofits}). All stars appear higher on this diagram at higher $Z$. As for A13 and 62024, \dmom{} and $\log L$ are equal within uncertainties between the two \kiwi{} runs, as are the rest of the parameters. Therefore the same optimal local minima were found for each run.

The case of B2 is different. While $L$ has remained the same between runs, \dmom{} is substantially higher in the higher $Z$ run, which is something one would not expect. This can be explained by the carbon abundances, $\epsilon_{\rm C}$, obtained by the two \kiwi{} fits. The lower $Z$ run found $\epsilon_{\rm C} = 7.8^{+0.5}_{-0.5}$, while the higher $Z$ run found $\epsilon_{\rm C} = 6.3^{+0.3}_{-0.4}$. To compensate for this decrease in $\epsilon_{\rm C}$, a larger \mdot{} was consequently preferred. Due to the modest SNR of the spectra of the stars in this sample ($\sim 5-10$ in the UV) and degeneracies among the free parameters, these permutations of \mdot{} and $\epsilon_{\rm C}$ are seen as equally optimal by \kiwi{} and even the photospheric C lines cannot tightly constrain the C abundance.

\begin{table*}
\centering
\caption{Best-fit parameters of A13, 62024, and B2, from \kiwi{} runs with $\SI{0.2}{\zsun}$.}
\label{tab:highz_ic1613_vals}
\begin{small}
\begin{tabular}
{l r@{}l r@{}l r@{}l r@{}l r@{}l r@{}l r@{}l r@{}l r@{}l r@{}l r@{}l r@{}l}
\midrule\midrule
Star  & \multicolumn{2}{c}{$T_{\rm eff}$} & \multicolumn{2}{c}{$\log g$} & \multicolumn{2}{c}{$Y_{\rm He}$} & \multicolumn{2}{c}{$\xi$} & \multicolumn{2}{c}{$\log \dot{M}$} & \multicolumn{2}{c}{$v_{\infty}$} & \multicolumn{2}{c}{$\beta$} & \multicolumn{2}{c}{$f_{\rm cl}$} & \multicolumn{2}{c}{$\epsilon_{\rm C}$} & \multicolumn{2}{c}{$\epsilon_{\rm N}$} & \multicolumn{2}{c}{$\epsilon_{\rm O}$} & \multicolumn{2}{c}{$\epsilon_{\rm Si}$} \\
  & \multicolumn{2}{c}{[kK]} & \multicolumn{2}{c}{$[{\rm cm\,s}^{-2}]$} & \multicolumn{2}{c}{[$N_{\rm He}/N_{\rm H}$]} & \multicolumn{2}{c}{$[{\rm km\,s}^{-1}]$} & \multicolumn{2}{c}{[$M_\odot\,{\rm yr}^{-1}$]} & \multicolumn{2}{c}{$[{\rm km\,s}^{-1}]$} & \multicolumn{2}{c}{} & \multicolumn{2}{c}{} & \multicolumn{2}{c}{} & \multicolumn{2}{c}{} & \multicolumn{2}{c}{} & \multicolumn{2}{c}{} \\
\midrule
A13 & $44.5$ & $^{+1.8}_{-1.5}$ & $3.95$ & $^{+0.10}_{-0.15}$ & $0.215$ & $^{+0.005}_{-0.025}$ & $19.1$ & $^{+1.6}_{-0.3}$ & $-6.46$ & $^{+0.10}_{-0.3}$ & $2100$ & $^{+50}_{-50}$ & $1.6$ & $^{+0.02}_{-0.08}$ & $5$ & $^{+12}_{-4}$ & $7.1$ & $^{+0.2}_{-0.1}$ & $7.4$ & $^{+0.5}_{-0.0}$ & $7.8$ & $^{+0.3}_{-0.0}$ & $8.0$ & $^{+0.1}_{-0.4}$  \\[3pt]
62024 & $39.0$ & $^{+2.0}_{-2.8}$ & $3.76$ & $^{+0.08}_{-0.34}$ & $0.15$ & $^{+0.09}_{-0.06}$ & $21.0$ & $^{+0.3}_{-7.5}$ & $-6.89$ & $^{+0.20}_{-0.2}$ & $1050$ & $^{+150}_{-180}$ & $2.17$ & $^{+0.05}_{-0.55}$ & $57$ & $^{+4}_{-47}$ & $6.0$ & $^{+0.3}_{-0.0}$ & $8.3$ & $^{+0.2}_{-0.7}$ & $6.8$ & $^{+1.4}_{-0.8}$ & $6.2$ & $^{+0.6}_{-0.2}$  \\[3pt]
B2 & $38.0$ & $^{+2.0}_{-1.5}$ & $3.74$ & $^{+0.08}_{-0.2}$ & $0.09$ & $^{+0.04}_{-0.03}$ & $18.5$ & $^{+2.8}_{-4.7}$ & $-7.2$ & $^{+0.15}_{-0.7}$ & $1400$ & $^{+420}_{-680}$ & $1.78$ & $^{+0.24}_{-0.48}$ & $1$ & $^{+54}_{-0}$ & $6.3$ & $^{+0.4}_{-0.3}$ & $7.1$ & $^{+0.6}_{-0.4}$ & $8.2$ & $^{+0.3}_{-1.2}$ & $6.5$ & $^{+0.5}_{-0.5}$  \\[3pt]
\midrule
\end{tabular}
\end{small}
\end{table*}

\end{appendix}
\end{document}